\documentclass[12pt, a4paper]{article}
\usepackage[T1]{fontenc}

\usepackage[a4paper, left=1in, top=1in, right=1in, bottom=.8in, nohead, includefoot]{geometry}
\usepackage{charter, graphicx, natbib, setspace, multirow, amsmath, amssymb, amsfonts, enumitem}

\usepackage{comment}
\usepackage[dvipsnames]{xcolor}
\usepackage{bm}
\usepackage{graphicx}
\usepackage{diagbox}
\usepackage{array}
\usepackage{setspace}
\usepackage{rotating}
\usepackage{multirow}
\usepackage{booktabs}
\usepackage{color}
\usepackage{setspace}
\usepackage{enumitem}
\usepackage{lscape}
\usepackage[colorlinks=true, urlcolor=blue, linkcolor=blue, citecolor=blue]{hyperref}
\usepackage{rotating}

\usepackage{sectsty}
\allsectionsfont{\normalsize}

\makeatletter
\renewcommand\theequation{\thesection.\arabic{equation}}
\@addtoreset{equation}{section}
\makeatother

\usepackage{caption}
\usepackage[labelformat=simple]{subcaption}
\makeatletter\let\p@subfigure\thefigure\makeatother

\usepackage{cleveref}
\crefname{chapter}{Chapter}{Chapters}
\crefname{section}{Section}{Sections}
\crefname{subsection}{Subsection}{Subsections}
\crefname{subsubsection}{Section}{Sections}
\crefname{figure}{Figure}{Figures}
\crefname{table}{Table}{Tables}
\crefname{equation}{Equation}{Equations}
\crefname{appendix}{Appendix}{Appendices}

\usepackage[title,titletoc]{appendix}
\makeatletter

\renewenvironment{appendices}{%
    \begin{oldappendices}%
    \renewcommand{\thefigure}{\ifnum \c@section>\z@ \thesection.\fi\@arabic\c@figure}%
    \@addtoreset{figure}{section}%
    \renewcommand{\thetable}{\ifnum \c@section>\z@ \thesection.\fi\@arabic\c@table}%
    \@addtoreset{table}{section}%
}{%
    \end{oldappendices}%
}\makeatother

\newcommand{\N}{\mathcal{N}}

\newcommand{\G}{\mathcal{G}}

\newcommand{\Pee}{\mathcal{P}}

\usepackage{colortbl}
\definecolor{greyc}{rgb}{0.9,0.9,0.9}
\definecolor{greenc}{rgb}{0.8,1,0.8}
\definecolor{redc}{rgb}{1,0.8,0.8}

\def\equationautorefname~#1\null{
Eq.~(#1)\null
}

\newcolumntype{d}[1]{D{.}{.}{#1}}

\title{\textbf{\Large Model instability in predictive exchange rate regressions} 
}
\usepackage{authblk}

\author[1,2]{Niko Hauzenberger\footnote{Corresponding author. Address: Vienna University of Economics and Business (WU), Welthandelsplatz 1, 1020 Wien, Austria. Email: \href{mailto:niko.hauzenberger@wu.ac.at}{niko.hauzenberger@wu.ac.at}.}}
\author[2]{Florian Huber
\footnote{The authors would like to thank Michael Pfarrhofer, Maximilian B{\"o}ck and Gregor Zens for valuable comments and suggestions, and excellent discussions on this topic. The authors gratefully acknowledge financial support from the Austrian National Bank, Jubilaeumsfond grant no. $17650$.}}
\affil[1]{WU Vienna University of Economics and Business}
\affil[2]{Salzburg Centre of European Union Studies (SCEUS), Paris Lodron University of Salzburg}



\begin{document}
\graphicspath{{Figs/}}
\maketitle

\begin{abstract}
\noindent
In this paper we aim to improve existing empirical exchange rate models by accounting for uncertainty with respect to the underlying structural representation. Within a flexible Bayesian non-linear time series framework, our modeling approach assumes  that different regimes are characterized by commonly used structural exchange rate models, with their evolution being driven by a Markov process. We assume a time-varying transition probability matrix with transition probabilities depending on a measure of the monetary policy stance of the central bank at the home and foreign country. We apply this model to a set of eight exchange rates against the US dollar.  In a forecasting exercise, we show that model evidence varies over time and a model approach that takes this empirical evidence seriously yields improvements in accuracy of density forecasts for most currency pairs considered. 
\end{abstract}

\begin{tabular}{p{1\hsize}p{1\hsize}} 
\end{tabular}

\newpage

\section{Introduction}
\label{sec:intro}
Since the end of the Bretton Woods system in 1971, economists have been confronted with the challenging issue of designing empirical models of bilateral exchange rates  which are also useful for forecasting applications. In a seminal contribution, \cite{meese1983empirical} provide some early evidence that exchange rates are difficult to predict, at least in the short-run. Using a set of theoretical models in the spirit of \cite{frankel1979mark, dornbusch1976expectations, hooper1982fluctuations} that guide the choice of covariates included in the forecasting regression, \cite{meese1983empirical} find that a simple random walk benchmark is difficult to outperform for most major exchange rate pairs. One reason for the dismal performance of most empirical and structural models is that, within a standard asset pricing framework, the high persistence of the underlying fundamentals in light of a  discount factor near unity translates into highly persistent exchange rates. One consequence is that a random walk appears to be a benchmark extremely hard to beat \citep[see][]{EngelWest2005}. 

Over the years, a plethora of alternative econometric techniques emerged that provide more sophisticated means for analyzing exchange rate data to successfully improving longer-term predictions. The literature on unit roots and cointegration, for example, opened the way for tools to explicitly discriminate between short-term movements of a given currency pair and its long-run behavior. \cite{mark1995exchange}, for instance, applies an error correction model to a set of four exchange rates against the US dollar. Within this error correction framework, the exchange rate is assumed to return to its long-run equilibrium value determined by a simple monetary model eventually, with short-run fluctuations determined by own lagged values of the exchange rate and its fundamentals.  The finding that exchange rates tend to be predictable in the medium- and long-run sparked a series of related contributions that corroborate this result for different periods and currency pairs \citep{groen2000monetary,mark2001nominal, rapach2002testing}.

More recently, several studies emphasized the usefulness of accounting for non-linearities in the underlying econometric models to provide more precise exchange rate predictions \citep[see, for example,][]{canova1993modelling, sarno2004monetary,mark2009changing,byrne2016exchange,huber2016forecasting,huber2017structural, tbvecm_huber}. These non-linearities may relate to movements in the error variances of the models or to changes in the regression coefficients over time. \cite{byrne2016exchange} assess whether exchange rate predictions can be improved by using time-varying parameter models for a set of competing models with variable choice guided by a set of theoretical models.

The majority of the above literature deals with the question on whether a given empirical model that is loosely based on an underlying structural model outperforms a set of competing models. However, another key source of non-linearities could stem from the fact that the underlying theoretical model changes over time, potentially jeopardizing the predictive fit of the econometric specification.\footnote{For recent contributions that deal with this issue, see \cite{wright2008bayesian, beckmann2016forecasting, byrne2018sources, korobilis2018learning}.} For instance, the recent success of Taylor rule based models \citep[see][]{EngelWest2006, molodtsova2008taylor, MolodtsovaPapell2009, molodtsova2011taylor} can be attributed to the fact that the involved central banks did actually follow a policy rule that is closely related to a Taylor rule. With short-term interest rates, however, reaching the zero lower bound (ZLB) and central banks starting to adopt unconventional monetary policy measures, the question arises whether a Taylor rule still proves to be an adequate exchange rate model. In fact, recent literature on  non-linear Taylor rules suggests that during the ZLB, Taylor rule based models loose their momentum against simple random walk specifications \citep{byrne2016exchange, huber2017structural}.

In this paper, we contribute to the literature by acknowledging this empirical evidence and propose a modeling framework that is capable of handling model instability over time in a flexible manner. We allow for dynamically switching between regimes, each incorporating different empirical exchange rate models. This is achieved by proposing a Markov switching (MS) regression model with each regime being characterized by different covariates arising from structural exchange rate models. In contrast to the existing literature, which relies on dynamic Bayesian model averaging techniques, our approach is an integrated modeling device. Through the introduction of time-varying transition probabilities, it allows to assess how the likelihood that a given model is adopted for each point in time depends on some signal variable.  As  signal variables,  we adopt the (lagged) interest rate of the home and foreign country. This specification is motivated by the observation that Taylor rule fundamentals are good predictors in periods of the great moderation (with policy rates being significantly larger than zero), but are known for their weak performance after the Great Recession (characterized by policy rates close to zero).

We assess the merits of the proposed approach using a forecasting exercise for eight different exchange rates against the US dollar. By considering the resulting regime allocation and the transition probabilities, we examine whether structural models indeed tend to change and how this is related to movements in policy rates. The findings indicate that allowing for time-varying probabilities is a key feature, pointing towards a strong relationship between policy rates and the underlying transition distribution of the Markov process. 
 In terms of forecasting, we find that our proposed model improve upon the random walk for selected currencies, both in terms of point and density predictions. The improvements for point forecasts are, however, muted.   Comparing different model features reveals that a model based on a set of predictors that is based on fundamentals from various structural models combined with shrinkage priors and non-linearities (in the form of Markov switching) is also competitive.

The remainder of this paper is organized as follows. \cref{sec:theory} discusses the four structural exchange rate models adopted while \cref{sec:instability} proposes the econometric framework.  The empirical application is presented in \cref{sec:empircial}. Finally, the last section summarizes and concludes the paper. A technical appendix provides details on the estimation algorithm adopted.


\section{Theoretical exchange rate models}
\label{sec:theory}
In this section, we briefly discuss the main theoretical underpinnings to be used to guide covariate inclusion in the empirical model as well as to structurally identify the different regimes considered in our non-linear regression framework.

The point of departure for the discussion is a set of macroeconomic and financial quantities stored in an $R$-dimensional matrix $\bm{X}_t$,
\begin{equation}\label{eq:X}
\begin{aligned}
\mathbf{X}_t =& \left(
\ooalign{
$\underbrace{i_{t-1}, i_{t-1}^{*}, \pi_{t}, \pi_{t}^{*}, x_{t},  x_{t}^{*}, q_t}_{\mathbf{X}_{1t}}, \overbrace{m_{t},  m_{t}^{*}, y_{t},  y_{t}^{*}, e_t}^{\mathbf{X}_{2t}}, p_{t}, p_{t}^{*}, \underbrace{i_{t},  i_{t}^{*}}_{\mathbf{X}_{4t}}$\cr
$\phantom{i_{t-1}, i_{t-1}^{*}, \pi_{t}, \pi_{t}^{*}, x_{t},  x_{t}^{*}, q_t, m_{t},  m_{t}^{*}, y_{t},  y_{t}^{*}, {}} {\underbrace{\phantom{e_t, p_t,  p_t^{*}}}_{\bm{X}_{3t}}} $\cr}\right)',
\end{aligned}
\end{equation}
with $i_{t-1}$ denoting the lagged short-term interest rate, $\pi_t$  inflation, $x_t$  output gap, $m_t$  money supply, $y_{t}$  income, $p_t$  price level, while  the real exchange rate is denoted by  $q_t$ and the exchange rate by  $e_t$.\footnote{Asterisks denote foreign quantities. Moreover, $y_t, m_t$, $p_t$, $q_t$ and $e_t$ are measured in  logarithms. For simplicity, we suppress subset-specific intercepts.}
The subsets of $\bm X_t, \bm X_{jt}~(j=1,\dots,4)$, represent the different structural models we are going to describe next. 


\subsection*{A  taxonomy of selected models of exchange rate determination}
In the following, we provide a brief taxonomy of the theoretical models considered, depending on the theory adapted, guiding the specific partitions of $\bm X_t$.
\begin{itemize}
\item Our starting point is the model based on \textit{Taylor rule fundamentals} \citep[see][for a recent forecasting study]{MolodtsovaPapell2009}. This specification assumes that the set of predictors is given by $\bm X_{1t}$  and thus includes the  lagged short-term interest rate,  inflation and the output gap of both the home and foreign country, and the real exchange rate. This model has proved to be successful in terms of describing exchange rate movements, both in-sample \citep{EngelWest2006} and out-of-sample \citep{molodtsova2008taylor, molodtsova2011taylor}. However, one critical assumption of this model is that the central bank at home and abroad is actively pursuing a Taylor rule-type monetary policy strategy. Especially during the  recent period of the ZLB, this assumption could be violated, effectively leading to an inferior model fit.


\item The second model considered is the \textit{long-run monetary model}. The monetary model assumes that the covariates are given by $\bm X_{2t}$ and  include data on domestic and foreign money supply as well as the cross-country differences in income for a given income elasticity. As mentioned by \cite{rapach2002testing}, the long-run monetary model simply states that the price level of  the home and foreign country is determined by the money supply and the level of production. Assuming purchasing power parities (PPP) and uncovered interest rate parity (UIP), one is able to relate the change of the exchange rate to supply and demand for money.  

\item Third, we consider a model based on PPP. This model is based on using $\bm X_{3t}$, leading to a regression model that includes domestic and foreign price indices. PPP originates from the theory of one price in goods markets, which in turn implies that the real exchange rate is supposed to revert to a long-run equilibrium level determined by relative prices. If this turns out to be true, the real exchange rate is a stationary process. However, \cite{sarno2005puzzle} highlights substantial persistence in real exchange rates. The convergence towards PPP is thus slow in the long run and  real exchange rates typically display pronounced deviations from their PPP-implied fundamentals in the short run. 

\item Finally, we also augment our forecasting regression with the UIP model. By selecting $\bm X_{4t}$, this model simply establishes a relationship between the change in the exchange rate and the interest rate differential between home and abroad. Following \cite{chinn2006uirp}, UIP implies a positive one-to-one relationship between the interest rate differential and changes in the exchange rate. A positive change in the interest differential may be potentially followed by both an immediate and persistent appreciation in the short run, implying that UIP does not hold immediately. Here, we follow \cite{MolodtsovaPapell2009}, who address the UIP puzzle by not placing any restrictions on the coefficients.\footnote{See, for instance, \cite{engel2014handbook, chinn2006uirp} who observe coefficients that are less than one or even smaller than zero.}
\end{itemize}
All these models have been shown to possess some merit in terms of predictive power. However, several recent studies find remarkable heterogeneity with respect to the fundamental model adopted \citep[see, inter alia,][]{wright2008bayesian, beckmann2016forecasting, byrne2018sources, korobilis2018learning}. 
In particular, \citet{korobilis2018learning} argue that an investor is capable to adjust his strategy by means of sequential learning and incorporating new information arriving in each point in time. The authors assume this behaviour is captured best by a dynamically changing set of fundamentals depending on historical forecast performance. 

We now turn to describing our model framework that allows for dealing with issues of model instability in an intuitive way.

\section{Controlling for model instability in empirical exchange rate models}
\label{sec:instability}
In this section, we propose a  model that controls for dynamic model instability by specifying  a non-linear econometric framework.  After  summarizing the model structure in  \cref{sec:model}, we highlight the prior setup adopted in \cref{sec:priors}.
\subsection{A Markov switching model specification}\label{sec:model}
We now turn to describing the proposed Markov switching model with time-varying transition probabilities (MS-TVP). The key feature of our proposed framework is that, in general, it allows for switching between the fundamentals implied by  $K$ competing theoretical exchange rate models. We assume that  exchange rate returns $\Delta e_{t}$ follow an MS-TVP model given by
\begin{equation}
\begin{aligned}
\Delta e_{t} =& \bm{X}'_{S_{t} t-1}\bm{\beta}_{S_{t}} + \eta_{t}.
\end{aligned}
\end{equation}
Hereby,  $S_{t} \in \{1, \dots , K \}$ follows a first-order Markov process, $\bm{\beta}_{k}$ represents a vector of dimension $M_k$ that collects the state-specific coefficients of state $S_{t} = k$ while  $\eta_{t} \sim \N (0, \sigma_{S_{t}}^2 )$ is a white noise shock with regime-specific variance $\sigma_{S_{t}}^2$. Note that, each $\bm{\beta}_{k}$ may exhibit different dimensions.  We depart from the traditional literature on Markov switching models \citep[see, among many others,][]{hamilton1994, engel1994can, filardo1994, amisano2013money,  kaufmann2015k, billio2016interconnections, huber2018markov, casarin2018bayesian} by assuming that the regimes are characterized by competing structural exchange rate models, implying that different fundamentals enter the predictive exchange rate regression at different points in time.

In the spirit of \cite{belmontekoop2014} and  \cite{fruhwirth2006} we introduce a selection matrix $\bm{D}_{S_t}$ that entails switching between $K$ alternative model specifications,
\begin{equation}
\begin{aligned}
\Delta e_{t} =& \bm{X}'_{t-1} \bm{D}_{{S_{t}}} \bm{\beta} + \eta_{t},\label{eq:model}
\end{aligned}
\end{equation}
with $\bm{X}_{t-1}$ denoting an $R$-dimensional vector of the full set of economic fundamentals. We define $\bm{\beta} = (\bm{\beta}_{1}',  \dots, \bm{\beta}_{K}')'$ as a stacked $M$-dimensional vector of regime-specific coefficients with $M = \sum_{j=1}^K M_j$, and  $\bm{\sigma}^2 = (\sigma^2_{1}, \dots, \sigma^2_{K})'$ collecting the $K$ state-specific variances.

The selection matrix $\bm{D}_{k}$ of state $S_t = k$ is a $R \times M$ dimensional matrix with binary indicators that allow for choosing $\bm{\beta}_{k}$ and $\bm{X}_{kt-1}$ while zeroing out the elements in $\bm \beta$  and $\bm X_{t-1}$ associated with the remaining models. For instance, we effectively obtain the model based on Taylor rule fundamentals, characterized through $S_{t}=1$, by setting 
\begin{equation}
\bm D_{1} = \begin{pmatrix}
\bm I_7 & \hdots & \bm 0_{7\times 2} \\
\vdots &  \ddots & \vdots \\
\bm 0_{2 \times 7} & \hdots & \bm 0_{2\times 2} \\
\end{pmatrix},\label{eq: select_example}
\end{equation}
whereby $\bm 0_{i \times j}$ is a $i \times j$-dimensional matrix of zeros. Multiplying $\bm \beta$ from the left with $\bm D_{1}$ yields
\begin{equation}
\bm D_{1} \bm \beta = (\bm \beta'_1, \bm 0_{5 \times 1}', \bm 0_{3 \times 1}', \bm 0_{2 \times 1}')'.
\end{equation}
From this discussion, it is clear that the matrix $\bm D_{S_{t}}$ effectively controls the prevailing structural exchange rate model and the set of covariates to include in the state-specific regression. Notice, that an MS kitchen-sink regression is obtained by defining $\bm D_{S_{t}}$ in such a way that in each states all economic indicators are included at all points in time. 



\subsection*{Time-varying transition probabilities}
Assuming constant transition probabilities  is a standard (and potentially restrictive) assumption in  Markov switching models \citep[for economically motivated examples, see][]{filardo1994, amisano2013money, kaufmann2015k}. Both, \cite{amisano2013money} and \cite{kaufmann2015k} propose treating the transition distributions as being dependent on additional covariates. Here, and since our model features $K$ regimes, we follow \cite{kaufmann2015k} and parameterize  the transition probabilities  by a multinomial logit specification.  Given the forecasting evidence provided in the literature quoted above, we assume that the transition probabilities depend on a measure of the monetary policy stance such as the policy rate. This captures the notion that if policy rates approach the ZLB, a Taylor rule-based model might become inadequate and the likelihood of a regime-shift  could increase.

Let  $\bm{\tilde{S}}_T = (S_1, \dots, S_T)'$  denote the full history of the state vector, the multinomial likelihood reads
\begin{equation*}
P(S_t = j | S_{t-1} = k, \bm{Z}_{t}, \bm{\gamma}) = p_{kj,t} = \frac{\exp \left(\bm{Z}_{t}'\bm{\gamma}_{kj} \right)}{1 + \sum_{l = 1}^{K-1} \exp \left(\bm{Z}_{t}'\bm{\gamma}_{kl}\right)},
\end{equation*}
with category-specific regression coefficients $\bm{\gamma}_{kj} = (\gamma_{0,kj}, \dots ,\gamma_{N,kj})'$, collected in $\bm \gamma$ for all $k$ and $j$. 
Moreover, we define $\bm{Z}_{t}$ as an $N$-dimensional set of covariates. This set of covariates is given by
\begin{equation}\label{eq:XLogit}
\bm Z_t = (1, \bm z'_t, \mathcal{I}[S_{t-1}=1],\dots, \mathcal{I}[S_{t-1}=K-1])',
\end{equation}
whereby $\bm z_t$ is a vector of covariates that determine the dynamics of the transition probabilities while $\mathcal{I}(\bullet)$ denotes an indicator function that equals one if its argument is true. This implies that we capture a first-order Markov structure by including the previous states as additional regressors. Moreover, $\gamma_{0,kj}$ represents the intercept of the reference state $S_{t-1} = K$, and thus captures the corresponding time-invariant state persistence. Consistent with \cite{amisano2013money}, we let the coefficients associated with $\bm z_t$ be regime-invariant. It is worth noting that, if coefficients of $\bm z_t$ are zero, we obtain a classic fixed transition probability Markov switching model. For further convenience, define  $\bm{\tilde{Z}}_{T} = (\bm Z_1, \dots, \bm Z_{T})$ and $\bm{\tilde{z}}_{T} = (\bm z_1, \dots, \bm z_{T})$  to be the history of $\bm Z_t$ and $\bm z_t$ up to time $T$. 

The specific choice of $\bm z_t$ proves to be an important modeling decision. As mentioned above, our goal is to include a measure of the (conventional) monetary policy stance to signal a potential transition from Taylor rule-type based policy making to discretionary monetary policy actions such as quantitative easing (QE). In our case, we assume two early warning indicators $\bm{z}_{t} = (\tilde{i}_{t-1}, \tilde{i}^*_{t-1})'$, the demeaned, lagged interest rate at home and abroad. Consequently, the three-dimensional coefficient vector $\bm{\gamma}_{kj}$ determines the sensitivity of the transition probability that drives the transition
from the \textit{k}th to the \textit{j}th state. The demeaned covariates imply the centered parameterization of \cite{kaufmann2015k}, since a covariate $\bm{\tilde{z}}_{T}$ can be rewritten as a linear combination of its time-varying component $(\bm{\tilde{z}}_{T} - \bar{z})$ and its mean $\bar{z}$, which affects the time-invariant average state persistence. Demeaning covariates ensure that the time-invariant part does not depend on the scale of $\bm z_t$.

\subsection{Prior specification and estimation strategy}\label{sec:priors}

Our approach is Bayesian and this implies that we have to carefully specify priors on the parameters of the model. Here, we follow \cite{george1993ssvs, george1997ssvs} and specify a mixture of Gaussians prior on 
$\beta_{ik}$, the $i$th element of $\bm{\beta}_{k}$. 
The prior is centered on the theoretically motivated restrictions, in order to test whether these restrictions hold true. The prior mean is stored in a $M_k$-dimensional vector $\underline{\bm \beta}_{k}$ and summarized in \autoref{tab:priormean}. A priori we assume a symmetric Taylor rule with same coefficients for the home and foreign country and do not consider interest smoothing \citep[see][for a detailed discussion]{MolodtsovaPapell2009}. For the remaining models we center them on the implied long-run fundamental value. 

Formally, this prior reads
\begin{equation}
\begin{aligned}
\beta_{ik}|\delta_{ik} &\sim \N \left(\underline{\beta}_{ik}, \tau_{ik,1}^2\right)\delta_{ik} +  \N \left(\underline{\beta}_{ik}, \tau_{ik,0}^2\right)(1- \delta_{ik}). 
\end{aligned}
\end{equation}
Here, we let $\tau_{ik,0}^2$ and $\tau_{ik,1}^2$ be prior variances (with $\tau_{ik,1}^2 \gg \tau_{ik,0}^2$), for $i = 1, \dots M_k$, and $\underline{\beta}_{ik}$ denotes the $i$th element of $\underline{\bm \beta}_{k}$. The first mixture component is referred to as the 'spike' component, tightly fixed around the prior mean $\underline{\beta}_{ik}$ while the second is called the 'slab' component, which translates into almost no prior influence. The indicator $\delta_{ik}$ serves to select the mixture component used. Following the semiautomatic approach of \cite{george2008semi}, moreover, we scale the prior variances, $\tau_{ik,0}^2$ and $\tau_{ik,1}^2$, with variances of the ordinary least square estimates of the underlying structural model of state $S_t = k$.

This modeling approach constitutes a data-driven way of assessing whether coefficients should be pushed towards theoretically motivated restrictions or allowed to be closely related to the corresponding maximum likelihood estimate.  Thus, if $\delta_{ik}= 0$, the posterior estimate of $\beta_{ik}$ is strongly pushed towards the prior restriction $\underline{\beta}_{ik}$ while in the opposite case only little prior information on  $\beta_{ik}$ is introduced.

\begin{table}[ht]
\centering
\scriptsize{
\begin{tabular}{rrrrrrrrrrrrrrrrrr}
  \toprule
&Intercept & $i_{t-1}$ & $i_{t-1}^{*}$& $\pi_{t}$ &$\pi_{t}^{*}$& $x_{t}$& $x_{t}^{*}$& $q_t$& $m_{t}$ & $m_{t}^{*}$ &$y_{t}$ & $y_{t}^{*}$ & $e_t$ & $p_{t}$ & $p_{t}^{*}$ & $i_{t}$  & $i_{t}^{*}$ \\ 
 \midrule
$\underline{\bm \beta}_1$ & 0 & 0 & 0& 1.5 & -1.5 & .5 & -.5 & 0 &&&&&&&&& \\ 
$\underline{\bm \beta}_2$ & 0 &&&&&&&& 1 & -1 & 1 & -1 & -1 &&&& \\ 
$\underline{\bm \beta}_3$ & 0 &&&&&&&&&&&& -1 & 1 & -1 && \\ 
$\underline{\bm \beta}_4$ & 0 &&&&&&&&&&&&&&& 1 & -1 \\ 
\bottomrule
\end{tabular}}
\caption{Prior mean $\underline{\bm \beta}_k$ for each state}\label{tab:priormean}
\end{table}

In what follows, we store all regime-specific indicators in a vector  $\bm{\delta}_{k}=(\delta_{1k}, \dots, \delta_{M_k})'$  that corresponds to the block of $\bm \beta$ associated with the $k$th structural model. Each element of the latent variable $\bm{\delta}_{k}$ is a priori independently Bernoulli distributed,
\begin{equation*}
\begin{aligned}
p(\delta_{ik} = 1| S_t = k) =& \underline{\omega}_{ik}, \\
p(\delta_{ik} = 0| S_t = k) =& 1 - \underline{\omega}_{ik},
\end{aligned}
\end{equation*}
for hyperparameters  $\underline{\omega}_{ik} \in \underline{\bm{\omega}}_{k}$, an $M_k$-dimensional vector, chosen by the researcher.
Again, as in the case of  $\bm{\delta}_{k}$, the dimensionality of $\underline{\bm{\omega}}_{k}$ and the elements directly corresponds to the coefficient vector $\bm{\beta}_{k}$.  A reasonable choice is $\underline{\omega}_{ik} = 0.5$, for all $i,k$, implying an equal prior probability of introducing significant prior information or using a relatively loose prior.\footnote{When considering a weakly informative coefficient prior, we define $\underline{\omega}_{ik}$ as being one for including all state-specific coefficients with certainty.} 

For the variances $\bm{\sigma}^2$, we assume an independent inverse Gamma prior for each element $\sigma^2_i~(i=1,\dots, K)$.  More specifically, we set  
\begin{equation*}
\sigma^2_i \sim \G^{-1}(a_0, A_0),
\end{equation*}
with $a_0$ and $A_0$ being scalars.
The specific values for $a_0$ and $A_0$ are chosen to be weakly informative with hyperparameters $a_0 = 0.01, A_0 = 0.01$.

The prior distribution on the initial state is set to $p(S_0 = k) = 1/K$, for all $k$. \citep{kaufmann2015k}. 
Finally, for the coefficients of the multinomial logit model, we adopt a weakly informative and symmetric prior across all states.
That is, 
\begin{equation*}
\bm{\gamma}_{kj} \sim \N(\bm{0}, \underline{\bm{V}}),
\end{equation*}
for all $k$ and $j = 1, \dots, K$ with $\underline{\bm{V}} = \zeta \bm{I}_K$, and $\zeta$ denoting a scalar. In the empirical application we set $\zeta = 100$. 

In a Bayesian framework, we combine the likelihood with the prior to obtain the posterior distribution. In our case, the joint posterior density is intractable. Fortunately, however, the full conditional posterior distributions take simple forms, permitting  Gibbs updating steps. The Markov chain Monte Carlo (MCMC) algorithm  is described in more detail in Appendix \autoref{append:mcmc}. In the empirical application, we repeat the algorithm 80,000 times, discard the first 30,000 draws as burn-in and define a thinning factor of ten, thus basing inference on 5,000 draws from the joint posterior. 

Before proceeding to the empirical application, a brief word on identification is in order. Identification is necessary for structural interpretation of the states, but is not relevant if interest centers exclusively on the predictive density of the model \citep{fruhwirth2001ident, fruhwirth2006}.\footnote{Markov switching models might suffer from identification problems due to the invariance of the likelihood with respect to permutations of the $K!$ possible labeling of the regimes, resulting in $K!$ modes.} Recall that in the present model, each regime is characterized by a different set of fundamentals, reflecting different theoretical exchange rate models. By exploiting the specific structure of the theoretical models we have imposed inequality constraints on the coefficients assuming which fundamentals enter  each regime. The only potential source of non-identifiability occurs in case of more than one state pointing towards a random walk. 
However, pushing coefficients in direction of theoretical  guided values is sufficient to disentangle regimes and fully identify the model. When considering the alternative specification, in which we always include all predictors, identification is certainly an issue.  Moreover, each state is implicitly centered on a random walk a priori. In this case, we apply a permutation sampling step and solely focus on predictive densities.

\section{Empirical application}\label{sec:empircial}
This section starts by briefly describing the dataset and forecasting design adopted in \cref{sec:data}. Then, we  discuss key in-sample features of the model in \cref{sec:insample}. Finally, \cref{sec: forecasts} presents the main forecasting results, discriminating between point and density forecasting performance of all models considered.
\subsection{Data, forecasting design, and competing models}\label{sec:data}
In this paper, our aim is to forecast bilateral exchange rates for Australia, Canada, Japan, Norway, South Korea, Sweden, Switzerland and the United Kingdom relative to the US dollar. We collect monthly data on nominal exchange rates, industrial productions, monetary aggregates, three-month money market rates and consumer price indices for countries under consideration. \autoref{tab:data1} depicts the data transformations of variables. \autoref{tab:data2} provides an overview of the data alongside information on time coverage and source of economic fundamentals. 
\begin{table}[!htbp] \centering \scriptsize
\begin{tabular}{llcc} 
\toprule
\textbf{Variable} & \multicolumn{1}{l}{\textbf{Description}} & \multicolumn{1}{c}{\textbf{Transformation}} &  \multicolumn{1}{c}{\textbf{Comments}} \\ 
\midrule
EXR & Nominal exchange rate & $\log$ difference & --- \\ 
IP & Industrial production & $\log$ & --- \\ 
M & Money aggregate & $\log$ & --- \\ 
$3$M-IR & $3$M Money market rate & --- & --- \\
CPI & Consumer price index & $\log$ & ---\\
\midrule
INF & Inflation & $\log$ differences of CPI & ---\\
REXR & Real exchange rate & $\log$(EXR) + $\log$(CPI*) - $\log$(CPI) & ---\\
IP-GAP & Output gap & HP filter& $\lambda = 14400$ for monthly data\\[1.2ex]
\bottomrule
\end{tabular} 
  	\caption{Transformation of variables.} 
  	\label{tab:data1} 
\end{table} 

\begin{table}[!htbp] \centering \scriptsize
\begin{tabular}{lllllll} 
\toprule 
\textbf{Country} & \multicolumn{1}{c}{\textbf{Coverage}} & \multicolumn{1}{l}{\textbf{EXR}} & \multicolumn{1}{l}{\textbf{IP (2010 = 100)}} & \multicolumn{1}{l}{\textbf{M}} & \multicolumn{1}{l}{\textbf{$3$M-IR,}} & \multicolumn{1}{l}{\textbf{CPI (2010 = 100)}}\\ 
\midrule
Australia (AU)    & $1975$M$06$:$2017$M$09$ & IFS  & OECD & M1, OECD & OECD & OECD \\ 
Canada (CA)       & $1973$M$01$:$2017$M$09$& IFS  & OECD & M1, OECD & OECD & OECD \\ 
Japan (JP)        & $1973$M$01$:$2017$M$03$& IFS  & OECD & M1, OECD& IFS & OECD\\ 
Norway (NO)       & $1979$M$01$:$2017$M$09$& IFS  & OECD & M1, OECD & OECD & OECD \\
South Korea (KR)  & $1981$M$01$:$2017$M$09$& IFS  & IFS & M1, OECD & IFS & OECD \\
Sweden (SE)       & $1973$M$01$:$2017$M$06$& IFS  & OECD & M3, OECD & IFS & OECD\\
Switzerland (CH)  & $1974$M$01$:$2017$M$09$ & IFS  & OECD & M1, OECD & OECD & OECD \\
United Kingdom (UK)& $1973$M$01$:$2017$M$02$ & IFS  & OECD & M0, IMF/FRED & IFS & OECD \\
United States (US) & $1973$M$01$:$2017$M$09$ &  & OECD & M1, OECD  & OECD & OECD\\
\bottomrule
\end{tabular} 
 \flushleft{\footnotesize{\textit{Note: All quantities are seasonally adjusted, except EXR and $3$M-IR. IP of Australia and Switzerland are interpolated to monthly frequency with cubic spline.}}}
  	\caption{Sources of economic fundamentals.} 
  	\label{tab:data2} 
\end{table} 

In order to assess whether time-varying transition probabilities improve predictive accuracy, the proposed model framework is  benchmarked with MS specifications with fixed transition probabilities (labeled MS-FT), as well as standard structural exchange rate models that are estimated under weakly informative priors (labeled linear). These linear benchmarks  are based on Taylor rule, monetary, PPP, and UIP fundamentals.
Therefore, in the forecasting exercise, the set of competing models is divided in the three overall classes: MS-TVP, MS-FT and linear. Moreover, we consider not only theoretically motivated MS-TVP and MS-FT specifications, but also models that include all macroeconomic indicators of $\bm X_{t-1}$ within each state (labeled kitchen-sink). For the kitchen-sink regressions, we consider different numbers of states, ranging from two to four regimes. Moreover, to allow for state-specific shrinkage in kitchen-sink regressions, the SSVS prior described above is centered on zero and different state-specific indicators are estimated. To assess the role of allowing for heteroscedasticity in forecasting exchange rates, we also consider MS-TVP specifications with state-specific variances. All models are then benchmarked to the random walk without drift.  

We evaluate  predictive accuracy by means of a recursive pseudo out-of-sample forecasting exercise. This implies choosing a initial estimation period that ranges from $t=1$ up to $t=T_0$ with the remaining periods used as a hold-out sample. In the present application, we estimate all models using data up to $2004$M$12$ and then proceed by computing $h$-step-ahead predictions for $t=T_0+1$. After obtaining draws from the corresponding predictive distributions, we consequently expand the initial estimation period by one month. This procedure is repeated until the end of the sample is reached.

To rank forecasts, we rely on cumulative squared forecast errors (CSFEs) to assess the quality of point forecasts. As point predictions, we take the posterior median of the predictive density. Turning to density forecasts, we follow \cite{geweke2010lps} and rely on the log predictive score (LPS) to measure density forecasting accuracy.  This has the advantage that,  conditional on the proposed model and data, uncertainty surrounding the parameters and latent quantities is integrated out. After obtaining the LPS, we compute  log predictive Bayes factors (LBFs) for the entire hold-out sample by computing the difference between the LPS of a given model relative to the random walk.

\subsection{Inspecting evidence for model instability}\label{sec:insample}
\begin{figure}[h]
	\centering
	\begin{subfigure}{0.49\textwidth}
		\includegraphics[width=\textwidth]{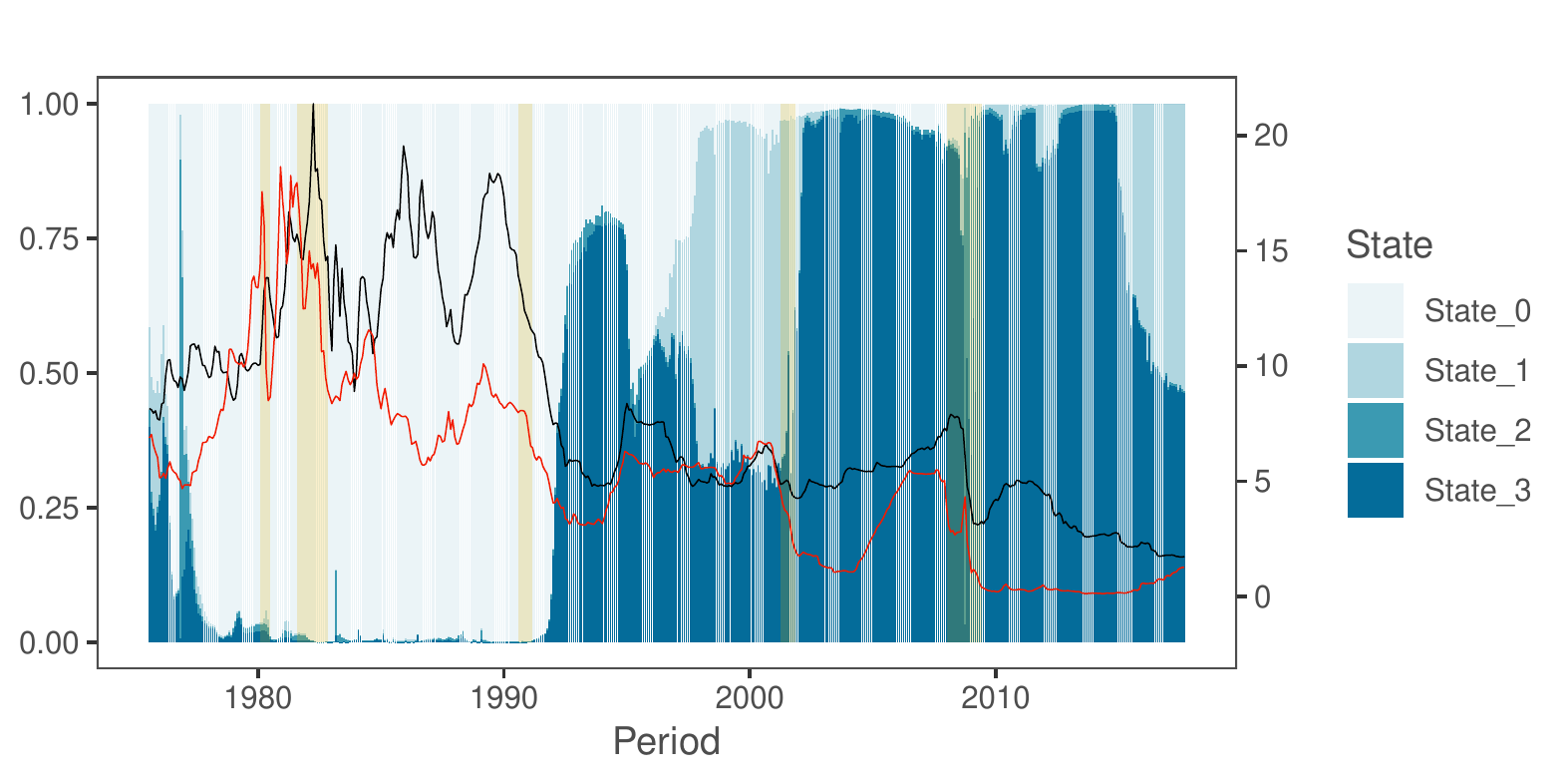}
  	  \caption{AU}
	\end{subfigure}
	\begin{subfigure}{0.49\textwidth}
		\includegraphics[width=\textwidth]{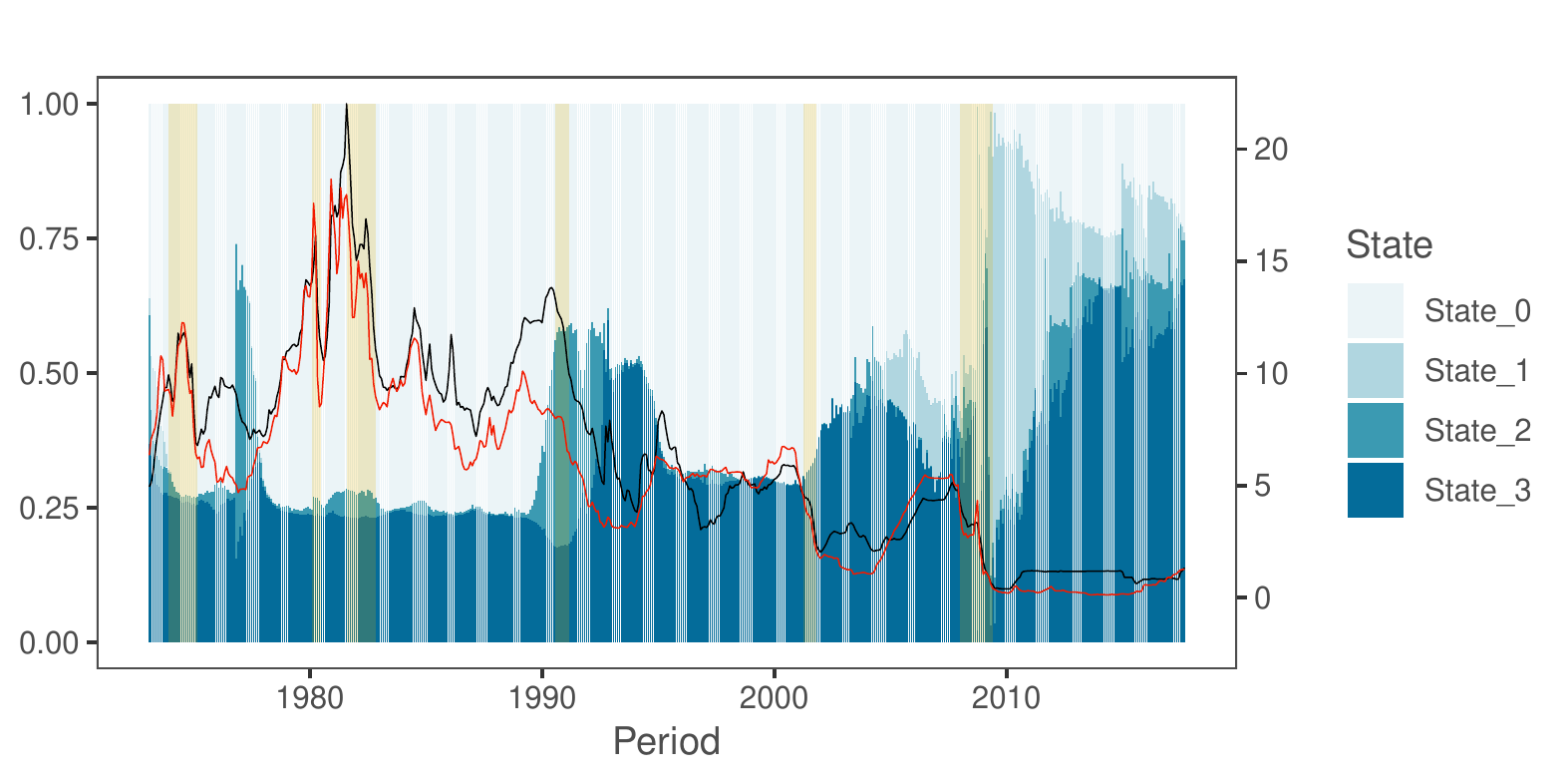}
  	  \caption{CA}
	\end{subfigure}
	\begin{subfigure}{0.49\textwidth}
		\includegraphics[width=\textwidth]{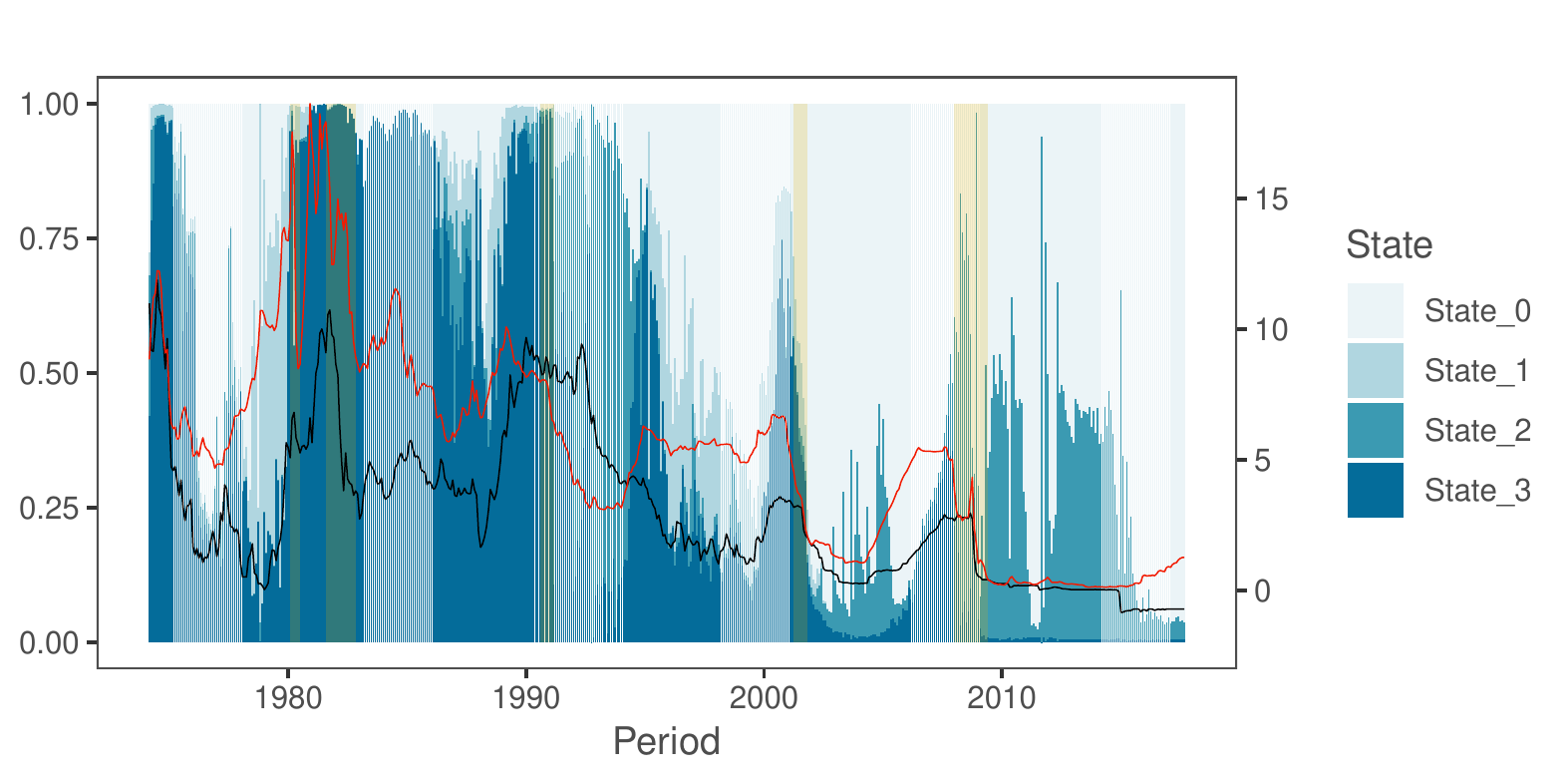}
  	  \caption{CH}
	\end{subfigure}
	\begin{subfigure}{0.49\textwidth}
		\includegraphics[width=\textwidth]{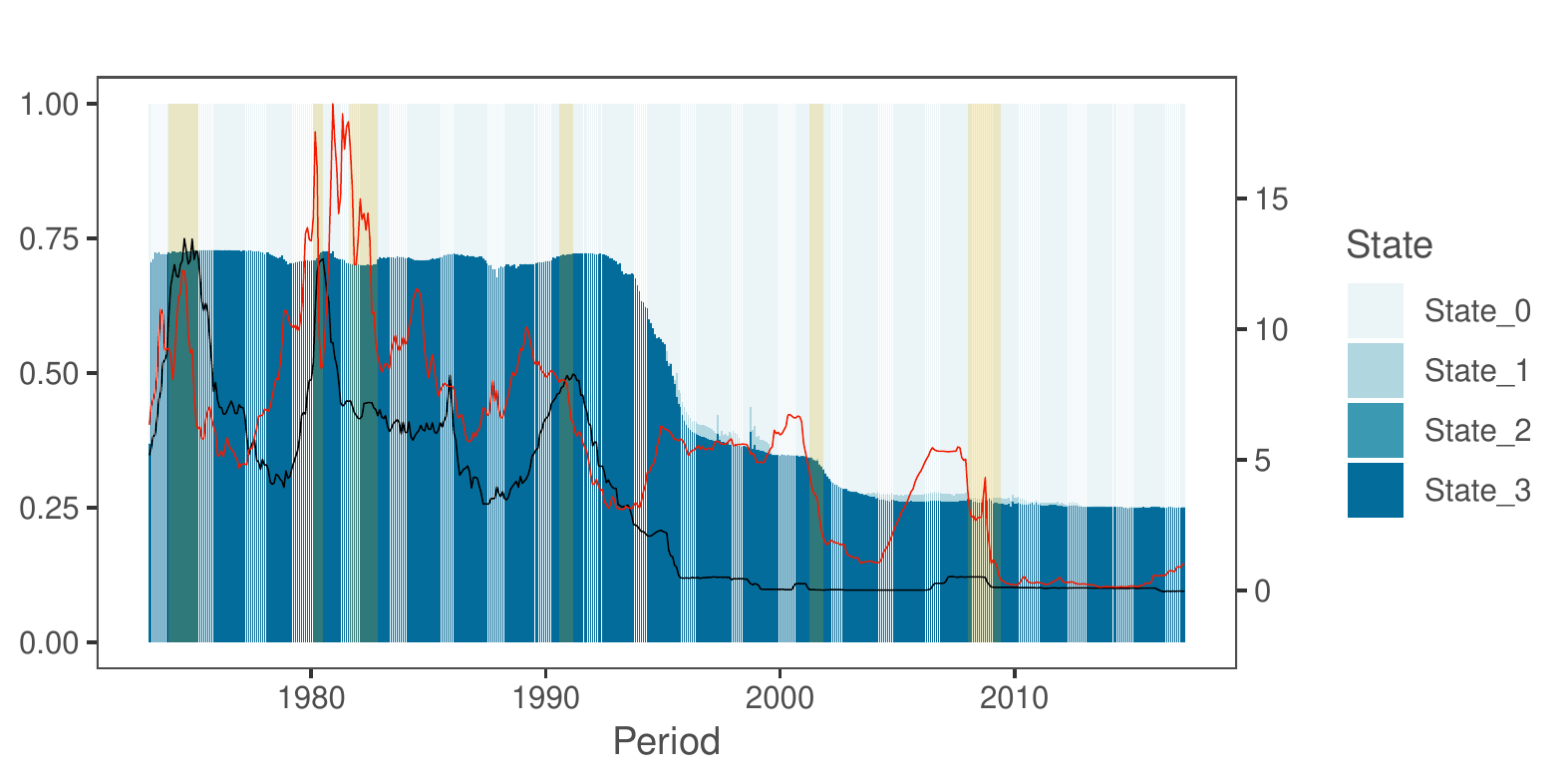}
  	  \caption{JP}
	\end{subfigure}
	\begin{subfigure}{0.49\textwidth}
		\includegraphics[width=\textwidth]{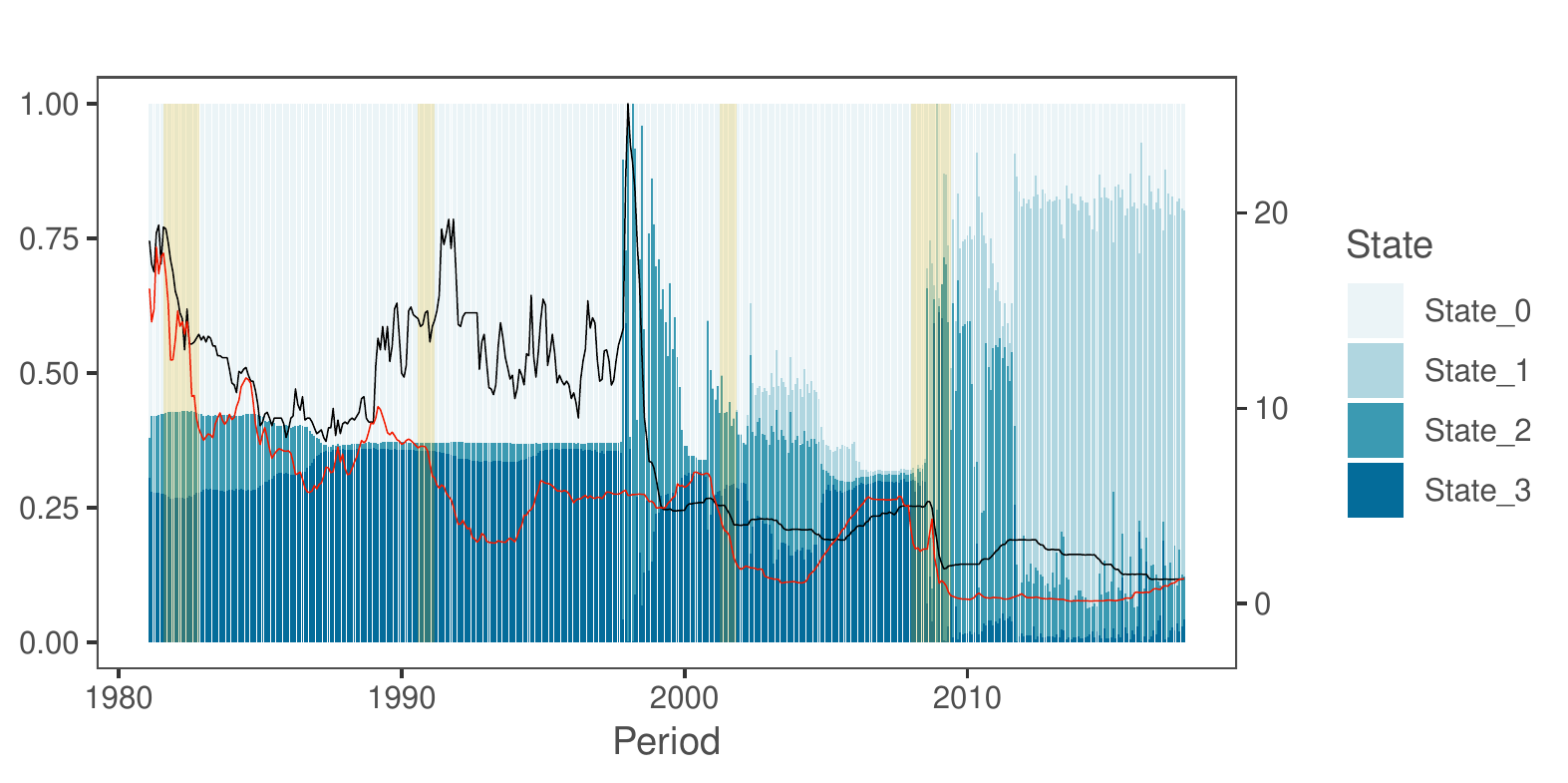}
  	  \caption{KR}
	\end{subfigure}
	\begin{subfigure}{0.49\textwidth}
		\includegraphics[width=\textwidth]{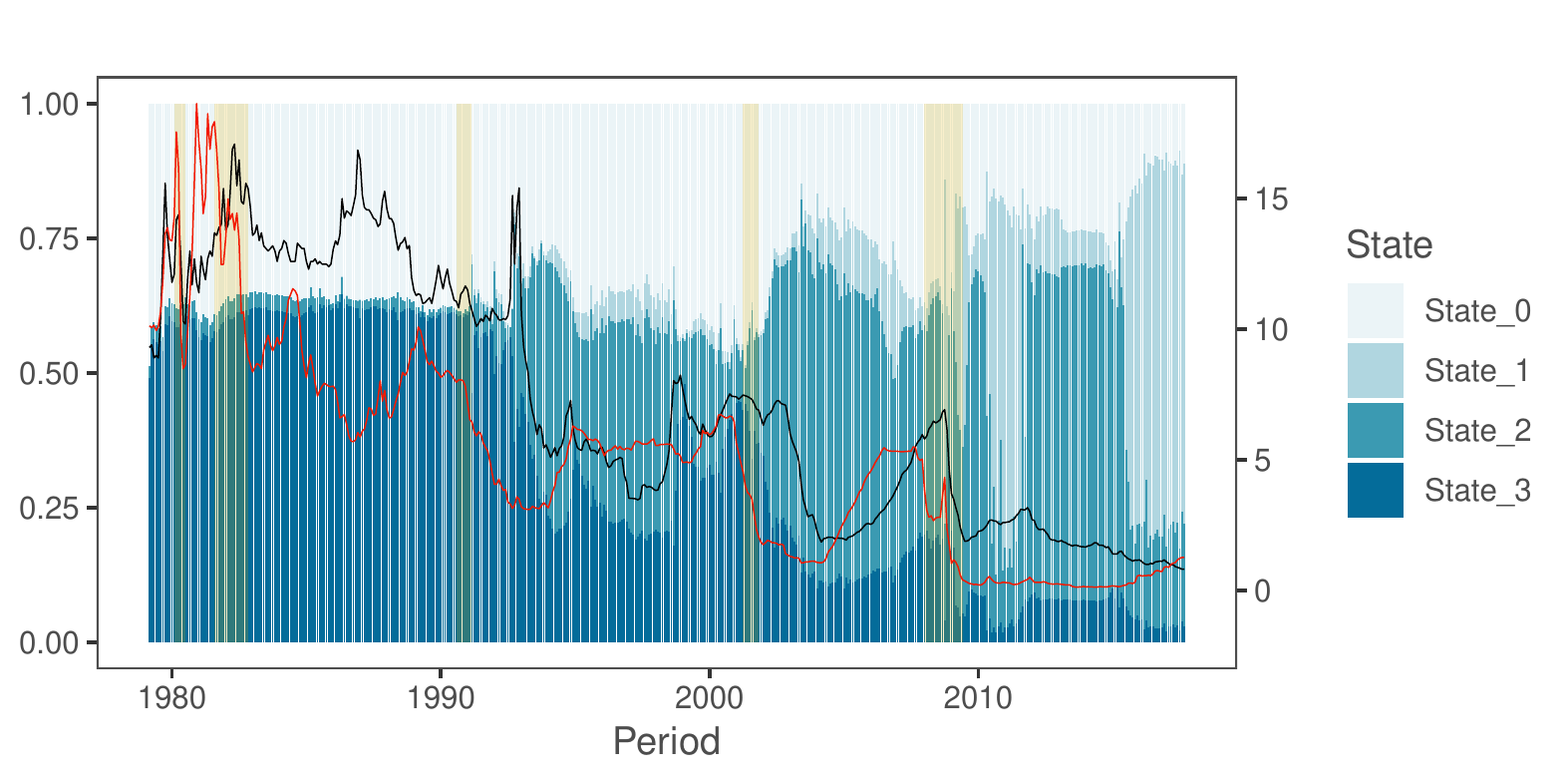}
  	  \caption{NO}
	\end{subfigure}
	\begin{subfigure}{0.49\textwidth}
		\includegraphics[width=\textwidth]{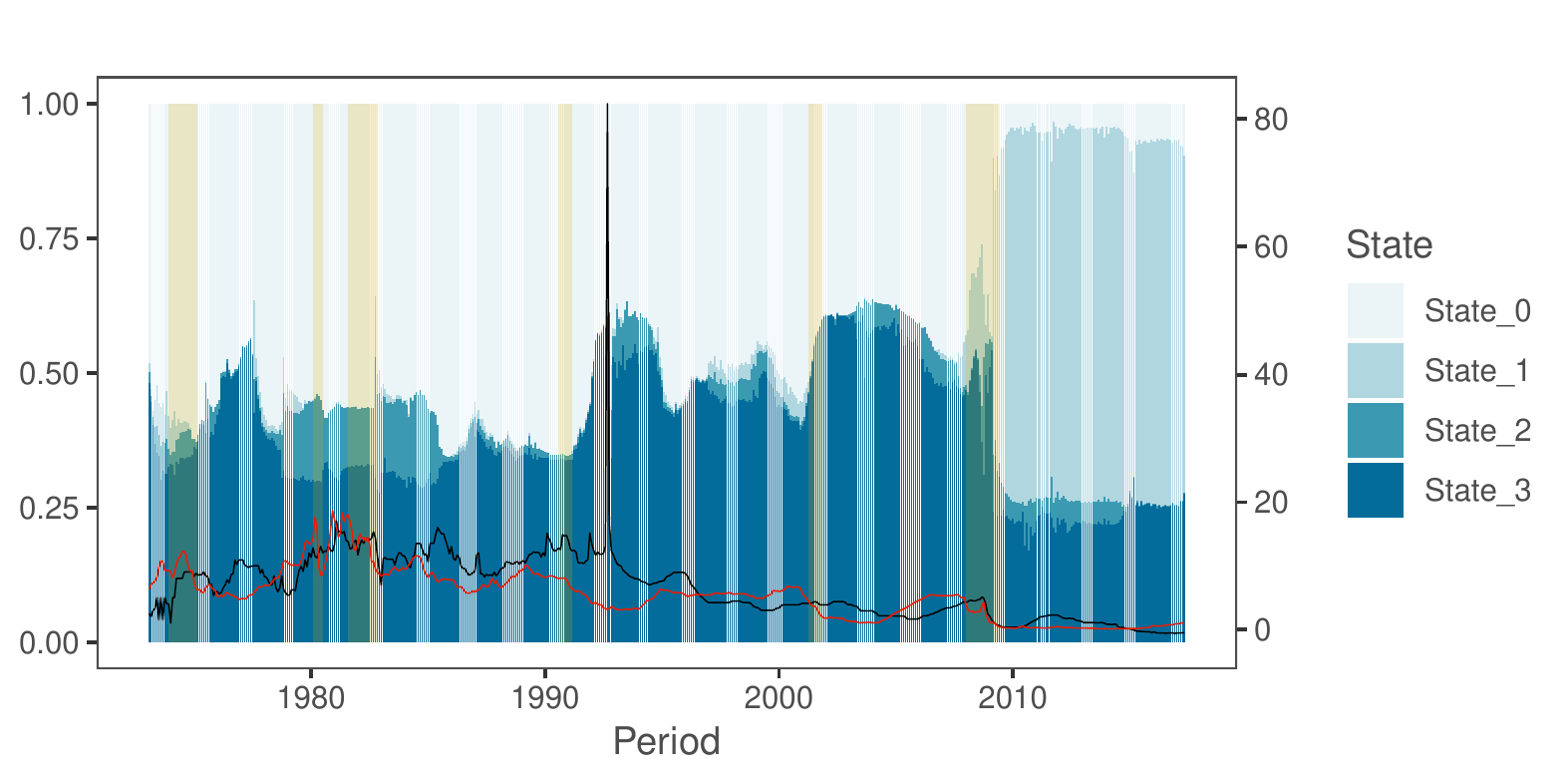}
  	  \caption{SE}
	\end{subfigure}
	\begin{subfigure}{0.49\textwidth}
		\includegraphics[width=\textwidth]{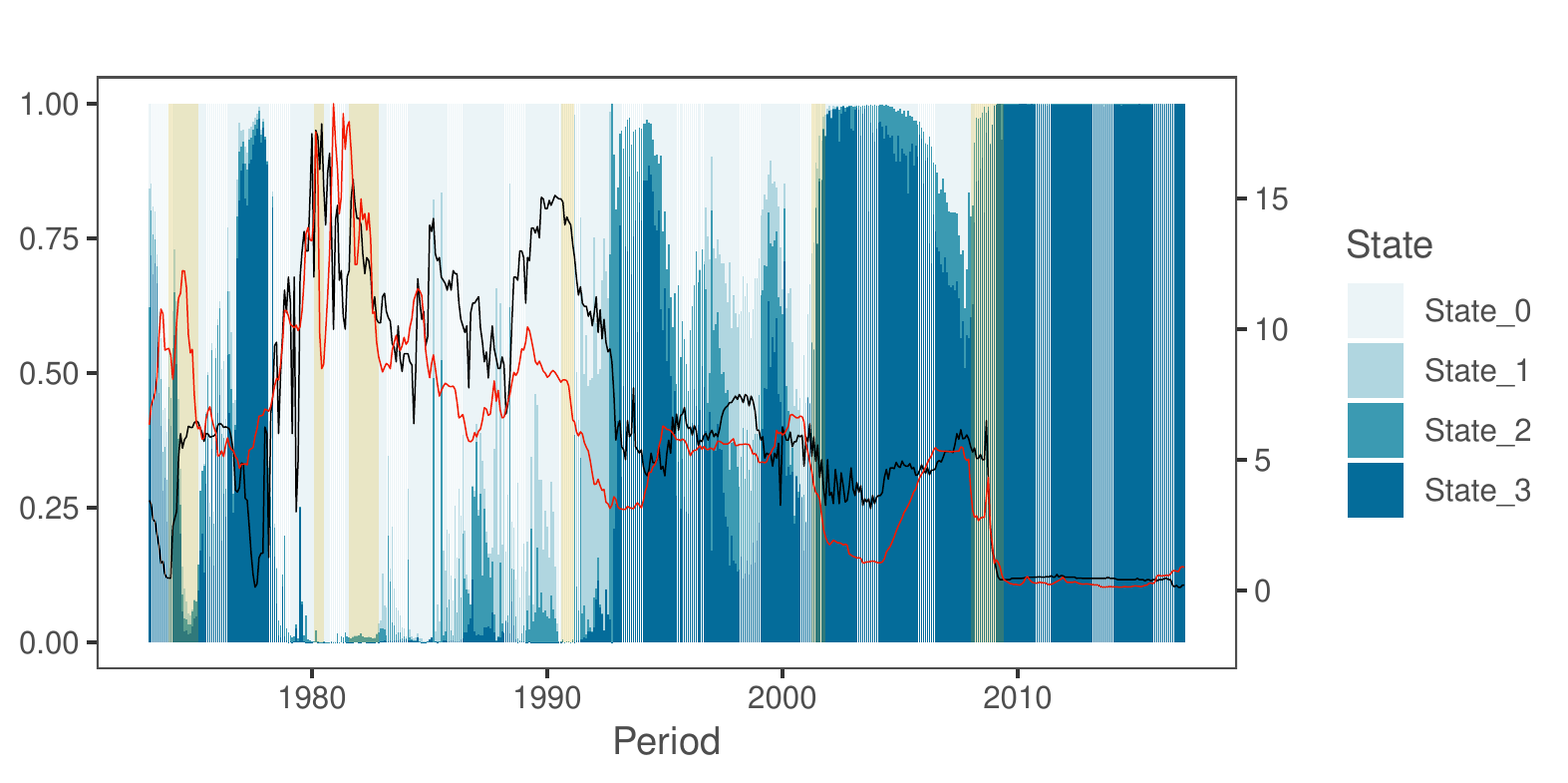}
  	  \caption{UK}
	\end{subfigure}			
	\caption{Mean posterior state probabilities for a weakly informative prior and a common variance across states. State $0$ indicates Taylor rule fundamentals, state $1$ monetary fundamentals, state $2$ PPP fundamentals and state $4$ UIP fundamentals. The vertical bars (yellow) indicate NBER recessions for the US. The black solid line depicts the country-specific interest rate and the red solid line the interest rate for the US. The left-axis shows the probabilities and the right-scale the values of interest rates.}
	  \label{fig:in-states}
\end{figure}

We now turn to assess whether our model proposed in \cref{sec:model} signals significant shifts in the underlying structural representation. \cref{fig:in-states} summarizes the mean of the filtered state probabilities for the eight  exchange rates considered.  In general, we observe that the regime dynamics across countries share one common feature. The models based on Taylor rule  (state 0) and the UIP fundamentals (state 4) appear to be the dominant states before the global financial crisis in 2008/2009. After that period, however, model evidence changes significantly for the majority of countries. More precisely, models based on monetary (state 2) and PPP (state 3) fundamentals tend to receive more posterior support.

Compared to the remaining currencies, the Swiss franc (see \autoref{fig:in-states}(c)) exhibits a somewhat higher regime-switching frequency. \cref{fig:in-states}, moreover, suggests that hitting the ZLB does shift filtered probabilities, supporting regimes other than Taylor rule fundamentals (state 0) for countries such as Australia, Canada, South Korea, Sweden, and the United Kingdom.\footnote{Notice that Australia never hit the ZLB during the  sample, but the foreign country the United States did. Moreover, it is worth emphasizing that Japan already hit the ZLB in the midst of the 1990s.} Countries such as Japan and Switzerland, on the other hand, indicate an opposite dynamic, namely a shift of probabilities towards the Taylor rule state. 

Taking a closer look at the United Kingdom, Taylor rule fundamentals are the predominant regime, reflecting the fact that these quantities tend to describe exchange rates well in times when the primary policy rule of the Bank of England is the Taylor rule. After this period, transition probabilities point towards a first shift during the crisis of the European Monetary System.  After the financial crisis, and upon hitting the ZLB, the model based on Taylor rule fundamentals receives only limited posterior support. It is noteworthy that after 2010, the short-term interest rate (both at home and abroad) is stuck at zero (and almost constant). This implies that the model based on interest fundamentals closely mimic a random walk with drift during this period, even without introducing shrinkage.  
\begin{figure}[h]
	\centering
	\begin{subfigure}{0.49\textwidth}
		\includegraphics[width=\textwidth]{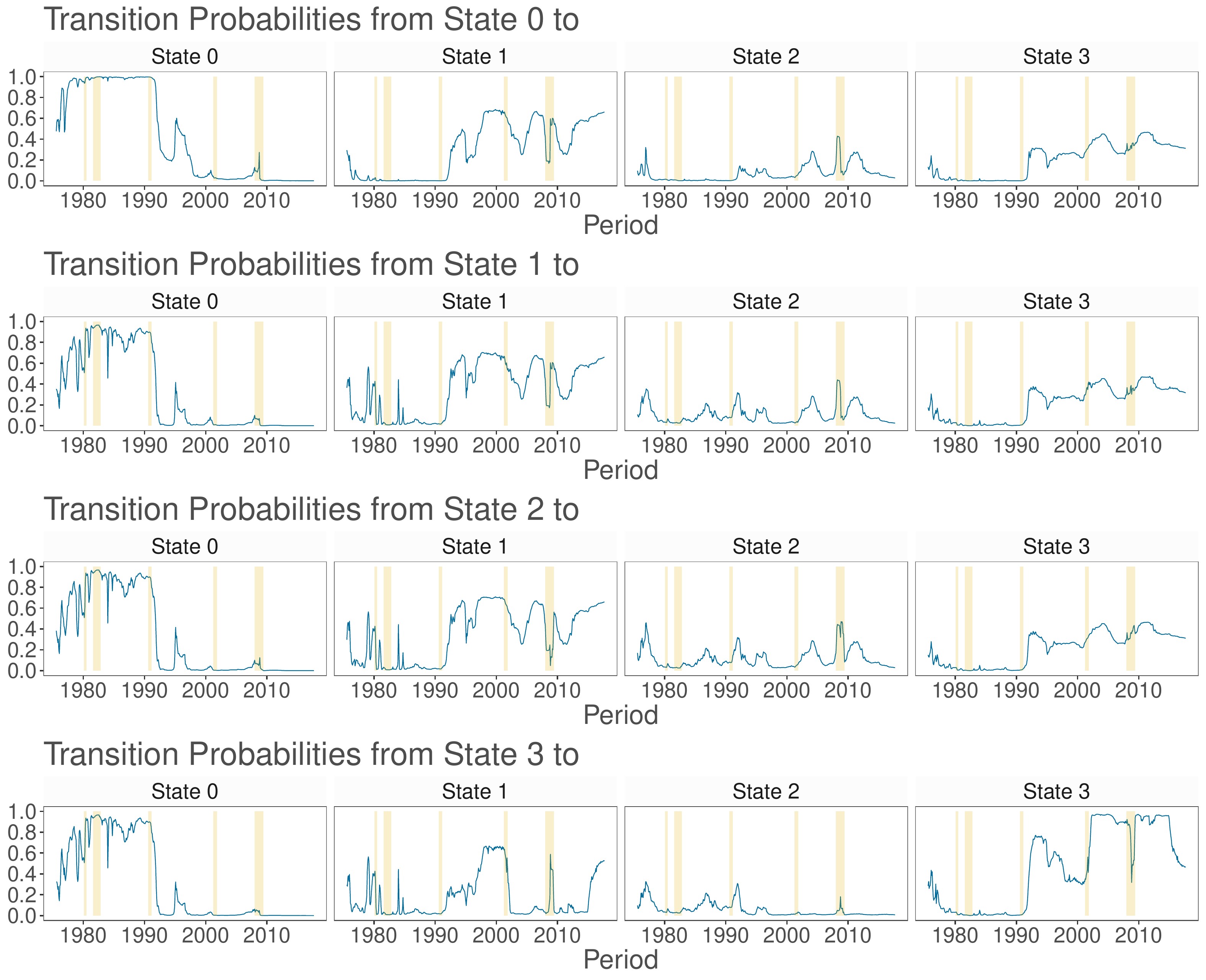}
	\caption{AU}	
	\end{subfigure}
	\begin{subfigure}{0.49\textwidth}
		\includegraphics[width=\textwidth]{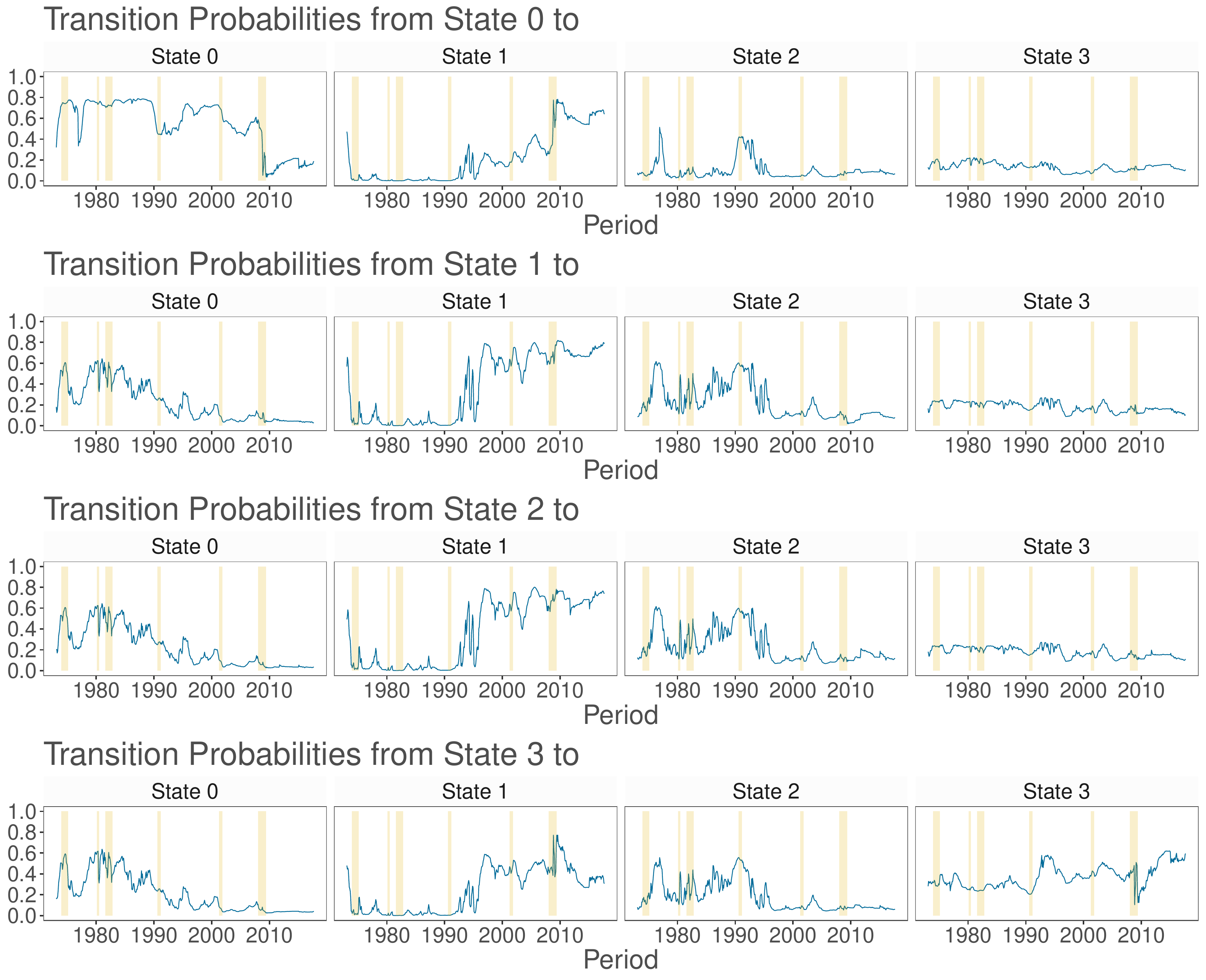}
	\caption{CA}	
	\end{subfigure}
	\begin{subfigure}{0.49\textwidth}
		\includegraphics[width=\textwidth]{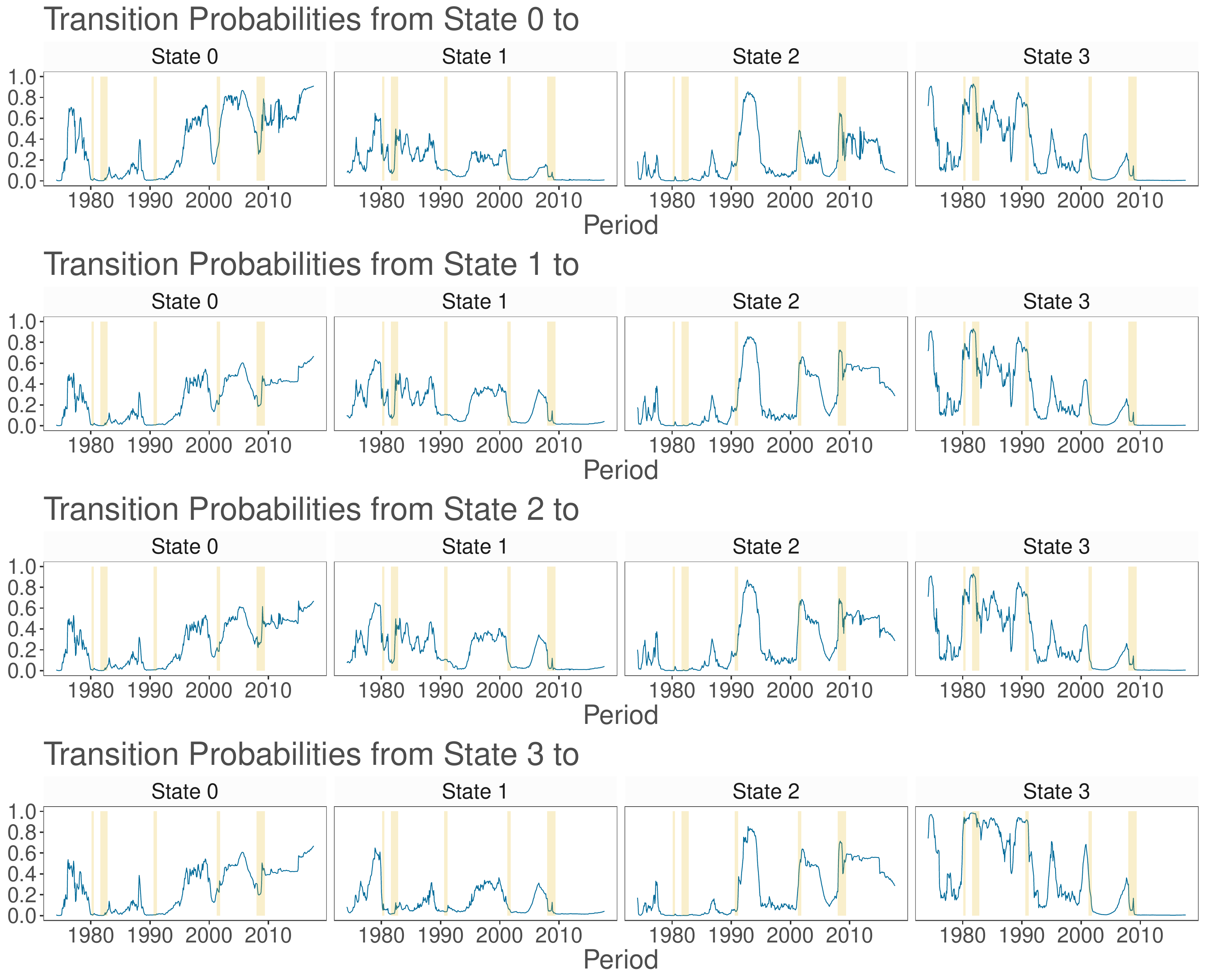}
	\caption{CH}	
	\end{subfigure}
	\begin{subfigure}{0.49\textwidth}
		\includegraphics[width=\textwidth]{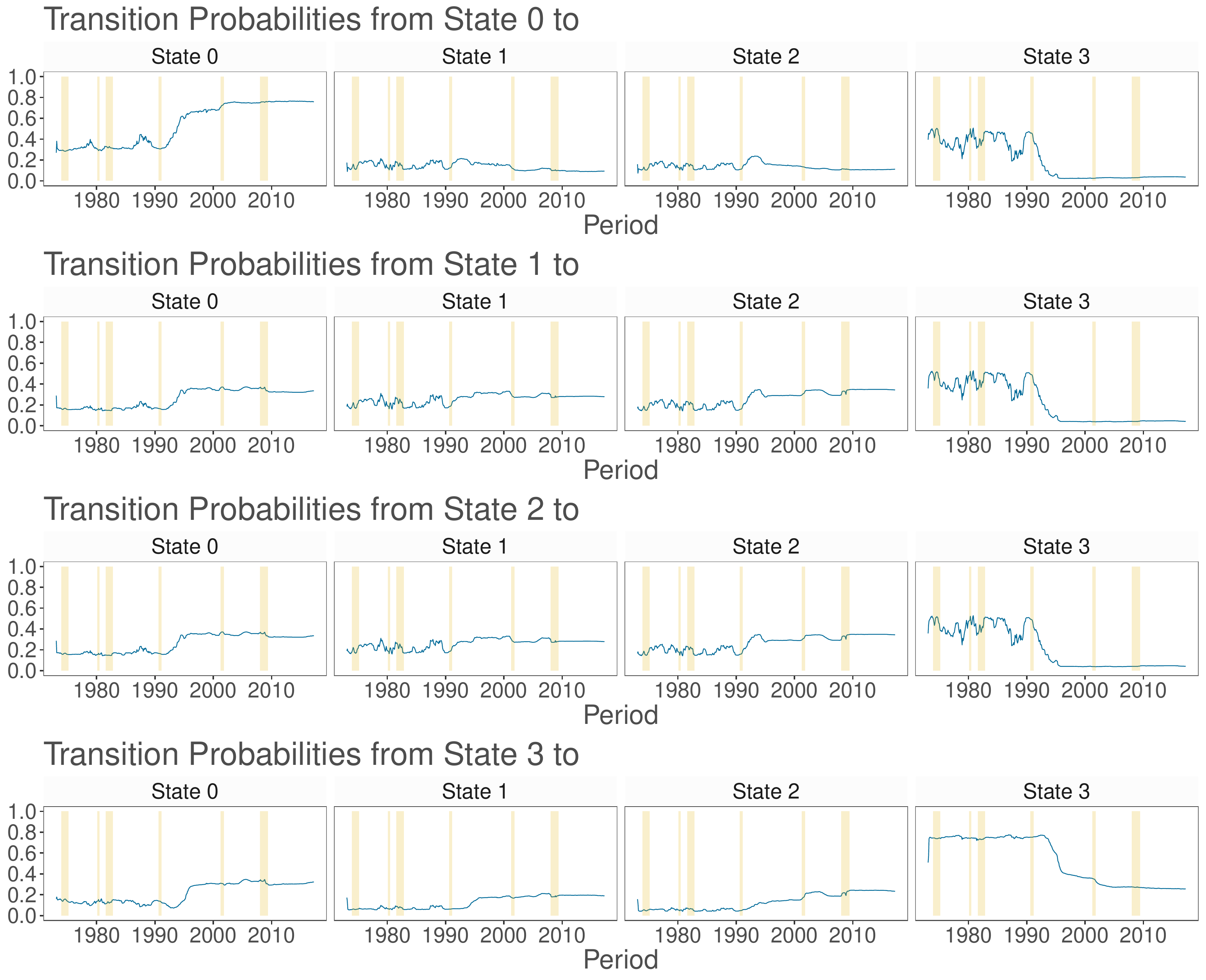}
	\caption{JP}	
	\end{subfigure}	
	\caption{The blue line depicts the posterior mean of time-varying transition probabilities for each state with a weakly informative prior and common variance across states. State $0$ indicates Taylor rule fundamentals, state $1$ monetary fundamentals, state $2$ PPP fundamentals and state $4$ interest rate fundamentals. The vertical bars (yellow) indicate NBER recessions for the US.}
	\label{fig:intp1}
\end{figure}	

\begin{figure}[h]
	\centering	
	\begin{subfigure}{0.49\textwidth}
\includegraphics[width=\textwidth]{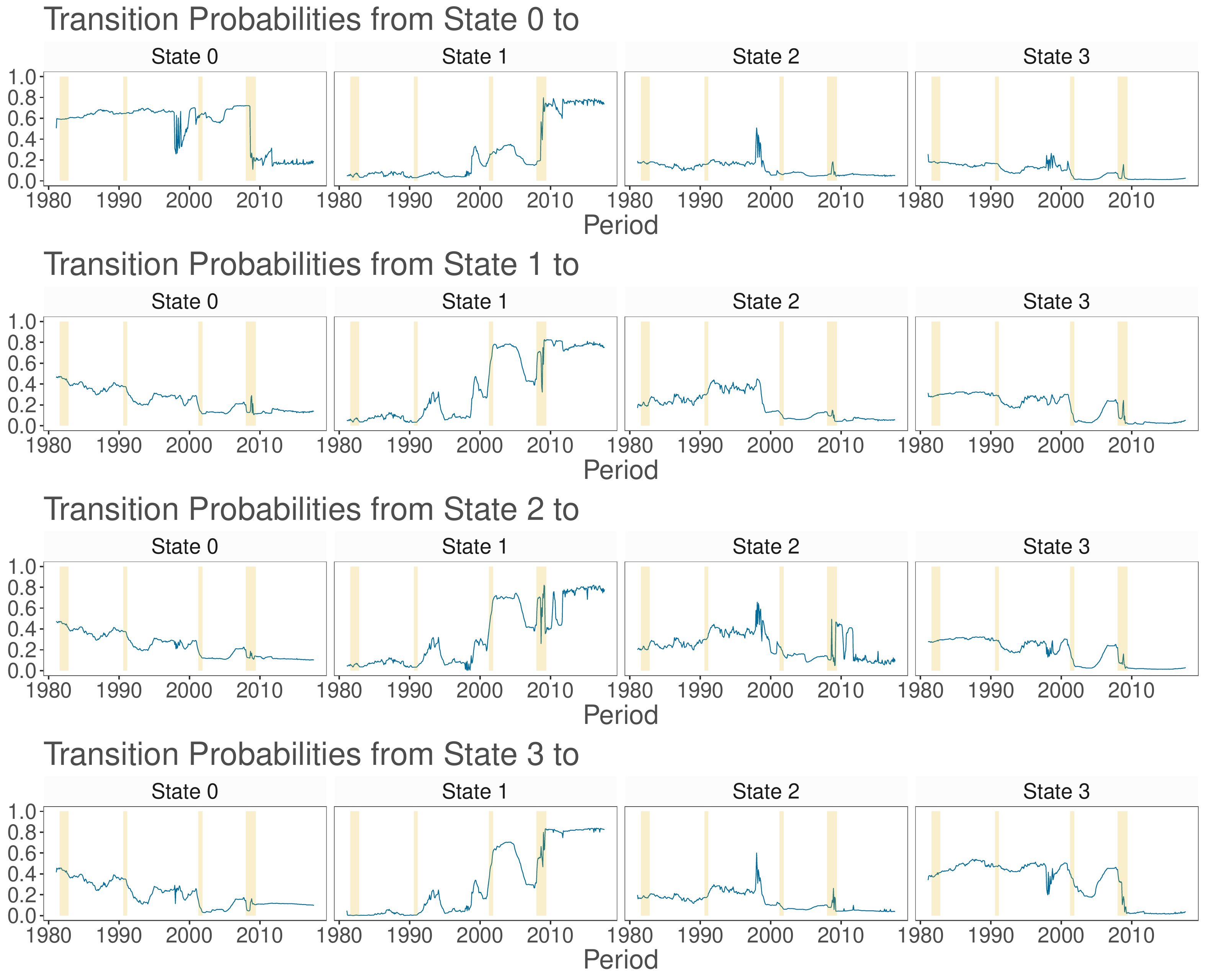}
	\caption{KR}	
	\end{subfigure}	
	\begin{subfigure}{0.49\textwidth}
		\includegraphics[width=\textwidth]{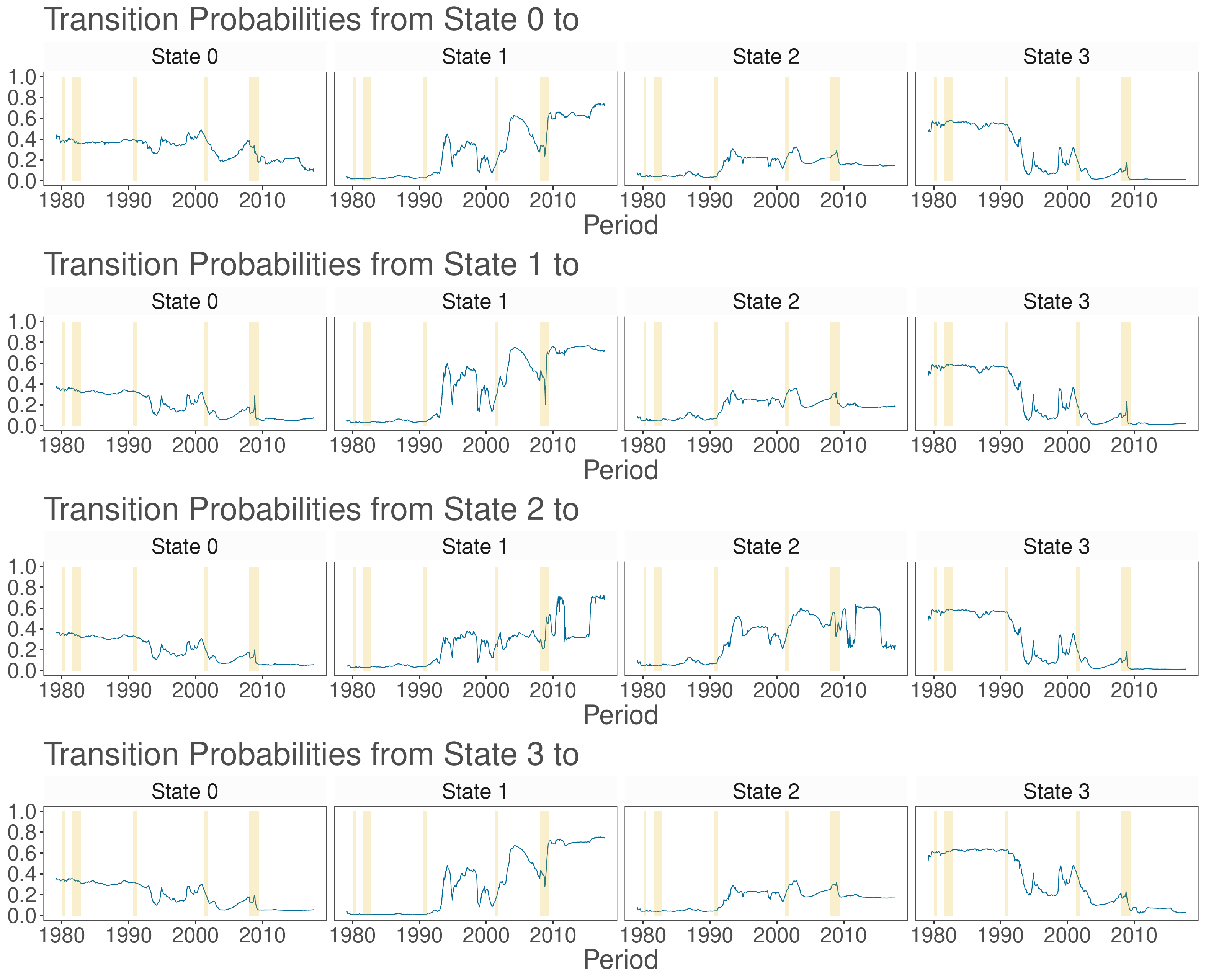}
	\caption{NO}	
	\end{subfigure}	
	\begin{subfigure}{0.49\textwidth}			\includegraphics[width=\textwidth]{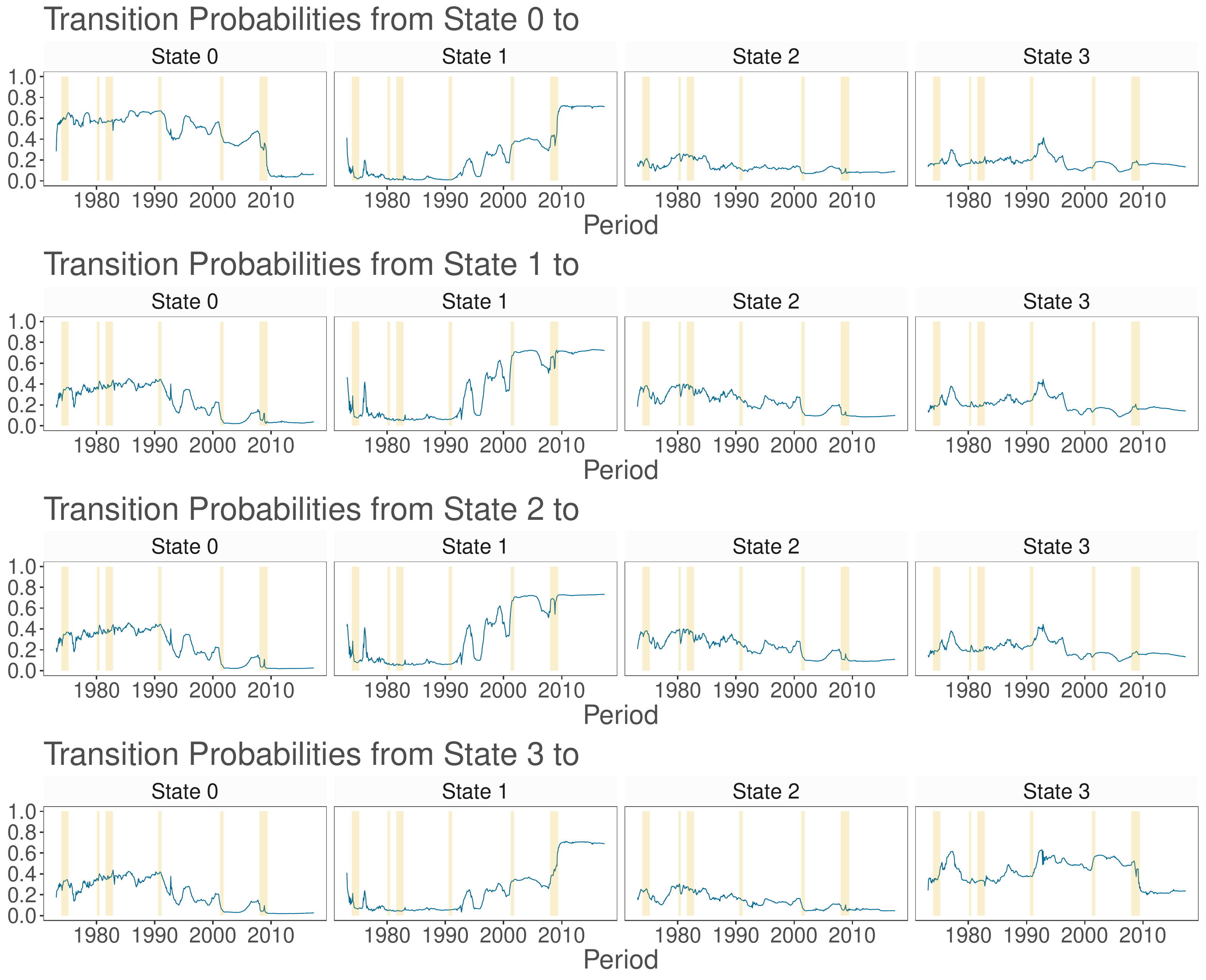}
	\caption{SE}	
	\end{subfigure}
	\begin{subfigure}{0.49\textwidth}
		\includegraphics[width=\textwidth]{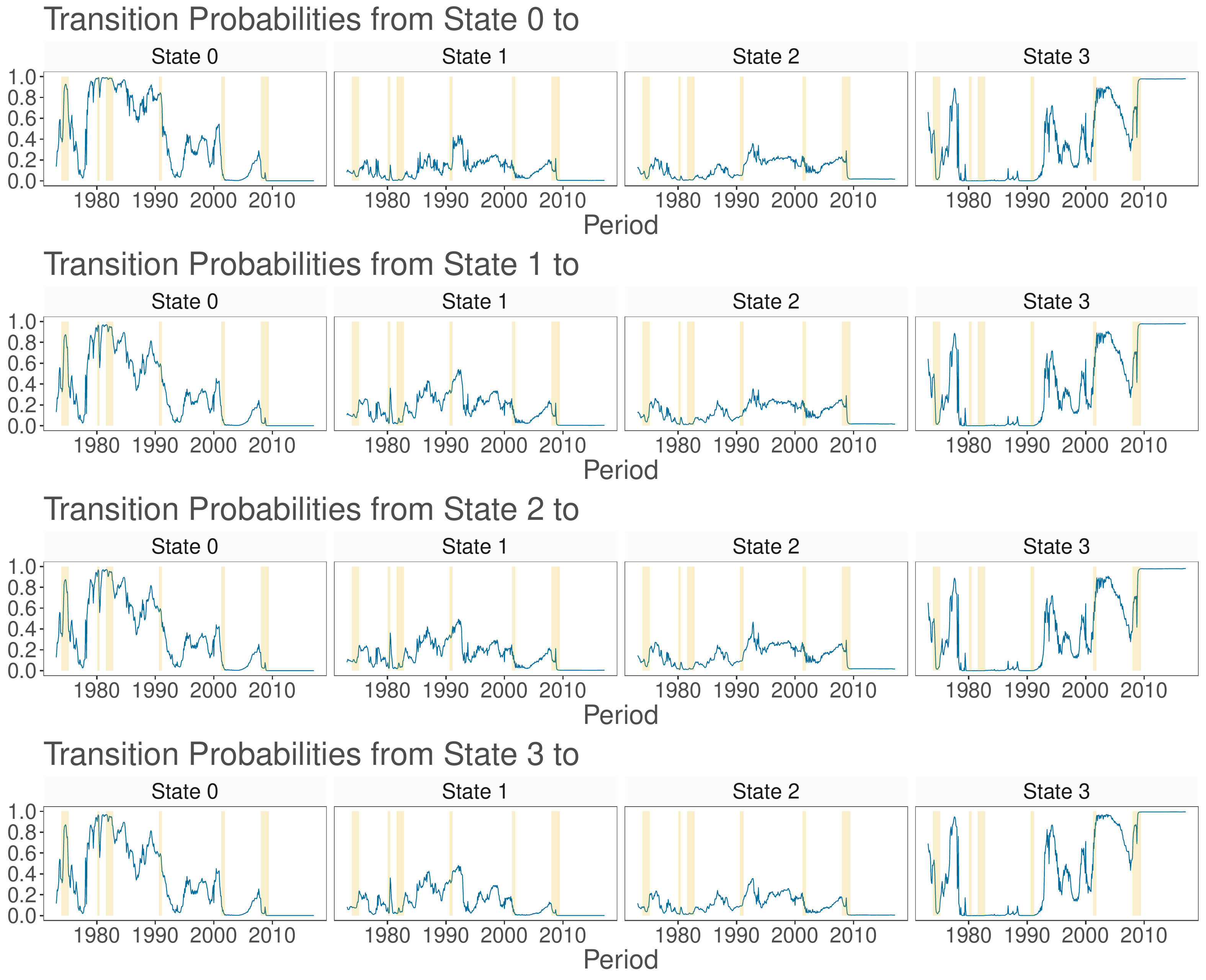}
	\caption{UK}	
	\end{subfigure}		
	\caption{The blue line depicts the posterior mean of time-varying transition probabilities for each state with a weakly informative prior and common variance across states. State $0$ indicates Taylor rule fundamentals, state $1$ monetary fundamentals, state $2$ PPP fundamentals and state $4$ interest rate fundamentals. The vertical bars (yellow) indicate NBER recessions for the US.}
	\label{fig:intp2}
\end{figure}

The transition probabilities, depicted in Figs.~\ref{fig:intp1} and \ref{fig:intp2}, generally track the movements in filtered state probabilities, providing considerable evidence of time-varying transition distributions. Our findings thus suggest that a measure of the monetary policy stance at home and abroad tends to drive transitions between structural models. This is consistent with our conjecture that during the period of the ZLB, using Taylor rule-based exchange rate models might be inappropriate, at least from an in-sample perspective.

\subsection{Forecasting results}\label{sec: forecasts}
In this section, interest centers on the predictive performance of our proposed MS-TVP specification. The discussion in the last section highlights that it proves to be important to allow for time-varying transition probabilities in-sample for several exchange rate pairs. This suggests that parameterizing the transition distributions with additional covariates helps to avoid situations where the model gets stuck within a certain state. \cite{amisano2013money} and \cite{kaufmann2015k} highlight this issue and point towards advantages of explaining the regime-switching behavior of the model as opposed to a model based on constant transition probabilities. The key question, however, is whether this additional flexibility also improves predictive performance. We answer this question using both, point and density forecasts.
 
\subsection*{Point forecasts}
\begin{landscape}
\begin{figure}[!h]
	\centering
	\begin{subfigure}{0.7\textwidth}
		\includegraphics[width=\textwidth]{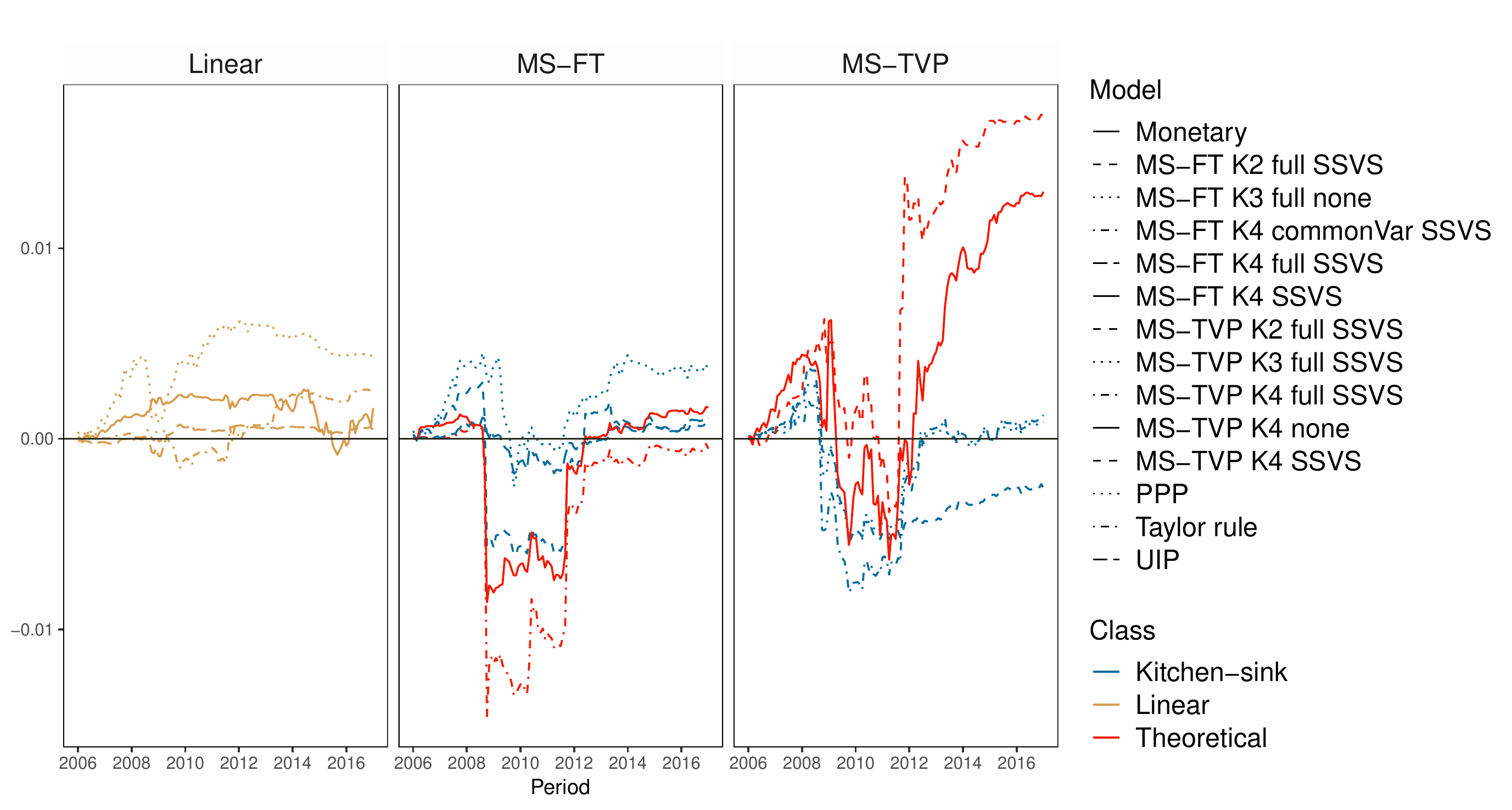}
	\caption{AU}
	\label{fig:outmse1AU}
	\end{subfigure}	
	\begin{subfigure}{0.7\textwidth}
		\includegraphics[width=\textwidth]{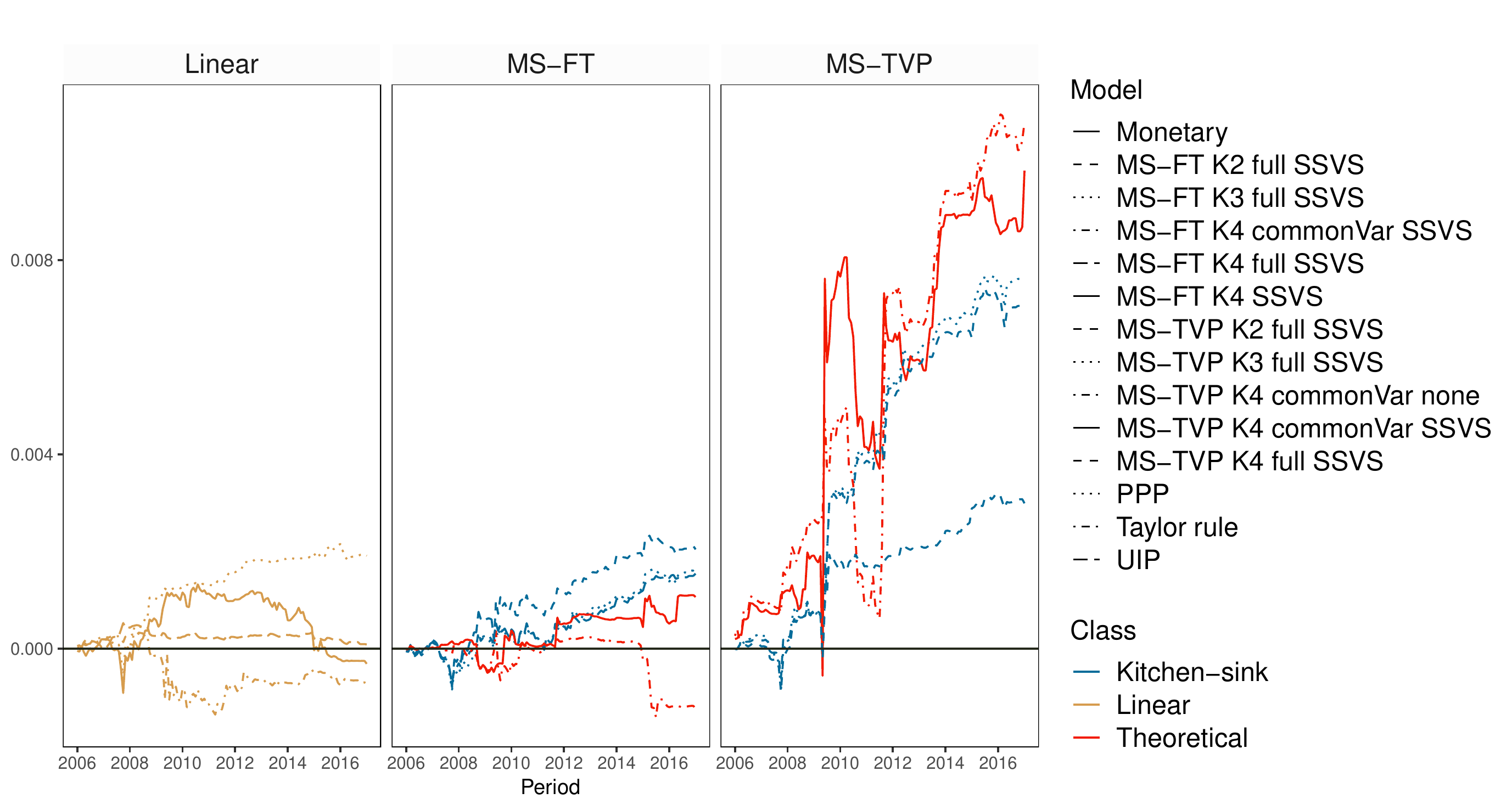}
	\caption{CA}
	\label{fig:outmse1CA}	
	\end{subfigure}
	\begin{subfigure}{0.7\textwidth}
		\includegraphics[width=\textwidth]{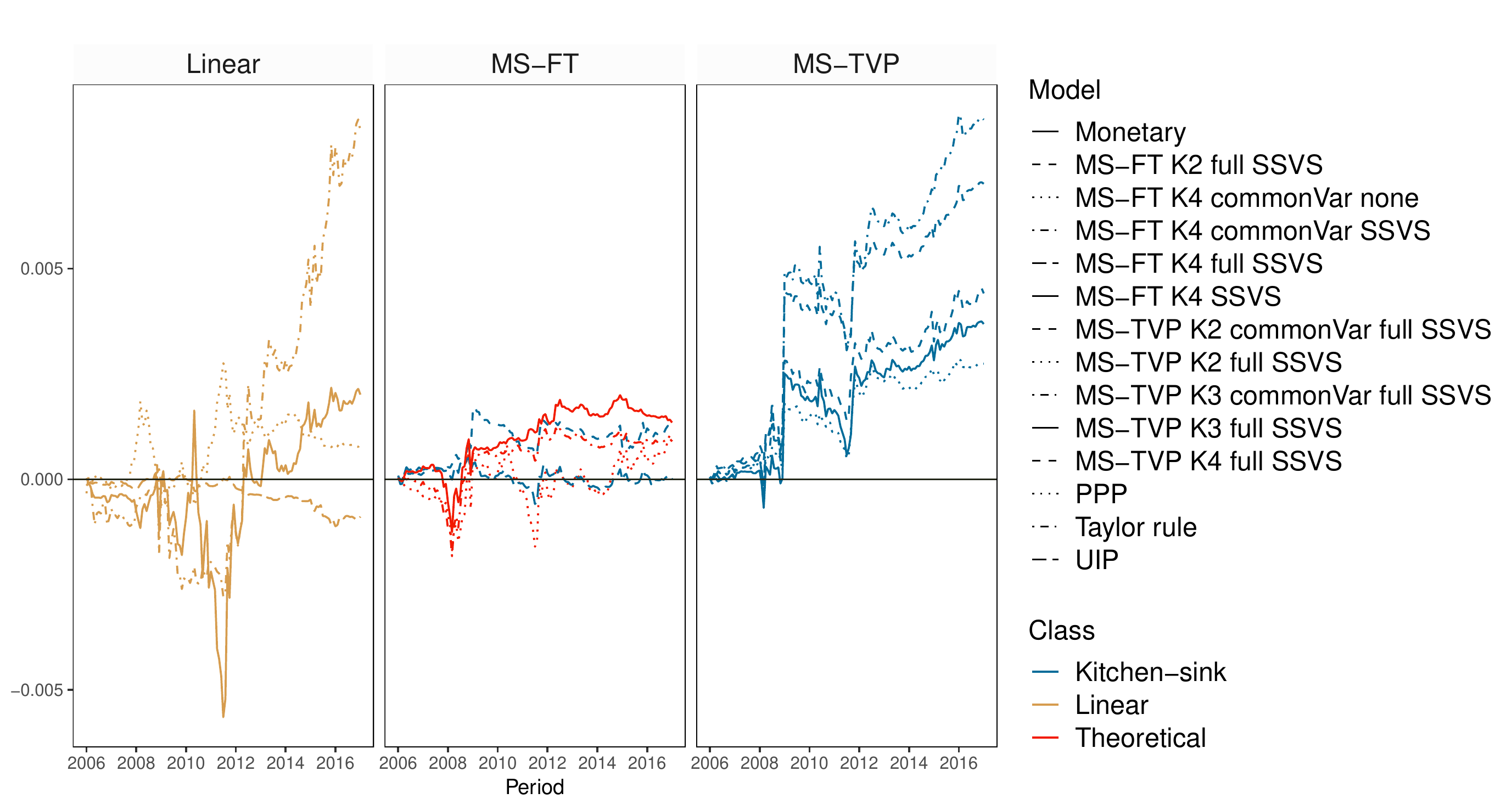}
	\caption{CH}
	\label{fig:outmse1CH}	
	\end{subfigure}	
	\begin{subfigure}{0.7\textwidth}
		\includegraphics[width=\textwidth]{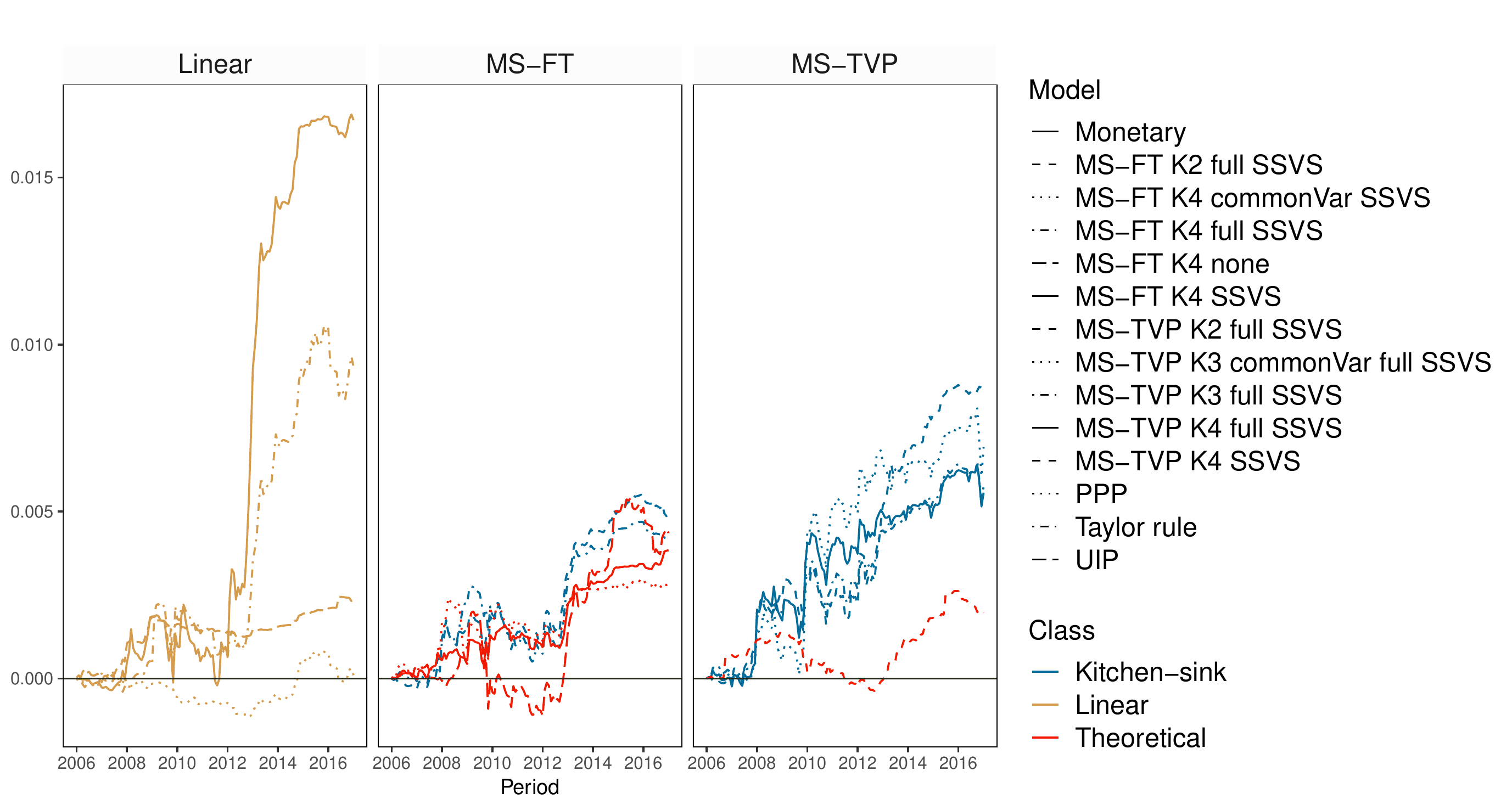}
	\caption{JP}
	\label{fig:outmse1JP}
	\end{subfigure}				
	\caption{Difference to the random walk of one-step-ahead CSFEs for the full hold-out sample for Australia, Canada, Switzerland and Japan. 'Linear' specifies the linear univariate exchange rate regressions. For the Markov switching models with time-varying transition probabilities ('MS-TVPs') and models with fixed transition probabilities ('MS-FT'), $K[2-4]$ specifies the number of states. We evaluate all models with a common state variance ('commonVar') and individual state variances, with both the theoretical state and the kitchen-sink ('full') state specification. Moreover, we estimate all Markov switching models with and without an SSVS prior. We consider the five best performing MS-TVP and and five best MS-FT models according to the CSFEs at the end of the hold-out-sample.}
	\label{fig:outmse1_1}
\end{figure}
\end{landscape}

\begin{landscape}
\begin{figure}[!h]
	\centering
	\begin{subfigure}{0.7\textwidth}
		\includegraphics[width=\textwidth]{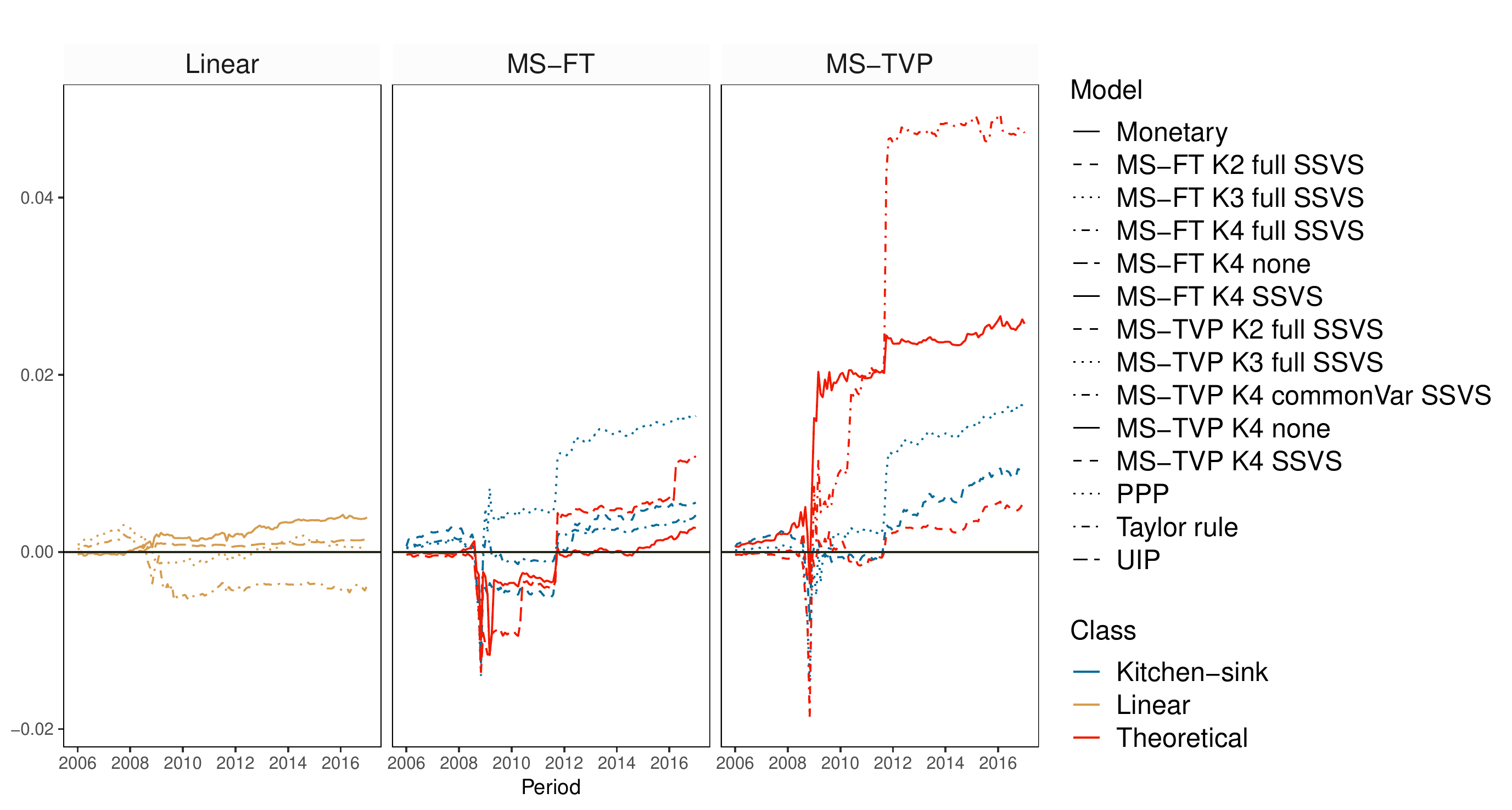}
	\caption{KR}
	\label{fig:outmse1KR}	
	\end{subfigure}
	\begin{subfigure}{0.7\textwidth}
		\includegraphics[width=\textwidth]{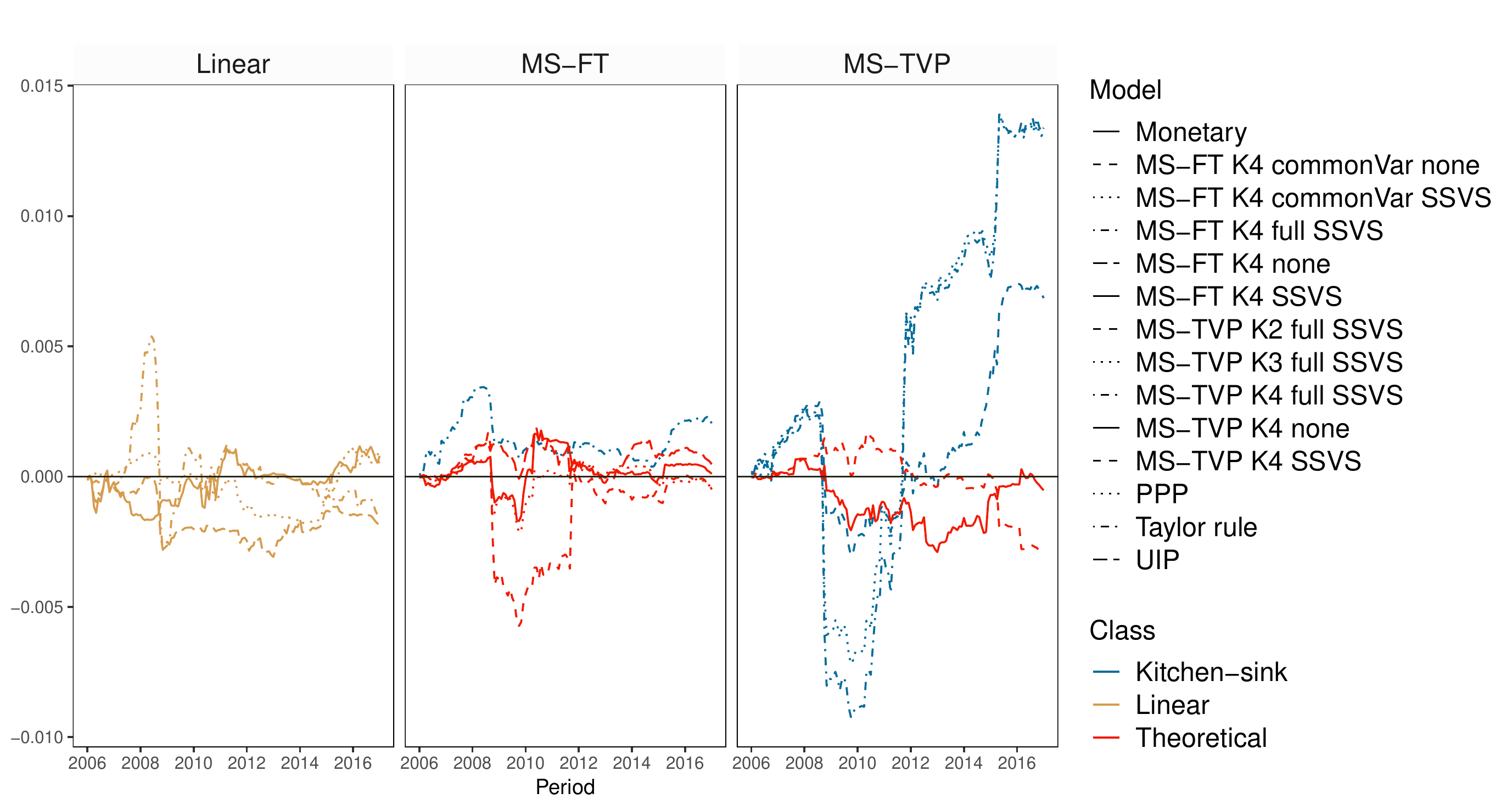}
	\caption{NO}
	\label{fig:out1NO}				
	\end{subfigure}
	\begin{subfigure}{0.7\textwidth}
		\includegraphics[width=\textwidth]{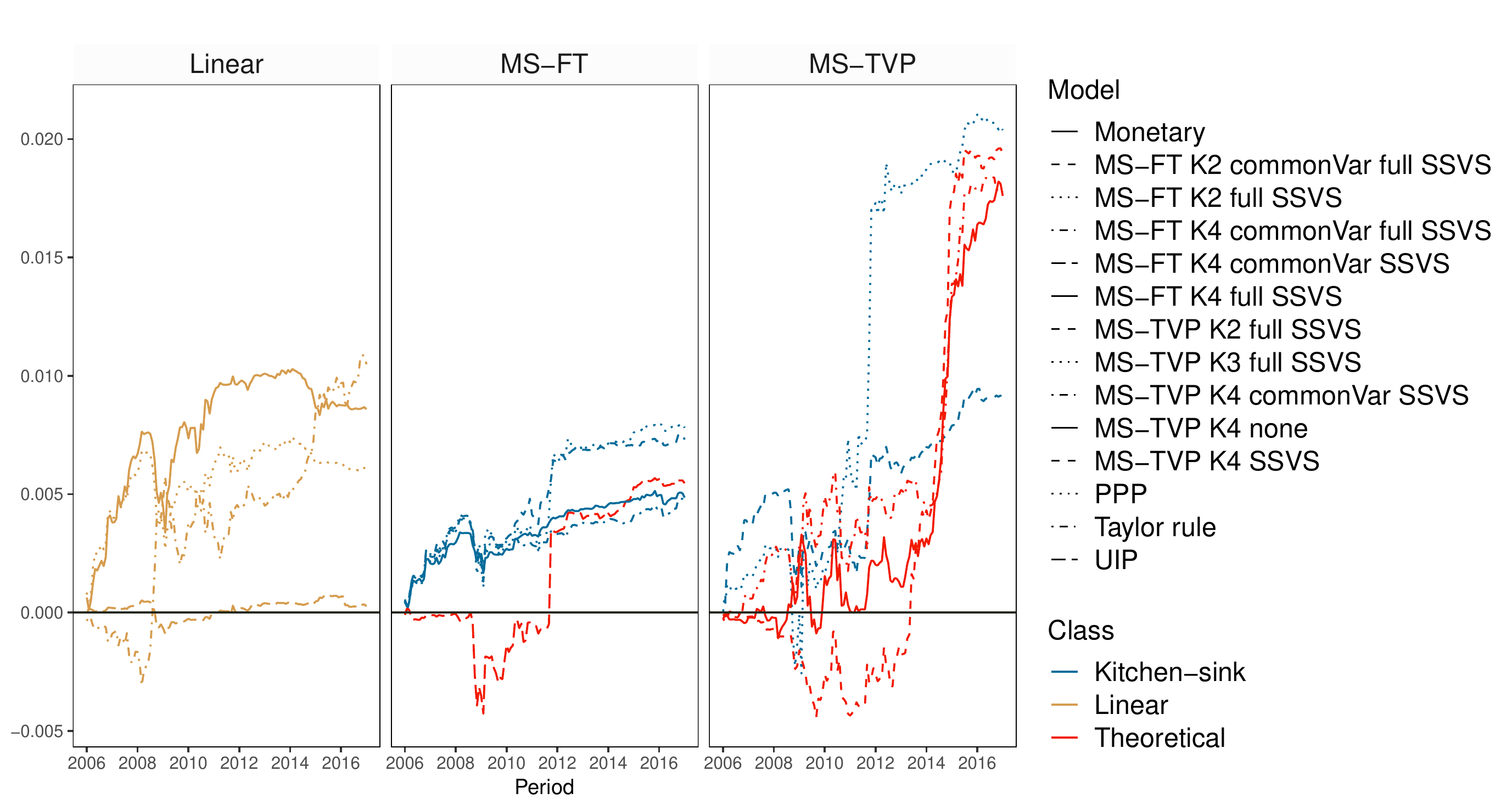}
	\caption{SE}
	\label{fig:outmse1SE}				
	\end{subfigure}	
	\begin{subfigure}{0.7\textwidth}
		\includegraphics[width=\textwidth]{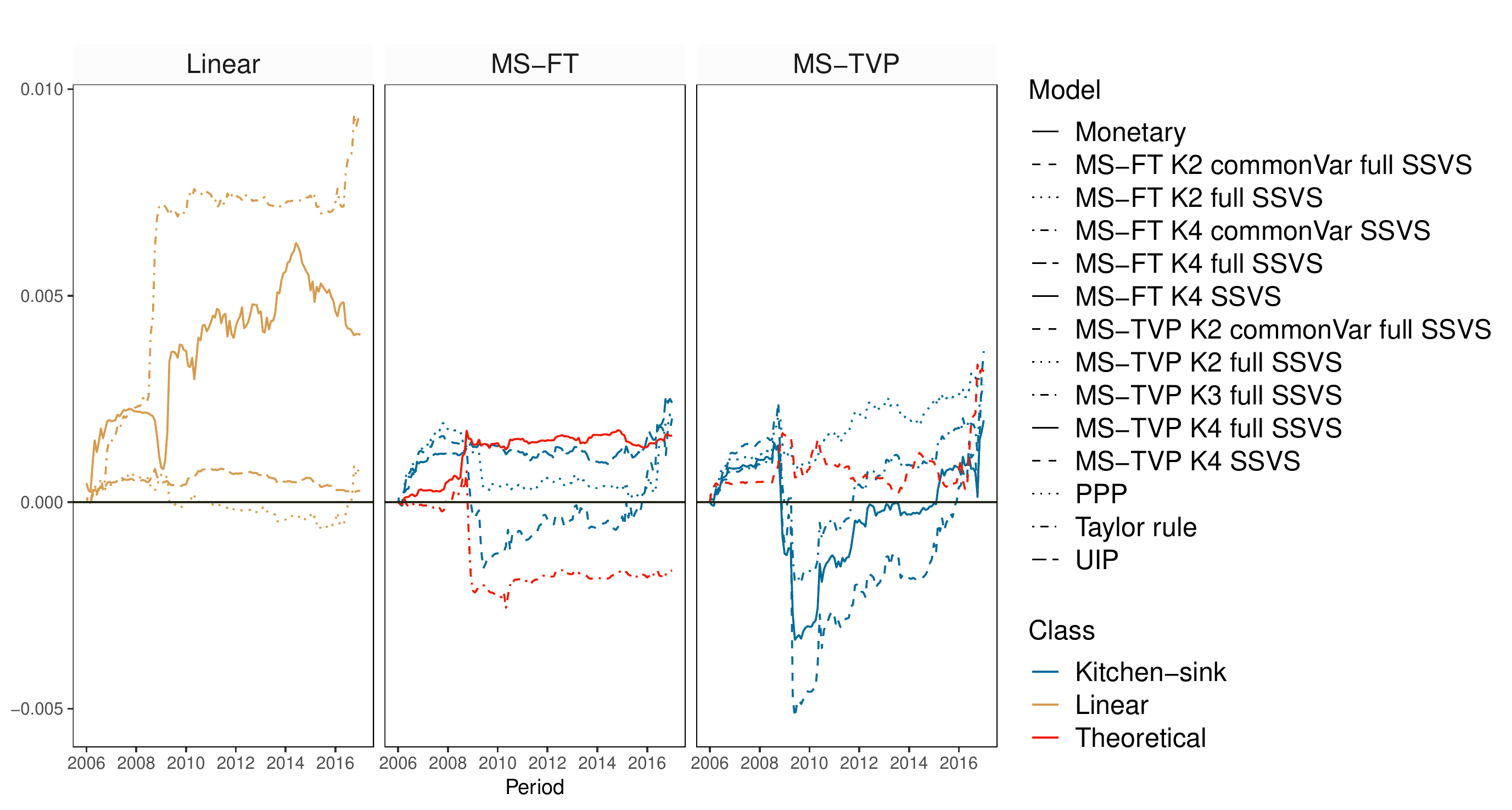}
	\caption{UK}
	\label{fig:outmse1UK}				
	\end{subfigure}		
	\caption{Difference to the random walk of one-step-ahead CSFEs for the full hold-out sample for South Korea, Norway, Sweden and the United Kingdom. 'Linear' specifies the linear univariate exchange rate regressions. For the Markov switching model with time-varying transition probabilities ('MS-TVPs') and models with fixed transition probabilities ('MS-FT'), $K[2-4]$ specifies the number of states. We evaluate all models with a common state variance ('commonVar') and individual state variances, with both the theoretical state and the kitchen-sink ('full') state specification. Moreover, we estimate all Markov switching models with and without an SSVS prior. We consider the five best performing MS-TVP and and five best MS-FT models according to the CSFEs at the end of the hold-out-sample.}
	\label{fig:outmse1_2}
\end{figure}
\end{landscape}

Figures \ref{fig:outmse1_1} and \ref{fig:outmse1_2} present the evolution of CSFEs of one-step-ahead forecasts for the best performing models across the considered model classes. We consider all linear predictive exchange rate regressions and the five best performing MS-TVP and MS-FT models, according the CSFEs at the end of the hold-out-sample. CSFEs of all models are presented relative to the CSFEs of the random walk benchmark. Thus, values below zero  indicate more accurate forecast relative to random walk predictions. Here, we focus on one-step-ahead forecasts since we find that models that perform well at the one-step-ahead horizon also do well for $h>1$ periods ahead.\footnote{Additional results for $h>1$ step ahead forecasts are provided on request.} When considering density forecasts, we report the results for higher order predictions as well.

Turning to the actual results, we observe pronounced differences across countries. For instance, in  Australia, Canada, Norway, Sweden and the United Kingdom, modeling non-linearities pays off, in particular during periods of financial turmoil, outperforming forecasts of linear models as well as the random walk benchmark. herefore, one interesting finding is that controlling for heteroscedasticity also tends to exert a positive effect on the point forecasting performance during volatile periods of the business cycle.  

By contrast, for South Korea and Switzerland, the random walk appears to be hard to beat. In general, we observe error ratios that are close to unity when averaged over the full hold-out sample. This indicates that including more information does not necessarily translate into improved point predictions relative to a simple no-change forecast for these two economies.  Again, we find some heterogeneity in relative forecasting performance over time.

Turning to the performance of the theoretically inspired MS-TVP specifications, we observe strong forecasting accuracy for Japan, Norway, South Korea, and the United Kingdom, at least for one specification (marked red in 
Figs. \ref{fig:out1_1} and \ref{fig:out1_2}). On the other hand, it appears that kitchen-sink specifications dominate theoretically inspired regimes for Australia, Canada, Switzerland and Sweden. In particular, this holds true for Switzerland as  indicated by the absence of a red colored line in \autoref{fig:out1CH} for MS-TVPs.

When comparing MS-TVP to MS-FT models, we observe that MS-FT specifications appear to be more robust over time in terms of CSFEs. This can be seen by noting that the forecast errors of the MS-FTs feature fewer  outliers. In general, we observe a better performance of MS-TVP models for Australia, Japan and Norway, at least for one model, compared to MS-FT models; although MS-FT models perform well  for the United Kingdom and Canada. 

Moreover, accuracy differences across simple linear specifications appears to be diverse. For the United Kingdom, Sweden and Japan, we observe an inferior predictive performance for at least two linear models. In particular, models based on  monetary fundamentals exhibit a  weak forecast performance relative to the remaining models under scrutiny. For Australia, Canada and South Korea, all linear models do well and show a similar point forecast performance as the random walk. When focusing on Taylor rule fundamentals, the linear regression performs well at the beginning of the hold-out-sample (see, for example, in the case of the United Kingdom), but exhibits a systematic loss of performance in periods after the financial crisis. This can be explained by the fact that Tayor rule based models build on the assumption that both central banks' monetary policy might be well described by a Taylor rule. However, after the financial crisis, interest rates hit the ZLB and central banks increasingly adopted non-standard policy measures. This, in turn, leads to a deteriorating  performance of this model class, effectively confirming findings reported in the recent literature  \citep[see, for example,][]{byrne2016exchange, molodtsova2013crisis}. 

\subsection*{Density forecasts}
\begin{landscape}
\begin{figure}[h]
	\centering
	\begin{subfigure}{0.7\textwidth}
		\includegraphics[width=\textwidth]{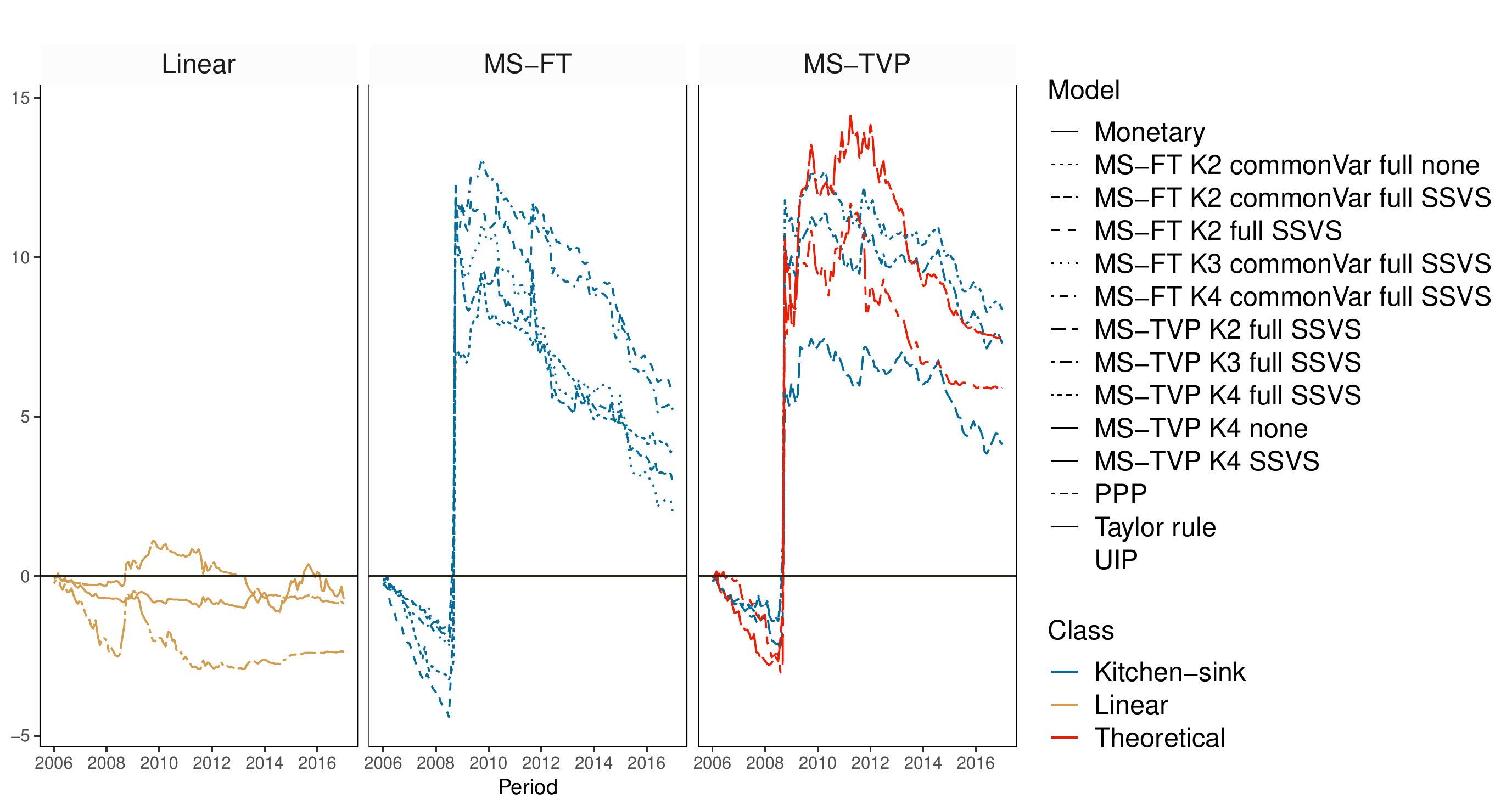}
	\caption{AU}
	\label{fig:out1AU}
	\end{subfigure}	
	\begin{subfigure}{0.7\textwidth}
		\includegraphics[width=\textwidth]{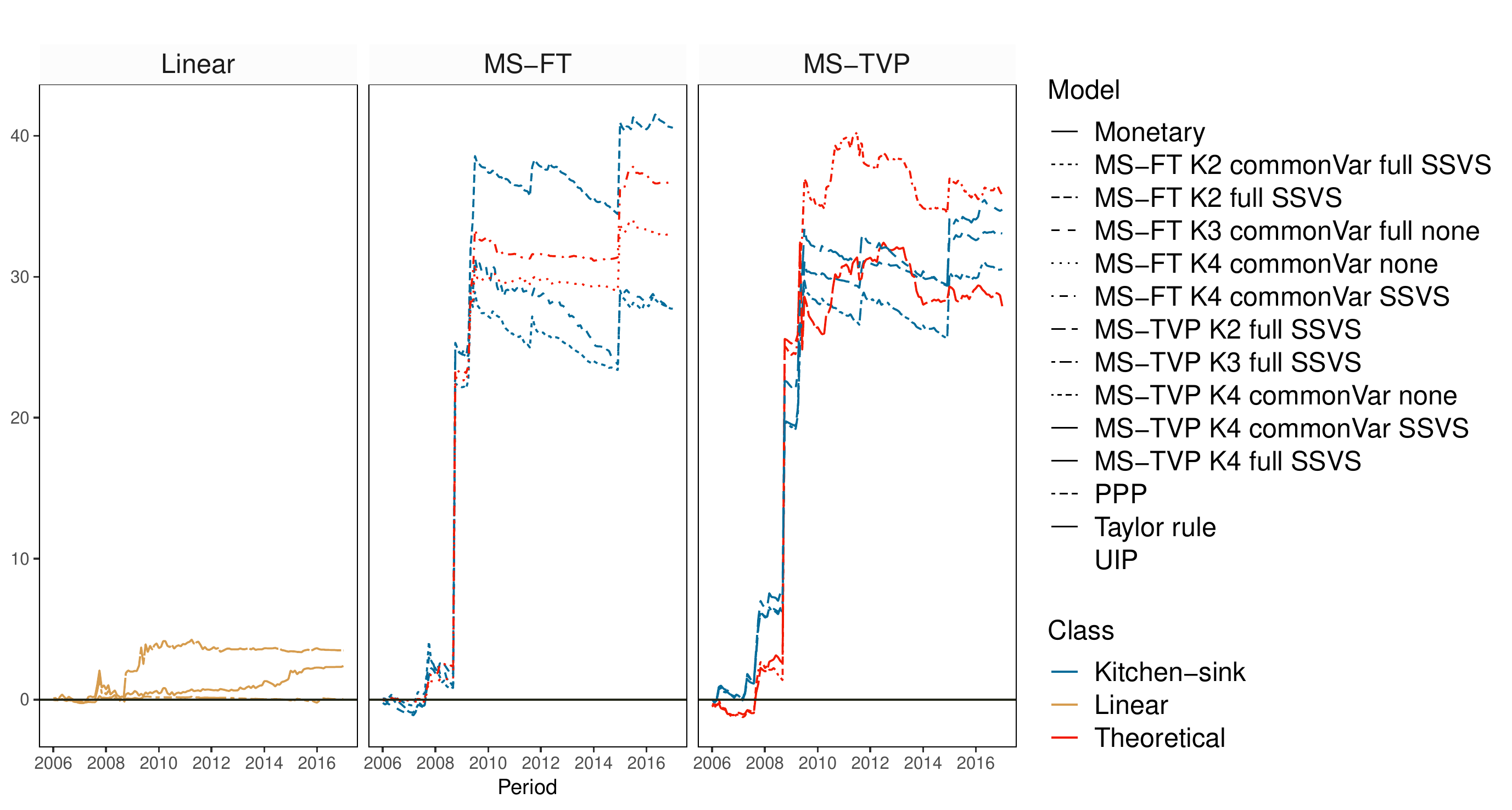}
	\caption{CA}
	\label{fig:out1CA}	
	\end{subfigure}
	\begin{subfigure}{0.7\textwidth}
		\includegraphics[width=\textwidth]{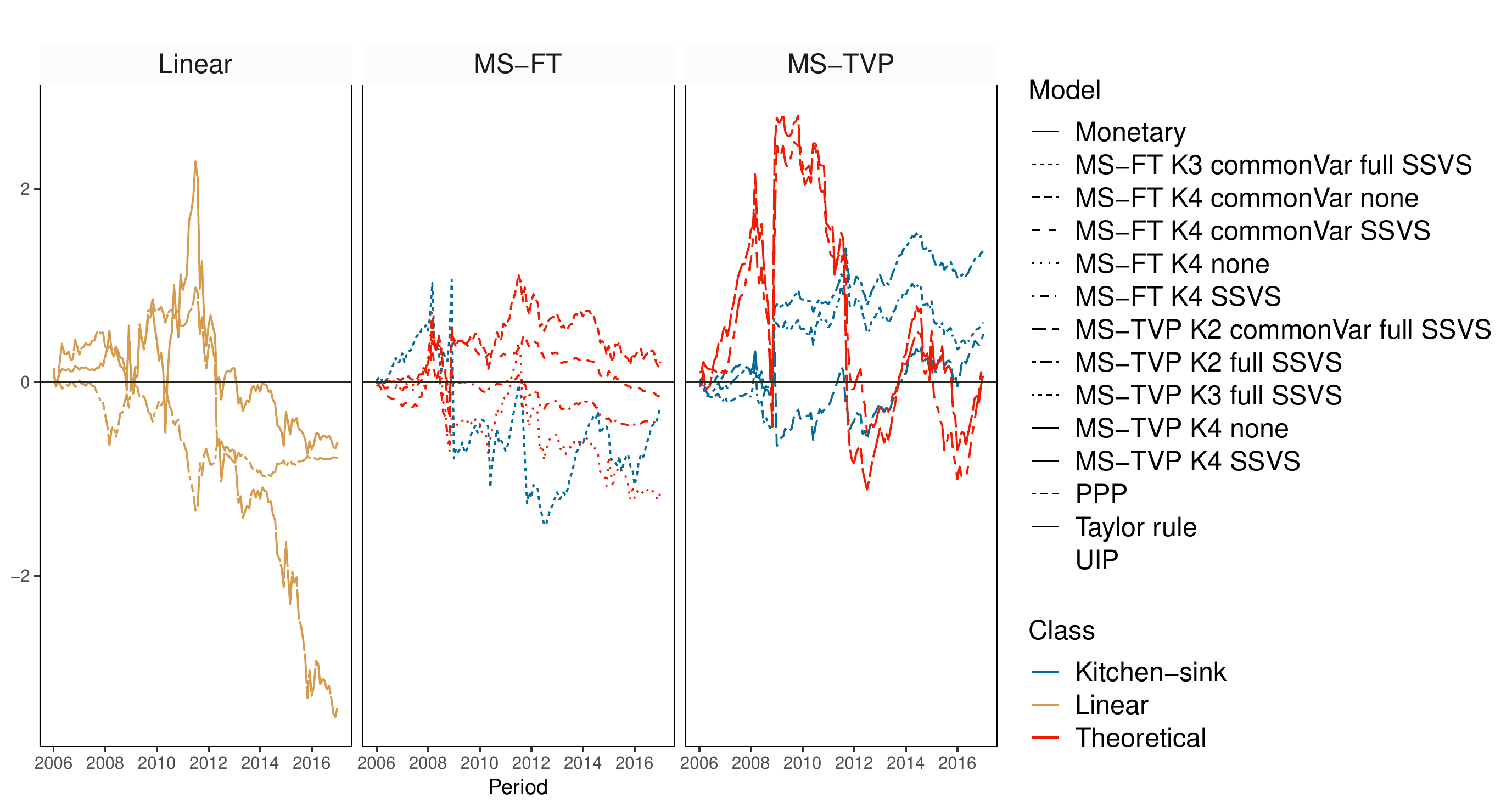}
	\caption{CH}
	\label{fig:out1CH}	
	\end{subfigure}	
	\begin{subfigure}{0.7\textwidth}
		\includegraphics[width=\textwidth]{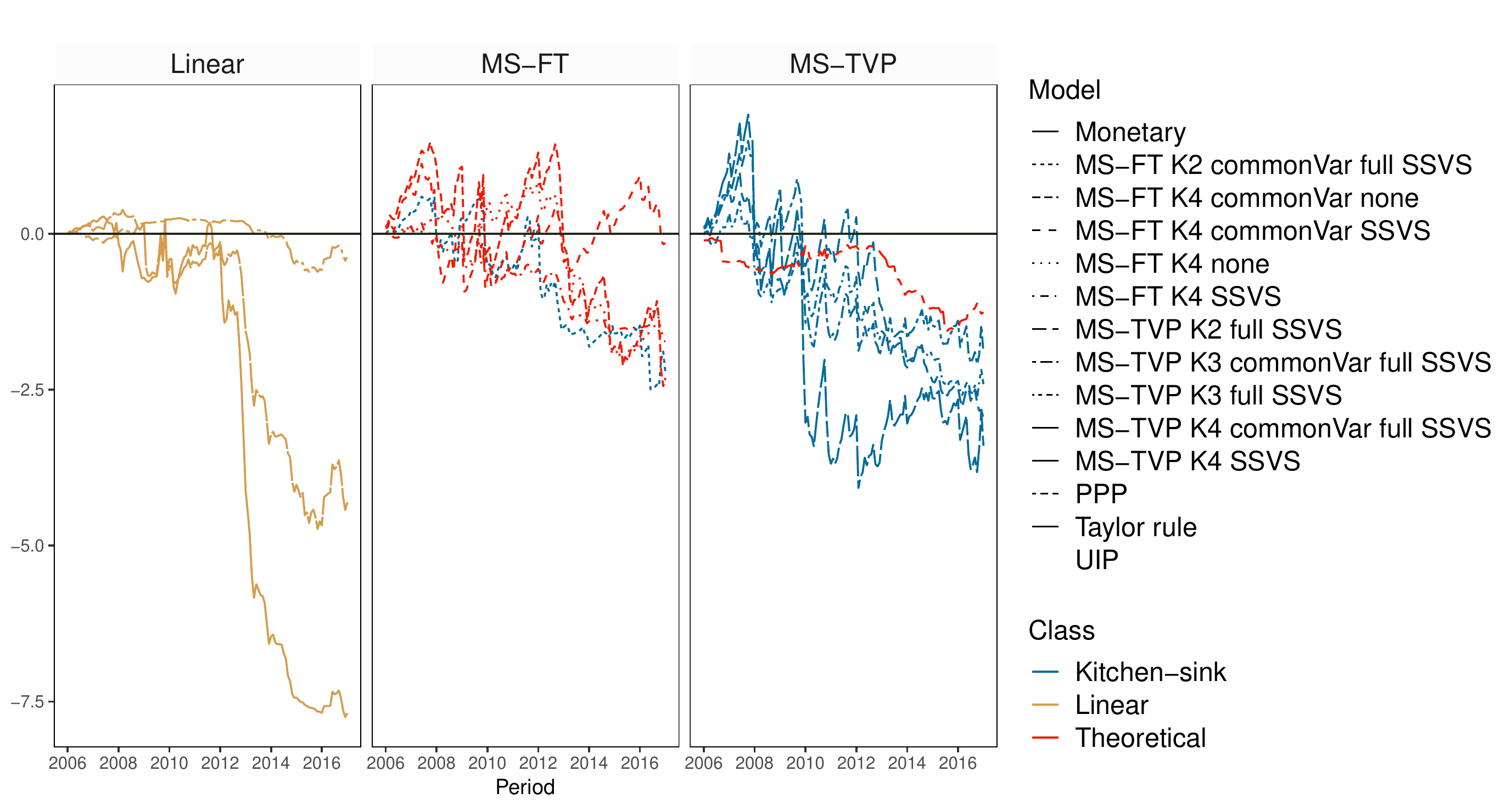}
	\caption{JP}
	\label{fig:out1JP}
	\end{subfigure}	
	\caption{Cumulative one-step-ahead LBFs (random walk benchmark) for the full hold-out sample for Australia, Canada, Switzerland and Japan. 'Linear' specifies the linear univariate exchange rate regressions. For the Markov switching models with time-varying transition probabilities ('MS-TVPs') and models with fixed transition probabilities ('MS-FT'), $K[2-4]$ specifies the number of states. We evaluate all models with a common state variance ('commonVar') and individual state variances, with both the theoretical state and the kitchen-sink ('full') state specification. Moreover, we estimate all Markov switching models with and without an SSVS prior. We consider the five best performing MS-TVP and and five best MS-FT models according to cumulative LBFs at the end of the hold-out.}
	\label{fig:out1_1}
\end{figure}
\end{landscape}

\begin{landscape}
\begin{figure}[h]
	\centering
	\begin{subfigure}{0.7\textwidth}
		\includegraphics[width=\textwidth]{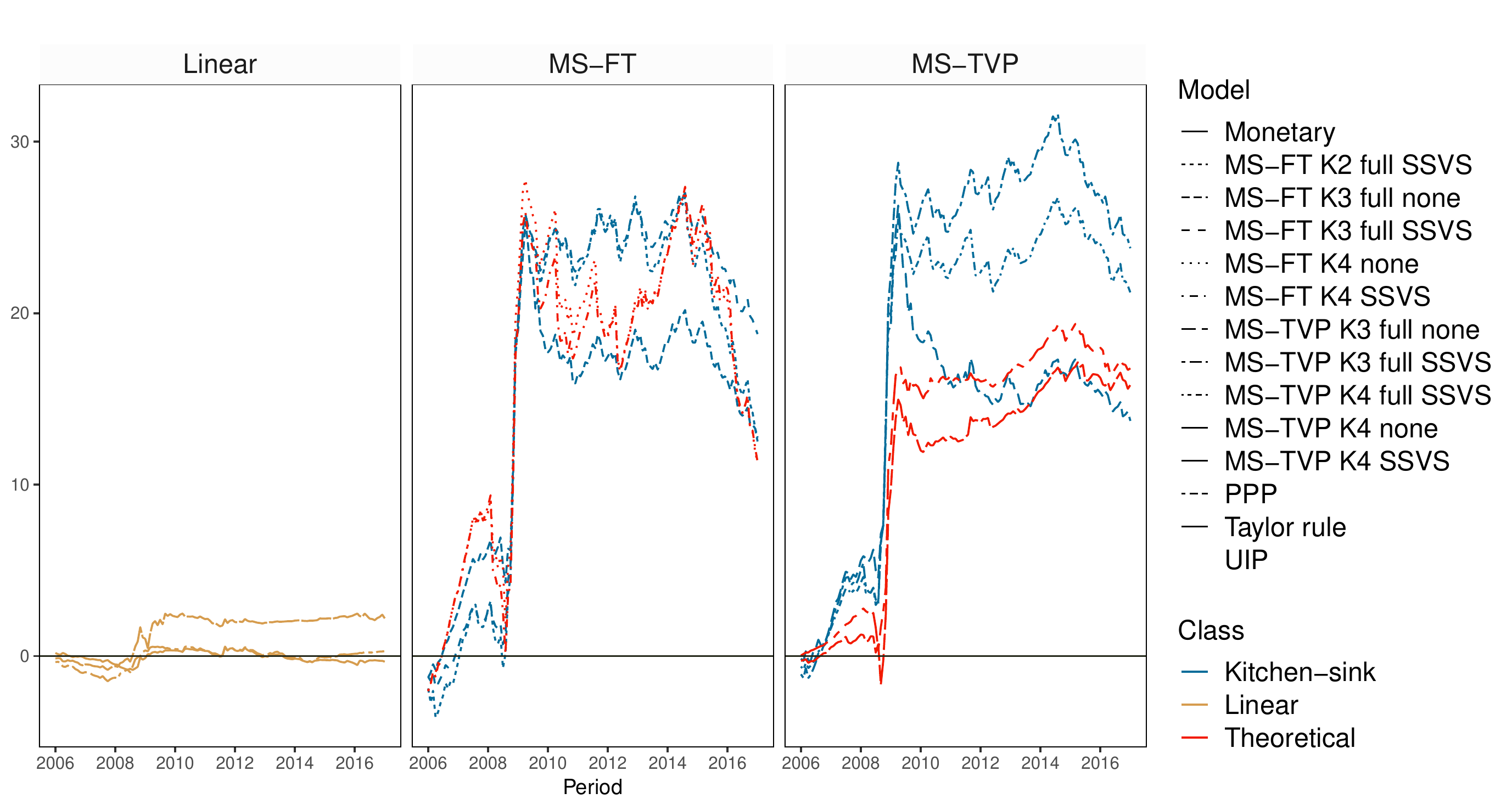}
	\caption{KR}
	\label{fig:out1KR}	
	\end{subfigure}
	\begin{subfigure}{0.7\textwidth}
		\includegraphics[width=\textwidth]{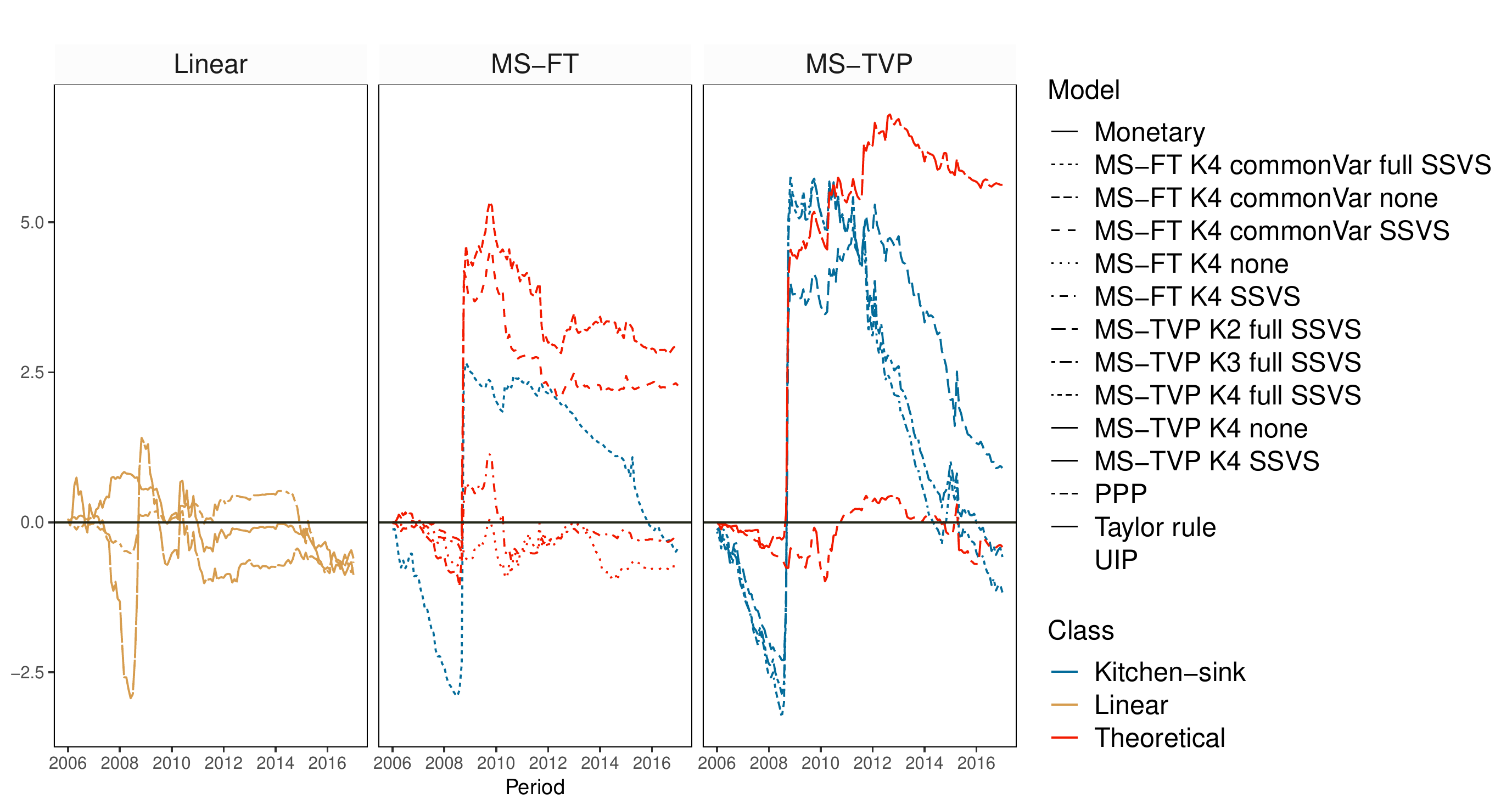}
	\caption{NO}
	\label{fig:out1NO}				
	\end{subfigure}
	\begin{subfigure}{0.7\textwidth}
		\includegraphics[width=\textwidth]{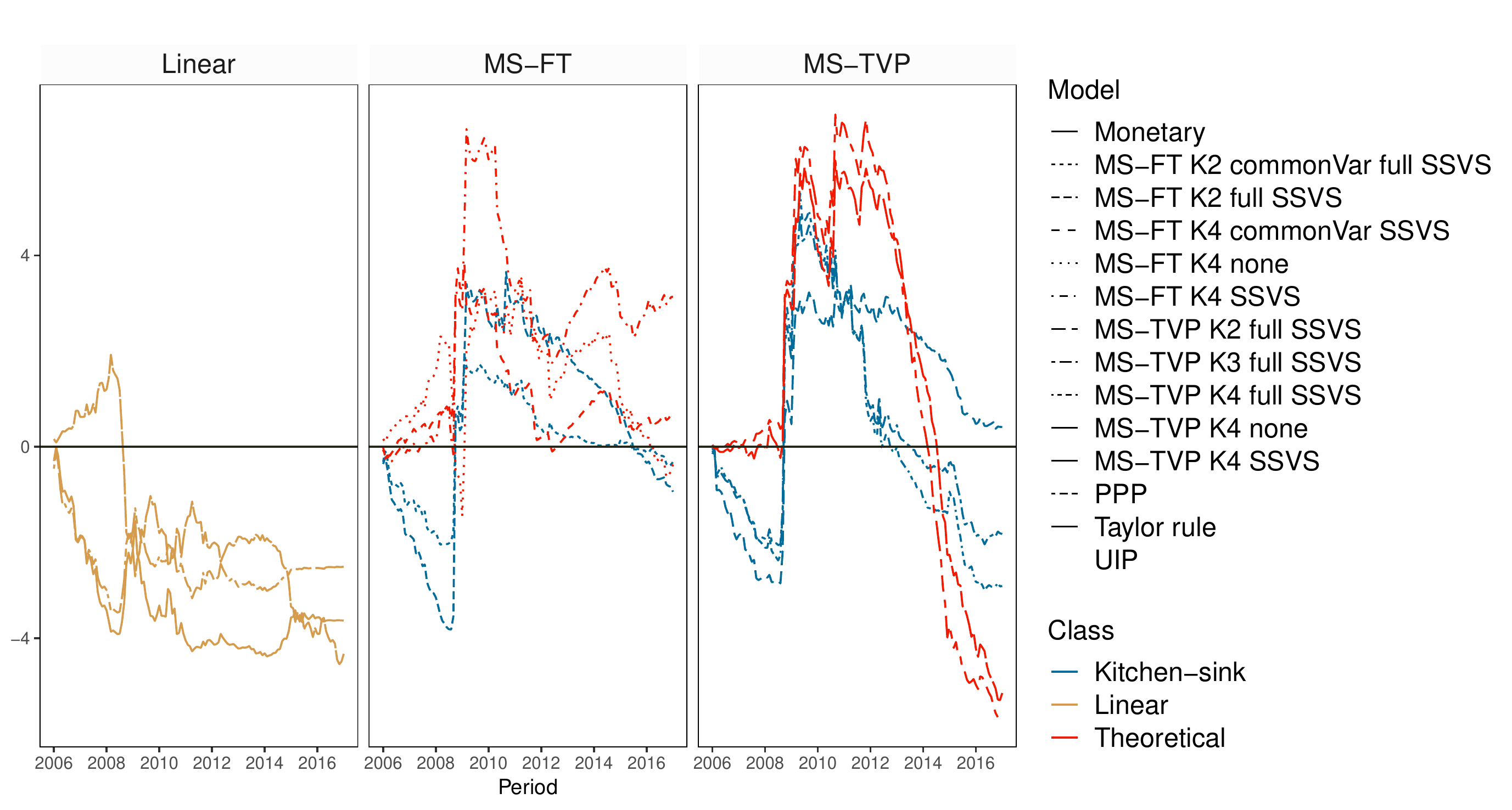}
	\caption{SE}
	\label{fig:out1SE}				
	\end{subfigure}	
	\begin{subfigure}{0.7\textwidth}
		\includegraphics[width=\textwidth]{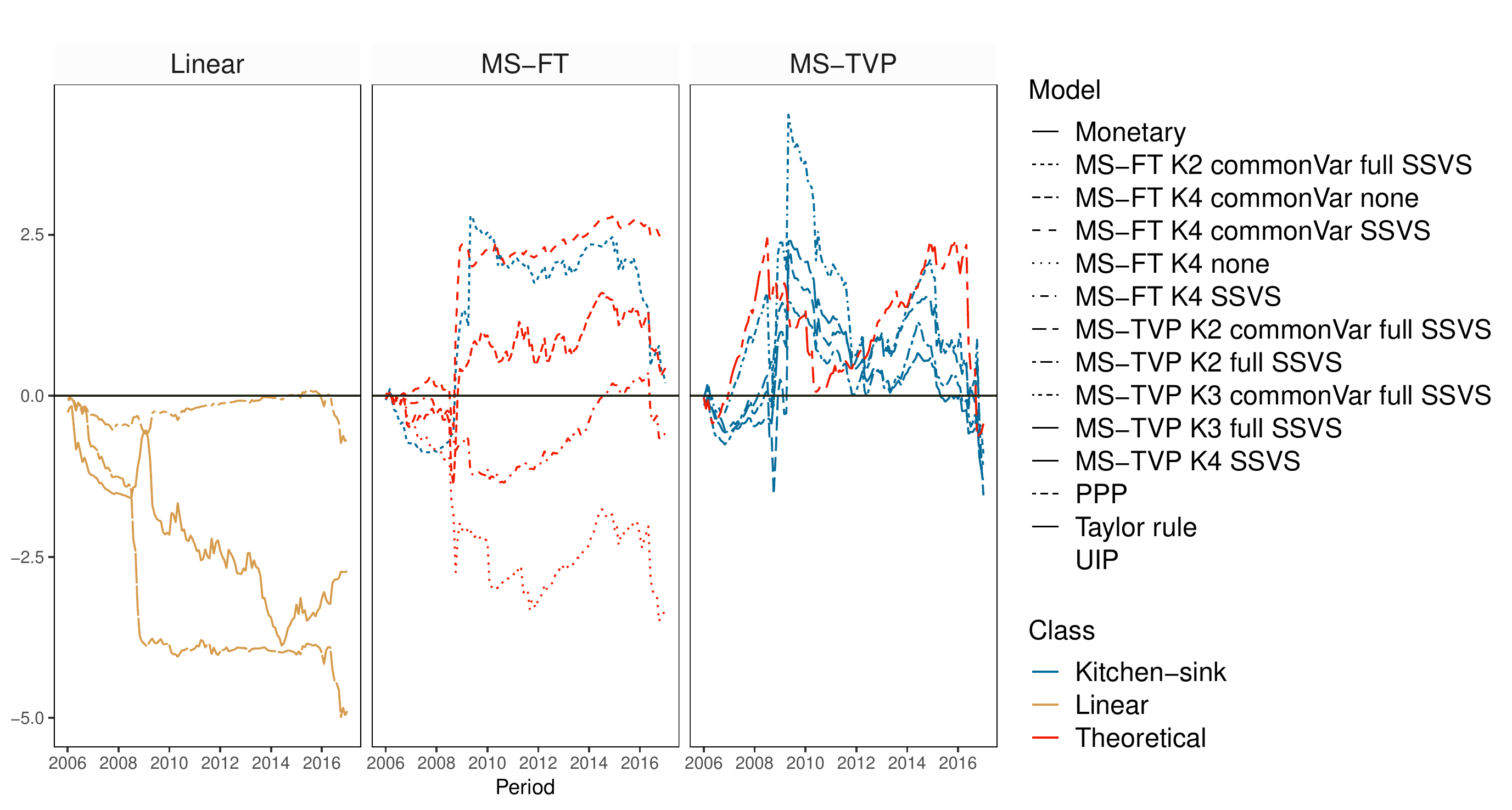}
	\caption{UK}
	\label{fig:out1UK}				
	\end{subfigure}		
	\caption{Cumulative one-step-ahead LBFs (random walk benchmark) for the full hold-out sample for South Korea, Norway, Sweden and the United Kingdom. 'Linear' specifies the linear univariate exchange rate regressions. For the Markov switching models with time-varying transition probabilities ('MS-TVPs') and models with fixed transition probabilities ('MS-FT'), $K[2-4]$ specifies the number of states. We evaluate all models with a common state variance ('commonVar') and individual state variances, with both the theoretical state and the kitchen-sink ('full') state specification. Moreover, we estimate all Markov switching models with and without an SSVS prior. We consider the five best performing MS-TVP and and five best MS-FT models cumulative LBFs at the end of the hold-out.}
	\label{fig:out1_2}
\end{figure}
\end{landscape}

Tables \ref{tab:out1} and \ref{tab:out2} depict a  summary of all models' LBFs for all currency pairs considered. Values highlighted in green point towards outperformance of a model relative to the random walk while red values signal a weaker predictive performance when benchmarked against the random walk. To provide a dynamic picture of  LBFs over time, Fig.s \ref{fig:out1_1} and \ref{fig:out1_2}, again, show the LBFs of all   linear predictive exchange rate regressions and the five best performing MS-TVP and MS-FT models.

In general, both tables attest non-linear specifications  good predictive power, while linear models display a somewhat weaker forecast performance. Furthermore, our results suggest that non-linear models that perform well in terms of point predictions also exhibit high predictive capabilities in terms of density forecasts. However, we find that predictive performance evolves differently for CSFEs and the LBFs. Density forecasts strengthen the argument in favor of non-linear models, as the performance gains of the MS models are even more sizable in periods of high exchange rate volatility while accuracy losses in tranquil times are rather muted. 

Although we do not observe a single dominant non-linear model across forecast horizons and countries, Tables \ref{tab:out1} and \ref{tab:out2} suggest that at least one non-linear specification outperforms the random walk and the linear counterparts as well.  One exception proves to be Japan, for which the best specification is either the random walk (for one-step ahead predictions) or the linear PPP model (for longer horizons). 

For all forecast horizons considered, non-linear specifications do well for Australia, Canada, South Korea, Sweden and the United Kingdom. Specifically, the MS-TVP kitchen-sink specification with four states coupled with state-specific variances, constituting the most flexible specification, has good predictive power for Australia, Canada and South Korea. For Australia, this model is the single best performing model across forecast horizons and for South Korea, it is the best for three- and twelve-step-ahead forecast and  the second best specification in terms of one-step ahead forecasts. Moreover, as shown in Figs. \ref{fig:out1_1} and \ref{fig:out1_2}, the poor performance of  linear exchange rate models is even more pronounced for density forecasts. In particular, the linear structural regression  display a sharp decline in predictive power after the financial crisis, corroborating findings in  \citet{byrne2016exchange, molodtsova2013crisis}.

Turning to the question whether allowing for  heteroscedasticty pays off in terms of density forecasting, we find substantial evidence that this additional flexibility proves to be important.  The gains in predictive accuracy of non-linear models can mainly be attributed to the more flexible variance specification of MS models.  This can also be seen by comparing   MS specifications with a common variance across states with MS models that feature individual state variances. For these models, we observe a slight accuracy premium relative to their homoscedastic counterparts. Allowing for state-specific variances thus appears to be  an important ingredient of a successful forecasting model. However,  this increased flexibility comes at a cost. Specifically, we observe that during normal periods, relative predictive accuracy declines steadily for several MS models (see, for example, \autoref{fig:out1SE}), in line with recent evidence provided in \citet{abbate2018point}. 

Contrasting MS-TVP with MS-FT models, Figs. \ref{fig:out1_1} and \ref{fig:out1_2} show that MS-TVP models yield more precise density forecasts  for Australia, Norway and South Korea and Switzerland, while yielding    an almost equivalent performance for Canada. Moreover, with time-varying transition probabilities, theoretically motivated specifications play a more important role than for MS-FT models. This points towards potential accuracy premia obtained by allowing for time-varying transition probabilities and thus sharpen inference surrounding the  regime allocation.

Theoretically motivated MS-TVP specifications exhibit good forecast performance for Australia, Canada and Norway. Considering the results in Sweden, theoretically motivated MS-TVPs perform well during prolonged periods of high exchange rate volatility. In periods of low volatility, however, this specification is slightly outperformed by competing specifications. By contrast, we find that for South Korea, MS-TVP kitchen-sink regressions improve upon our proposed MS-TVP model that allow for switching across structural exchange rate regressions. For the remaining countries in Figs. \ref{fig:out1_1} and \ref{fig:out1_2}, no clear pattern emerges when comparing both model types.  For example, using structural MS-TVP specifications yields strong increases in predictive power during the financial crisis but a weaker performance afterwards. For kitchen-sink MS-TVP regressions, we find no gain during the crisis but, at the same time, no subsequent loss in the aftermath of the crisis.

Finally, we assess whether using shrinkage priors on the coefficients improves forecasts. Tables \ref{tab:out1} and \ref{tab:out2}  indicate that shrinkage generally translates into better results in pairwise comparisons with the corresponding non-shrinkage counterpart. This observation, however, is not consistent across all models, countries and forecast horizons considered. In particular, using a kitchen-sink regression without shrinkage leads to poor forecast performance, as already shown by \citet{liwang2015, wright2008bayesian}. Turning to theoretically motivated MS models provides mixed insights on whether using shrinkage is useful.   We conjecture that this stems from the fact that adopting Markov switching specifications with theoretically defined regimes already introduces a certain amount of regularization that helps avoid overfitting.

\begin{table}[!tbp]
{\tiny
\begin{center}
\begin{tabular}{llllclrrrlrrr}
\toprule
\multicolumn{1}{l}{\bfseries }&\multicolumn{3}{c}{\bfseries Specification}&\multicolumn{1}{c}{\bfseries }&\multicolumn{8}{c}{\bfseries Log predicitive Bayes factor}\tabularnewline
\cline{2-4} \cline{6-13}
\multicolumn{1}{l}{}&\multicolumn{1}{c}{States}&\multicolumn{1}{c}{Shrinkage}&\multicolumn{1}{c}{Variance}&\multicolumn{1}{c}{}&\multicolumn{1}{c}{}&\multicolumn{1}{c}{1-step}&\multicolumn{1}{c}{3-step}&\multicolumn{1}{c}{12-step}&\multicolumn{1}{c}{}&\multicolumn{1}{c}{1-step}&\multicolumn{1}{c}{3-step}&\multicolumn{1}{c}{12-step}\tabularnewline
\midrule
{\scshape }&&&&&&&&&&&&\tabularnewline
   ~~&\cellcolor{white}   &\cellcolor{white}   &\cellcolor{white}   &   &\cellcolor{white}   AU&\cellcolor{white}   &\cellcolor{white}   &\cellcolor{white}   &\cellcolor{white}   CA&\cellcolor{white}   &\cellcolor{white}   &\cellcolor{white}   \tabularnewline
\midrule
{\scshape MS-TVP theoretical}&&&&&&&&&&&&\tabularnewline
   ~~&\cellcolor{white}   K = 4&\cellcolor{white}   None&\cellcolor{white}   Common&   &\cellcolor{white}   &\cellcolor{redc}   -16.208&\cellcolor{redc}   -11.860&\cellcolor{redc}    -7.476&\cellcolor{white}   &\cellcolor{greenc}    35.802&\cellcolor{greenc}    28.588&\cellcolor{greenc}    22.695\tabularnewline
   ~~&\cellcolor{white}   &\cellcolor{white}   None&\cellcolor{white}   State-specific&   &\cellcolor{white}   &\cellcolor{greenc}     7.449&\cellcolor{redc}    -0.068&\cellcolor{redc}    -0.953&\cellcolor{white}   &\cellcolor{greenc}     1.911&\cellcolor{greenc}     6.161&\cellcolor{greenc}    28.652\tabularnewline
   ~~&\cellcolor{white}   &\cellcolor{white}   SSVS&\cellcolor{white}   Common&   &\cellcolor{white}   &\cellcolor{redc}    -6.908&\cellcolor{redc}    -3.737&\cellcolor{greenc}     2.113&\cellcolor{white}   &\cellcolor{greenc}    27.912&\cellcolor{greenc}    24.035&\cellcolor{greenc}    27.860\tabularnewline
   ~~&\cellcolor{white}   &\cellcolor{white}   SSVS&\cellcolor{white}   State-specific&   &\cellcolor{white}   &\cellcolor{greenc}     5.904&\cellcolor{redc}    -0.597&\cellcolor{greenc}     4.272&\cellcolor{white}   &\cellcolor{greenc}     3.577&\cellcolor{greenc}    23.006&\cellcolor{greenc}    10.700\tabularnewline
\midrule
{\scshape MS-FT theoretical}&&&&&&&&&&&&\tabularnewline
   ~~&\cellcolor{white}   K = 4&\cellcolor{white}   None&\cellcolor{white}   Common&   &\cellcolor{white}   &\cellcolor{redc}    -1.941&\cellcolor{redc}    -2.809&\cellcolor{redc}    -1.232&\cellcolor{white}   &\cellcolor{greenc}    32.992&\cellcolor{greenc}    29.238&\cellcolor{greenc}    29.081\tabularnewline
   ~~&\cellcolor{white}   &\cellcolor{white}   None&\cellcolor{white}   State-specific&   &\cellcolor{white}   &\cellcolor{greenc}     1.693&\cellcolor{greenc}     1.070&\cellcolor{redc}    -2.181&\cellcolor{white}   &\cellcolor{greenc}     8.760&\cellcolor{greenc}     7.406&\cellcolor{greenc}     1.590\tabularnewline
   ~~&\cellcolor{white}   &\cellcolor{white}   SSVS&\cellcolor{white}   Common&   &\cellcolor{white}   &\cellcolor{greenc}     1.696&\cellcolor{greenc}     0.602&\cellcolor{greenc}     0.386&\cellcolor{white}   &\cellcolor{greenc}    36.737&\cellcolor{greenc}    28.763&\cellcolor{greenc}    30.674\tabularnewline
   ~~&\cellcolor{white}   &\cellcolor{white}   SSVS&\cellcolor{white}   State-specific&   &\cellcolor{white}   &\cellcolor{greenc}     1.261&\cellcolor{greenc}     0.144&\cellcolor{redc}    -2.762&\cellcolor{white}   &\cellcolor{greenc}     7.856&\cellcolor{greenc}     9.720&\cellcolor{greenc}     7.995\tabularnewline
\midrule
{\scshape MS-TVP kitchen-sink}&&&&&&&&&&&&\tabularnewline
   ~~&\cellcolor{white}   K = 2&\cellcolor{white}   SSVS&\cellcolor{white}   Common&   &\cellcolor{white}   &\cellcolor{redc}   -22.236&\cellcolor{redc}   -12.630&\cellcolor{redc}   -26.266&\cellcolor{white}   &\cellcolor{redc}   -42.757&\cellcolor{redc}   -38.942&\cellcolor{redc}   -43.791\tabularnewline
   ~~&\cellcolor{white}   &\cellcolor{white}   SSVS&\cellcolor{white}   State-specific&   &\cellcolor{white}   &\cellcolor{greenc}     4.142&\cellcolor{greenc}     5.954&\cellcolor{greenc}     2.386&\cellcolor{white}   &\cellcolor{greenc}    33.067&\cellcolor{greenc}    33.136&\cellcolor{greenc}    31.270\tabularnewline
   ~~&\cellcolor{white}   K = 3&\cellcolor{white}   SSVS&\cellcolor{white}   Common&   &\cellcolor{white}   &\cellcolor{redc}   -19.408&\cellcolor{redc}   -19.043&\cellcolor{redc}   -24.598&\cellcolor{white}   &\cellcolor{redc}   -47.925&\cellcolor{redc}   -47.945&\cellcolor{redc}   -50.501\tabularnewline
   ~~&\cellcolor{white}   &\cellcolor{white}   SSVS&\cellcolor{white}   State-specific&   &\cellcolor{white}   &\cellcolor{greenc}     7.306&\cellcolor{greenc}     6.406&\cellcolor{greenc}     5.417&\cellcolor{white}   &\cellcolor{greenc}    30.543&\cellcolor{greenc}    31.877&\cellcolor{greenc}    24.891\tabularnewline
   ~~&\cellcolor{white}   K = 4&\cellcolor{white}   SSVS&\cellcolor{white}   Common&   &\cellcolor{white}   &\cellcolor{redc}   -23.884&\cellcolor{redc}   -29.087&\cellcolor{redc}   -29.255&\cellcolor{white}   &\cellcolor{redc}   -50.257&\cellcolor{redc}   -57.303&\cellcolor{redc}   -53.021\tabularnewline
   ~~&\cellcolor{white}   &\cellcolor{white}   SSVS&\cellcolor{white}   State-specific&   &\cellcolor{white}   &\cellcolor{greenc}     \textbf{8.342} &\cellcolor{greenc}     \textbf{7.407} &\cellcolor{greenc}     \textbf{6.263} &\cellcolor{white}   &\cellcolor{greenc}    34.765&\cellcolor{greenc}    35.336&\cellcolor{greenc}    29.015\tabularnewline
\midrule
{\scshape MS-FT kitchen-sink}&&&&&&&&&&&&\tabularnewline
   ~~&\cellcolor{white}   K = 2&\cellcolor{white}   SSVS&\cellcolor{white}   Common&   &\cellcolor{white}   &\cellcolor{greenc}     2.911&\cellcolor{greenc}     0.250&\cellcolor{greenc}     0.201&\cellcolor{white}   &\cellcolor{greenc}    27.733&\cellcolor{greenc}    27.812&\cellcolor{greenc}    24.748\tabularnewline
   ~~&\cellcolor{white}   &\cellcolor{white}   SSVS&\cellcolor{white}   State-specific&   &\cellcolor{white}   &\cellcolor{greenc}     5.765&\cellcolor{greenc}     4.801&\cellcolor{greenc}     3.528&\cellcolor{white}   &\cellcolor{greenc}    \textbf{40.572} &\cellcolor{greenc}    \textbf{39.997} &\cellcolor{greenc}    \textbf{37.250} \tabularnewline
   ~~&\cellcolor{white}   K = 3&\cellcolor{white}   SSVS&\cellcolor{white}   Common&   &\cellcolor{white}   &\cellcolor{greenc}     1.962&\cellcolor{greenc}     3.487&\cellcolor{greenc}     1.319&\cellcolor{white}   &\cellcolor{redc}   -31.072&\cellcolor{redc}   -25.389&\cellcolor{redc}   -23.545\tabularnewline
   ~~&\cellcolor{white}   &\cellcolor{white}   SSVS&\cellcolor{white}   State-specific&   &\cellcolor{white}   &\cellcolor{greenc}     1.856&\cellcolor{greenc}     1.061&\cellcolor{greenc}     0.368&\cellcolor{white}   &\cellcolor{greenc}    25.606&\cellcolor{greenc}    24.982&\cellcolor{greenc}    22.641\tabularnewline
   ~~&\cellcolor{white}   K = 4&\cellcolor{white}   SSVS&\cellcolor{white}   Common&   &\cellcolor{white}   &\cellcolor{greenc}     5.192&\cellcolor{greenc}     3.376&\cellcolor{greenc}     3.624&\cellcolor{white}   &\cellcolor{greenc}    24.224&\cellcolor{greenc}    27.691&\cellcolor{greenc}    19.228\tabularnewline
   ~~&\cellcolor{white}   &\cellcolor{white}   SSVS&\cellcolor{white}   State-specific&   &\cellcolor{white}   &\cellcolor{redc}   -16.540&\cellcolor{redc}   -17.012&\cellcolor{redc}   -17.715&\cellcolor{white}   &\cellcolor{greenc}    15.123&\cellcolor{greenc}    14.930&\cellcolor{greenc}    12.654\tabularnewline
\midrule
{\scshape Linear}&&&&&&&&&&&&\tabularnewline
   ~~&\cellcolor{white}   Taylor rule&\cellcolor{white}   &\cellcolor{white}   &   &\cellcolor{white}   &\cellcolor{redc}    -0.866&\cellcolor{redc}    -1.661&\cellcolor{redc}    -0.592&\cellcolor{white}   &\cellcolor{greenc}     3.496&\cellcolor{greenc}     3.324&\cellcolor{redc}    -0.164\tabularnewline
   ~~&\cellcolor{white}   Monetary&\cellcolor{white}   &\cellcolor{white}   &   &\cellcolor{white}   &\cellcolor{redc}    -0.709&\cellcolor{redc}    -0.052&\cellcolor{redc}    -1.914&\cellcolor{white}   &\cellcolor{greenc}     2.379&\cellcolor{greenc}     1.594&\cellcolor{greenc}     0.007\tabularnewline
   ~~&\cellcolor{white}   PPP&\cellcolor{white}   &\cellcolor{white}   &   &\cellcolor{white}   &\cellcolor{redc}    -2.362&\cellcolor{redc}    -2.061&\cellcolor{redc}    -3.797&\cellcolor{white}   &\cellcolor{greenc}     0.050&\cellcolor{greenc}     0.347&\cellcolor{redc}    -1.133\tabularnewline
   ~~&\cellcolor{white}   UIP&\cellcolor{white}   &\cellcolor{white}   &   &\cellcolor{white}   &\cellcolor{redc}    -0.389&\cellcolor{redc}    -0.286&\cellcolor{redc}    -0.580&\cellcolor{white}   &\cellcolor{redc}    -0.479&\cellcolor{greenc}     0.478&\cellcolor{greenc}     0.207\tabularnewline
\midrule
\midrule
{\scshape }&&&&&&&&&&&&\tabularnewline
   ~~&\cellcolor{white}   &\cellcolor{white}   &\cellcolor{white}   &   &\cellcolor{white}   CH&\cellcolor{white}   &\cellcolor{white}   &\cellcolor{white}   &\cellcolor{white}   JP&\cellcolor{white}   &\cellcolor{white}   &\cellcolor{white}   \tabularnewline
\midrule
{\scshape MS-TVP theoretical}&&&&&&&&&&&&\tabularnewline
   ~~&\cellcolor{white}   K = 4&\cellcolor{white}   None&\cellcolor{white}   Common&   &\cellcolor{white}   &\cellcolor{redc}    -6.457&\cellcolor{redc}    -4.135&\cellcolor{redc}    -2.080&\cellcolor{white}   &\cellcolor{redc}   -11.153&\cellcolor{redc}   -10.498&\cellcolor{redc}    -6.965\tabularnewline
   ~~&\cellcolor{white}   &\cellcolor{white}   None&\cellcolor{white}   State-specific&   &\cellcolor{white}   &\cellcolor{greenc}     0.081&\cellcolor{redc}    -3.537&\cellcolor{greenc}     \textbf{0.655} &\cellcolor{white}   &\cellcolor{redc}    -5.688&\cellcolor{redc}    -7.339&\cellcolor{redc}    -1.672\tabularnewline
   ~~&\cellcolor{white}   &\cellcolor{white}   SSVS&\cellcolor{white}   Common&   &\cellcolor{white}   &\cellcolor{redc}    -4.213&\cellcolor{redc}    -6.261&\cellcolor{redc}    -0.866&\cellcolor{white}   &\cellcolor{redc}    -5.443&\cellcolor{redc}    -4.734&\cellcolor{redc}    -0.660\tabularnewline
   ~~&\cellcolor{white}   &\cellcolor{white}   SSVS&\cellcolor{white}   State-specific&   &\cellcolor{white}   &\cellcolor{greenc}     0.176&\cellcolor{redc}    -2.195&\cellcolor{redc}    -1.363&\cellcolor{white}   &\cellcolor{redc}    -1.258&\cellcolor{redc}    -3.183&\cellcolor{redc}    -0.842\tabularnewline
\midrule
{\scshape MS-FT theoretical}&&&&&&&&&&&&\tabularnewline
   ~~&\cellcolor{white}   K = 4&\cellcolor{white}   None&\cellcolor{white}   Common&   &\cellcolor{white}   &\cellcolor{greenc}     0.210&\cellcolor{redc}    -1.905&\cellcolor{redc}    -2.192&\cellcolor{white}   &\cellcolor{redc}    -2.310&\cellcolor{redc}    -3.072&\cellcolor{redc}    -0.849\tabularnewline
   ~~&\cellcolor{white}   &\cellcolor{white}   None&\cellcolor{white}   State-specific&   &\cellcolor{white}   &\cellcolor{redc}    -1.142&\cellcolor{redc}    -1.636&\cellcolor{redc}    -1.787&\cellcolor{white}   &\cellcolor{redc}    -1.599&\cellcolor{redc}    -1.536&\cellcolor{redc}    -3.688\tabularnewline
   ~~&\cellcolor{white}   &\cellcolor{white}   SSVS&\cellcolor{white}   Common&   &\cellcolor{white}   &\cellcolor{redc}    -0.136&\cellcolor{redc}    -0.471&\cellcolor{redc}    -2.668&\cellcolor{white}   &\cellcolor{redc}    -0.151&\cellcolor{redc}    -1.152&\cellcolor{redc}    -1.759\tabularnewline
   ~~&\cellcolor{white}   &\cellcolor{white}   SSVS&\cellcolor{white}   State-specific&   &\cellcolor{white}   &\cellcolor{redc}    -0.385&\cellcolor{redc}    -0.431&\cellcolor{redc}    -3.329&\cellcolor{white}   &\cellcolor{redc}    -1.722&\cellcolor{redc}    -0.936&\cellcolor{redc}    -1.896\tabularnewline
\midrule
{\scshape MS-TVP kitchen-sink}&&&&&&&&&&&&\tabularnewline
   ~~&\cellcolor{white}   K = 2&\cellcolor{white}   SSVS&\cellcolor{white}   Common&   &\cellcolor{white}   &\cellcolor{greenc}     0.497&\cellcolor{redc}    -2.012&\cellcolor{redc}    -2.175&\cellcolor{white}   &\cellcolor{redc}    -9.272&\cellcolor{redc}    -2.967&\cellcolor{redc}    -1.222\tabularnewline
   ~~&\cellcolor{white}   &\cellcolor{white}   SSVS&\cellcolor{white}   State-specific&   &\cellcolor{white}   &\cellcolor{greenc}     \textbf{1.353} &\cellcolor{greenc}     \textbf{0.408} &\cellcolor{redc}    -0.569&\cellcolor{white}   &\cellcolor{redc}    -2.964&\cellcolor{redc}    -2.992&\cellcolor{redc}    -0.679\tabularnewline
   ~~&\cellcolor{white}   K = 3&\cellcolor{white}   SSVS&\cellcolor{white}   Common&   &\cellcolor{white}   &\cellcolor{redc}    -0.213&\cellcolor{redc}    -2.987&\cellcolor{redc}    -4.000&\cellcolor{white}   &\cellcolor{redc}    -1.906&\cellcolor{redc}    -2.174&\cellcolor{greenc}     0.331\tabularnewline
   ~~&\cellcolor{white}   &\cellcolor{white}   SSVS&\cellcolor{white}   State-specific&   &\cellcolor{white}   &\cellcolor{greenc}     0.619&\cellcolor{redc}    -0.520&\cellcolor{redc}    -1.308&\cellcolor{white}   &\cellcolor{redc}    -2.411&\cellcolor{redc}    -2.367&\cellcolor{redc}    -1.370\tabularnewline
   ~~&\cellcolor{white}   K = 4&\cellcolor{white}   SSVS&\cellcolor{white}   Common&   &\cellcolor{white}   &\cellcolor{redc}    -1.158&\cellcolor{redc}    -5.658&\cellcolor{redc}    -5.427&\cellcolor{white}   &\cellcolor{redc}    -3.400&\cellcolor{redc}    -3.803&\cellcolor{redc}    -1.843\tabularnewline
   ~~&\cellcolor{white}   &\cellcolor{white}   SSVS&\cellcolor{white}   State-specific&   &\cellcolor{white}   &\cellcolor{redc}    -0.543&\cellcolor{redc}    -1.922&\cellcolor{redc}    -2.322&\cellcolor{white}   &\cellcolor{redc}    -3.669&\cellcolor{redc}    -3.662&\cellcolor{redc}    -3.095\tabularnewline
\midrule
{\scshape MS-FT kitchen-sink}&&&&&&&&&&&&\tabularnewline
   ~~&\cellcolor{white}   K = 2&\cellcolor{white}   SSVS&\cellcolor{white}   Common&   &\cellcolor{white}   &\cellcolor{redc}    -2.365&\cellcolor{redc}    -2.906&\cellcolor{redc}    -2.920&\cellcolor{white}   &\cellcolor{redc}    -2.230&\cellcolor{redc}    -3.487&\cellcolor{redc}    -2.038\tabularnewline
   ~~&\cellcolor{white}   &\cellcolor{white}   SSVS&\cellcolor{white}   State-specific&   &\cellcolor{white}   &\cellcolor{redc}    -5.686&\cellcolor{redc}    -6.616&\cellcolor{redc}    -6.344&\cellcolor{white}   &\cellcolor{redc}    -6.240&\cellcolor{redc}    -8.560&\cellcolor{redc}    -8.543\tabularnewline
   ~~&\cellcolor{white}   K = 3&\cellcolor{white}   SSVS&\cellcolor{white}   Common&   &\cellcolor{white}   &\cellcolor{redc}    -0.256&\cellcolor{redc}    -3.726&\cellcolor{redc}    -4.255&\cellcolor{white}   &\cellcolor{redc}    -3.233&\cellcolor{redc}    -4.880&\cellcolor{redc}    -3.300\tabularnewline
   ~~&\cellcolor{white}   &\cellcolor{white}   SSVS&\cellcolor{white}   State-specific&   &\cellcolor{white}   &\cellcolor{redc}    -6.504&\cellcolor{redc}    -9.566&\cellcolor{redc}   -10.908&\cellcolor{white}   &\cellcolor{redc}   -12.533&\cellcolor{redc}   -13.159&\cellcolor{redc}   -13.005\tabularnewline
   ~~&\cellcolor{white}   K = 4&\cellcolor{white}   SSVS&\cellcolor{white}   Common&   &\cellcolor{white}   &\cellcolor{redc}    -1.772&\cellcolor{redc}    -3.870&\cellcolor{redc}    -4.908&\cellcolor{white}   &\cellcolor{redc}    -8.006&\cellcolor{redc}    -7.445&\cellcolor{redc}    -6.180\tabularnewline
   ~~&\cellcolor{white}   &\cellcolor{white}   SSVS&\cellcolor{white}   State-specific&   &\cellcolor{white}   &\cellcolor{redc}   -27.363&\cellcolor{redc}   -27.798&\cellcolor{redc}   -27.909&\cellcolor{white}   &\cellcolor{redc}   -27.305&\cellcolor{redc}   -26.146&\cellcolor{redc}   -25.993\tabularnewline
\midrule
{\scshape Linear}&&&&&&&&&&&&\tabularnewline
   ~~&\cellcolor{white}   Taylor rule&\cellcolor{white}   &\cellcolor{white}   &   &\cellcolor{white}   &\cellcolor{redc}    -3.356&\cellcolor{redc}    -2.715&\cellcolor{redc}    -4.107&\cellcolor{white}   &\cellcolor{redc}    -4.302&\cellcolor{redc}    -3.359&\cellcolor{redc}    -3.218\tabularnewline
   ~~&\cellcolor{white}   Monetary&\cellcolor{white}   &\cellcolor{white}   &   &\cellcolor{white}   &\cellcolor{redc}    -0.617&\cellcolor{redc}    -2.009&\cellcolor{redc}    -3.243&\cellcolor{white}   &\cellcolor{redc}    -7.684&\cellcolor{redc}    -6.439&\cellcolor{redc}    -1.597\tabularnewline
   ~~&\cellcolor{white}   PPP&\cellcolor{white}   &\cellcolor{white}   &   &\cellcolor{white}   &\cellcolor{redc}    -0.787&\cellcolor{redc}    -1.344&\cellcolor{redc}    -1.349&\cellcolor{white}   &\cellcolor{redc}    -0.369&\cellcolor{greenc}     \textbf{0.193}&\cellcolor{greenc}     \textbf{0.769}\tabularnewline
   ~~&\cellcolor{white}   UIP&\cellcolor{white}   &\cellcolor{white}   &   &\cellcolor{white}   &\cellcolor{greenc}     0.090&\cellcolor{greenc}     0.002&\cellcolor{greenc}     0.035&\cellcolor{white}   &\cellcolor{redc}    -0.507&\cellcolor{redc}    -0.548&\cellcolor{redc}    -0.559\tabularnewline
\bottomrule
\end{tabular}
\caption{Cumulative one-, three-, and twelve-step-ahead LBFs (random walk benchmark) at the end of the full hold-out sample summarized for Australia, Canada, Switzerland and Japan. Values highlighted green are greater than zero, values highlighted red are smaller than zero, indicating a better or a weaker performance compared to the random walk. Best model in bold.}
\label{tab:out1}\end{center}}
\end{table}

\begin{table}[!tbp]
{\tiny
\begin{center}
\begin{tabular}{llllclrrrlrrr}
\toprule
\multicolumn{1}{l}{\bfseries }&\multicolumn{3}{c}{\bfseries Specification}&\multicolumn{1}{c}{\bfseries }&\multicolumn{8}{c}{\bfseries Log predicitive Bayes factor}\tabularnewline
\cline{2-4} \cline{6-13}
\multicolumn{1}{l}{}&\multicolumn{1}{c}{States}&\multicolumn{1}{c}{Shrinkage}&\multicolumn{1}{c}{Variance}&\multicolumn{1}{c}{}&\multicolumn{1}{c}{}&\multicolumn{1}{c}{1-step}&\multicolumn{1}{c}{3-step}&\multicolumn{1}{c}{12-step}&\multicolumn{1}{c}{}&\multicolumn{1}{c}{1-step}&\multicolumn{1}{c}{3-step}&\multicolumn{1}{c}{12-step}\tabularnewline
\midrule
{\scshape }&&&&&&&&&&&&\tabularnewline
   ~~&\cellcolor{white}   &\cellcolor{white}   &\cellcolor{white}   &   &\cellcolor{white}   KR&\cellcolor{white}   &\cellcolor{white}   &\cellcolor{white}   &\cellcolor{white}   NO&\cellcolor{white}   &\cellcolor{white}   &\cellcolor{white}   \tabularnewline
\midrule
{\scshape MS-TVP theoretical}&&&&&&&&&&&&\tabularnewline
   ~~&\cellcolor{white}   K = 4&\cellcolor{white}   None&\cellcolor{white}   Common&   &\cellcolor{white}   &\cellcolor{redc}    -91.338&\cellcolor{redc}    -73.679&\cellcolor{redc}    -40.795&\cellcolor{white}   &\cellcolor{redc}    -3.684&\cellcolor{redc}    -2.929&\cellcolor{redc}    -0.518\tabularnewline
   ~~&\cellcolor{white}   &\cellcolor{white}   None&\cellcolor{white}   State-specific&   &\cellcolor{white}   &\cellcolor{greenc}     15.820&\cellcolor{greenc}      8.041&\cellcolor{greenc}      5.458&\cellcolor{white}   &\cellcolor{greenc}     \textbf{5.623} &\cellcolor{redc}    -0.229&\cellcolor{redc}    -0.430\tabularnewline
   ~~&\cellcolor{white}   &\cellcolor{white}   SSVS&\cellcolor{white}   Common&   &\cellcolor{white}   &\cellcolor{redc}    -48.414&\cellcolor{redc}    -51.446&\cellcolor{redc}    -43.495&\cellcolor{white}   &\cellcolor{redc}    -4.393&\cellcolor{redc}    -5.983&\cellcolor{greenc}     \textbf{2.784} \tabularnewline
   ~~&\cellcolor{white}   &\cellcolor{white}   SSVS&\cellcolor{white}   State-specific&   &\cellcolor{white}   &\cellcolor{greenc}     16.783&\cellcolor{greenc}     12.134&\cellcolor{greenc}     10.815&\cellcolor{white}   &\cellcolor{redc}    -0.411&\cellcolor{greenc}     \textbf{2.961} &\cellcolor{redc}    -0.638\tabularnewline
\midrule
{\scshape MS-FT theoretical}&&&&&&&&&&&&\tabularnewline
   ~~&\cellcolor{white}   K = 4&\cellcolor{white}   None&\cellcolor{white}   Common&   &\cellcolor{white}   &\cellcolor{greenc}      7.901&\cellcolor{redc}    -30.386&\cellcolor{greenc}      4.123&\cellcolor{white}   &\cellcolor{greenc}     2.898&\cellcolor{greenc}     0.733&\cellcolor{redc}    -1.624\tabularnewline
   ~~&\cellcolor{white}   &\cellcolor{white}   None&\cellcolor{white}   State-specific&   &\cellcolor{white}   &\cellcolor{greenc}     11.263&\cellcolor{greenc}     12.679&\cellcolor{greenc}      8.562&\cellcolor{white}   &\cellcolor{redc}    -0.737&\cellcolor{redc}    -1.887&\cellcolor{redc}    -6.043\tabularnewline
   ~~&\cellcolor{white}   &\cellcolor{white}   SSVS&\cellcolor{white}   Common&   &\cellcolor{white}   &\cellcolor{redc}     -5.469&\cellcolor{greenc}      0.267&\cellcolor{greenc}      2.419&\cellcolor{white}   &\cellcolor{greenc}     2.278&\cellcolor{greenc}     2.075&\cellcolor{redc}    -1.958\tabularnewline
   ~~&\cellcolor{white}   &\cellcolor{white}   SSVS&\cellcolor{white}   State-specific&   &\cellcolor{white}   &\cellcolor{greenc}     11.321&\cellcolor{greenc}     13.302&\cellcolor{greenc}     13.524&\cellcolor{white}   &\cellcolor{redc}    -0.273&\cellcolor{greenc}     0.219&\cellcolor{redc}    -1.010\tabularnewline
\midrule
{\scshape MS-TVP kitchen-sink}&&&&&&&&&&&&\tabularnewline
   ~~&\cellcolor{white}   K = 2&\cellcolor{white}   SSVS&\cellcolor{white}   Common&   &\cellcolor{white}   &\cellcolor{redc}    -39.401&\cellcolor{redc}    -84.104&\cellcolor{redc}    -51.234&\cellcolor{white}   &\cellcolor{redc}    -4.853&\cellcolor{redc}    -5.579&\cellcolor{redc}    -7.278\tabularnewline
   ~~&\cellcolor{white}   &\cellcolor{white}   SSVS&\cellcolor{white}   State-specific&   &\cellcolor{white}   &\cellcolor{greenc}     12.215&\cellcolor{greenc}      9.430&\cellcolor{greenc}     14.090&\cellcolor{white}   &\cellcolor{greenc}     0.908&\cellcolor{greenc}     0.514&\cellcolor{redc}    -1.961\tabularnewline
   ~~&\cellcolor{white}   K = 3&\cellcolor{white}   SSVS&\cellcolor{white}   Common&   &\cellcolor{white}   &\cellcolor{redc}   -114.149&\cellcolor{redc}   -156.993&\cellcolor{redc}   -126.614&\cellcolor{white}   &\cellcolor{redc}   -12.394&\cellcolor{redc}   -10.818&\cellcolor{redc}   -14.418\tabularnewline
   ~~&\cellcolor{white}   &\cellcolor{white}   SSVS&\cellcolor{white}   State-specific&   &\cellcolor{white}   &\cellcolor{greenc}     \textbf{23.788} &\cellcolor{greenc}     22.774&\cellcolor{greenc}     25.185&\cellcolor{white}   &\cellcolor{redc}    -0.577&\cellcolor{redc}    -0.994&\cellcolor{redc}    -3.072\tabularnewline
   ~~&\cellcolor{white}   K = 4&\cellcolor{white}   SSVS&\cellcolor{white}   Common&   &\cellcolor{white}   &\cellcolor{redc}   -187.724&\cellcolor{redc}   -190.133&\cellcolor{redc}   -209.206&\cellcolor{white}   &\cellcolor{redc}   -16.165&\cellcolor{redc}   -17.181&\cellcolor{redc}   -19.057\tabularnewline
   ~~&\cellcolor{white}   &\cellcolor{white}   SSVS&\cellcolor{white}   State-specific&   &\cellcolor{white}   &\cellcolor{greenc}     21.200&\cellcolor{greenc}     \textbf{25.655} &\cellcolor{greenc}     \textbf{26.937} &\cellcolor{white}   &\cellcolor{redc}    -1.168&\cellcolor{redc}    -1.968&\cellcolor{redc}    -3.554\tabularnewline
\midrule
{\scshape MS-FT kitchen-sink}&&&&&&&&&&&&\tabularnewline
   ~~&\cellcolor{white}   K = 2&\cellcolor{white}   SSVS&\cellcolor{white}   Common&   &\cellcolor{white}   &\cellcolor{redc}    -57.328&\cellcolor{redc}    -73.540&\cellcolor{redc}    -42.487&\cellcolor{white}   &\cellcolor{redc}    -0.957&\cellcolor{redc}    -1.397&\cellcolor{redc}    -1.684\tabularnewline
   ~~&\cellcolor{white}   &\cellcolor{white}   SSVS&\cellcolor{white}   State-specific&   &\cellcolor{white}   &\cellcolor{greenc}     12.749&\cellcolor{greenc}     11.300&\cellcolor{greenc}     15.030&\cellcolor{white}   &\cellcolor{redc}    -1.861&\cellcolor{redc}    -1.944&\cellcolor{redc}    -5.495\tabularnewline
   ~~&\cellcolor{white}   K = 3&\cellcolor{white}   SSVS&\cellcolor{white}   Common&   &\cellcolor{white}   &\cellcolor{redc}    -28.715&\cellcolor{redc}    -72.196&\cellcolor{redc}    -64.311&\cellcolor{white}   &\cellcolor{redc}    -1.758&\cellcolor{redc}    -0.587&\cellcolor{redc}    -8.577\tabularnewline
   ~~&\cellcolor{white}   &\cellcolor{white}   SSVS&\cellcolor{white}   State-specific&   &\cellcolor{white}   &\cellcolor{greenc}     18.781&\cellcolor{greenc}     16.662&\cellcolor{greenc}     17.812&\cellcolor{white}   &\cellcolor{redc}   -11.315&\cellcolor{redc}   -12.182&\cellcolor{redc}   -14.188\tabularnewline
   ~~&\cellcolor{white}   K = 4&\cellcolor{white}   SSVS&\cellcolor{white}   Common&   &\cellcolor{white}   &\cellcolor{redc}    -50.802&\cellcolor{greenc}      3.179&\cellcolor{greenc}      5.573&\cellcolor{white}   &\cellcolor{redc}    -0.430&\cellcolor{redc}    -0.716&\cellcolor{redc}    -2.885\tabularnewline
   ~~&\cellcolor{white}   &\cellcolor{white}   SSVS&\cellcolor{white}   State-specific&   &\cellcolor{white}   &\cellcolor{greenc}      5.515&\cellcolor{greenc}      3.421&\cellcolor{greenc}      3.481&\cellcolor{white}   &\cellcolor{redc}   -21.445&\cellcolor{redc}   -22.107&\cellcolor{redc}   -23.478\tabularnewline
\midrule
{\scshape Linear}&&&&&&&&&&&&\tabularnewline
   ~~&\cellcolor{white}   Taylor rule&\cellcolor{white}   &\cellcolor{white}   &   &\cellcolor{white}   &\cellcolor{greenc}      2.200&\cellcolor{greenc}      0.357&\cellcolor{greenc}      0.169&\cellcolor{white}   &\cellcolor{redc}    -0.601&\cellcolor{greenc}     0.605&\cellcolor{redc}    -3.410\tabularnewline
   ~~&\cellcolor{white}   Monetary&\cellcolor{white}   &\cellcolor{white}   &   &\cellcolor{white}   &\cellcolor{redc}     -0.324&\cellcolor{redc}     -1.030&\cellcolor{redc}     -1.239&\cellcolor{white}   &\cellcolor{redc}    -0.875&\cellcolor{redc}    -0.115&\cellcolor{redc}    -2.687\tabularnewline
   ~~&\cellcolor{white}   PPP&\cellcolor{white}   &\cellcolor{white}   &   &\cellcolor{white}   &\cellcolor{greenc}      0.289&\cellcolor{redc}     -1.328&\cellcolor{redc}     -0.407&\cellcolor{white}   &\cellcolor{redc}    -0.656&\cellcolor{redc}    -0.071&\cellcolor{redc}    -1.127\tabularnewline
   ~~&\cellcolor{white}   UIP&\cellcolor{white}   &\cellcolor{white}   &   &\cellcolor{white}   &\cellcolor{redc}     -0.276&\cellcolor{redc}     -0.083&\cellcolor{redc}     -0.190&\cellcolor{white}   &\cellcolor{greenc}     0.042&\cellcolor{redc}    -0.087&\cellcolor{redc}    -0.354\tabularnewline
\midrule
\midrule
{\scshape }&&&&&&&&&&&&\tabularnewline
   ~~&\cellcolor{white}   &\cellcolor{white}   &\cellcolor{white}   &   &\cellcolor{white}   SE&\cellcolor{white}   &\cellcolor{white}   &\cellcolor{white}   &\cellcolor{white}   UK&\cellcolor{white}   &\cellcolor{white}   &\cellcolor{white}   \tabularnewline
\midrule
{\scshape MS-TVP theoretical}&&&&&&&&&&&&\tabularnewline
   ~~&\cellcolor{white}   K = 4&\cellcolor{white}   None&\cellcolor{white}   Common&   &\cellcolor{white}   &\cellcolor{redc}    -8.982&\cellcolor{redc}    -9.025&\cellcolor{redc}    -2.870&\cellcolor{white}   &\cellcolor{redc}    -5.982&\cellcolor{redc}    -1.321&\cellcolor{greenc}     5.162\tabularnewline
   ~~&\cellcolor{white}   &\cellcolor{white}   None&\cellcolor{white}   State-specific&   &\cellcolor{white}   &\cellcolor{redc}    -5.147&\cellcolor{redc}    -0.361&\cellcolor{redc}    -2.487&\cellcolor{white}   &\cellcolor{redc}    -3.696&\cellcolor{greenc}     0.197&\cellcolor{greenc}     3.626\tabularnewline
   ~~&\cellcolor{white}   &\cellcolor{white}   SSVS&\cellcolor{white}   Common&   &\cellcolor{white}   &\cellcolor{redc}    -6.630&\cellcolor{redc}    -5.702&\cellcolor{greenc}     \textbf{1.461} &\cellcolor{white}   &\cellcolor{redc}    -1.987&\cellcolor{greenc}     1.101&\cellcolor{greenc}     \textbf{6.036} \tabularnewline
   ~~&\cellcolor{white}   &\cellcolor{white}   SSVS&\cellcolor{white}   State-specific&   &\cellcolor{white}   &\cellcolor{redc}    -5.582&\cellcolor{greenc}     1.893&\cellcolor{redc}    -0.237&\cellcolor{white}   &\cellcolor{redc}    -0.432&\cellcolor{greenc}     1.954&\cellcolor{greenc}     4.312\tabularnewline
\midrule
{\scshape MS-FT theoretical}&&&&&&&&&&&&\tabularnewline
   ~~&\cellcolor{white}   K = 4&\cellcolor{white}   None&\cellcolor{white}   Common&   &\cellcolor{white}   &\cellcolor{redc}    -1.387&\cellcolor{redc}    -1.828&\cellcolor{redc}    -2.744&\cellcolor{white}   &\cellcolor{greenc}     0.430&\cellcolor{greenc}     2.637&\cellcolor{greenc}     0.912\tabularnewline
   ~~&\cellcolor{white}   &\cellcolor{white}   None&\cellcolor{white}   State-specific&   &\cellcolor{white}   &\cellcolor{redc}    -0.370&\cellcolor{greenc}     1.454&\cellcolor{redc}    -2.202&\cellcolor{white}   &\cellcolor{redc}    -3.299&\cellcolor{redc}    -1.581&\cellcolor{redc}    -0.467\tabularnewline
   ~~&\cellcolor{white}   &\cellcolor{white}   SSVS&\cellcolor{white}   Common&   &\cellcolor{white}   &\cellcolor{greenc}     0.694&\cellcolor{greenc}     1.635&\cellcolor{greenc}     0.271&\cellcolor{white}   &\cellcolor{greenc}     \textbf{2.522} &\cellcolor{greenc}     \textbf{3.532} &\cellcolor{greenc}     1.887\tabularnewline
   ~~&\cellcolor{white}   &\cellcolor{white}   SSVS&\cellcolor{white}   State-specific&   &\cellcolor{white}   &\cellcolor{greenc}     \textbf{3.139} &\cellcolor{greenc}     \textbf{3.433} &\cellcolor{greenc}     0.806&\cellcolor{white}   &\cellcolor{redc}    -0.595&\cellcolor{greenc}     0.148&\cellcolor{redc}    -0.734\tabularnewline
\midrule
{\scshape MS-TVP kitchen-sink}&&&&&&&&&&&&\tabularnewline
   ~~&\cellcolor{white}   K = 2&\cellcolor{white}   SSVS&\cellcolor{white}   Common&   &\cellcolor{white}   &\cellcolor{redc}   -15.887&\cellcolor{redc}   -25.834&\cellcolor{redc}   -10.620&\cellcolor{white}   &\cellcolor{redc}    -1.545&\cellcolor{greenc}     0.370&\cellcolor{redc}    -0.034\tabularnewline
   ~~&\cellcolor{white}   &\cellcolor{white}   SSVS&\cellcolor{white}   State-specific&   &\cellcolor{white}   &\cellcolor{greenc}     0.412&\cellcolor{redc}    -0.519&\cellcolor{redc}    -0.114&\cellcolor{white}   &\cellcolor{redc}    -1.090&\cellcolor{greenc}     2.290&\cellcolor{greenc}     2.428\tabularnewline
   ~~&\cellcolor{white}   K = 3&\cellcolor{white}   SSVS&\cellcolor{white}   Common&   &\cellcolor{white}   &\cellcolor{redc}   -16.742&\cellcolor{redc}   -19.861&\cellcolor{redc}   -13.622&\cellcolor{white}   &\cellcolor{redc}    -0.896&\cellcolor{greenc}     1.430&\cellcolor{greenc}     0.443\tabularnewline
   ~~&\cellcolor{white}   &\cellcolor{white}   SSVS&\cellcolor{white}   State-specific&   &\cellcolor{white}   &\cellcolor{redc}    -1.821&\cellcolor{redc}    -0.810&\cellcolor{redc}    -2.402&\cellcolor{white}   &\cellcolor{redc}    -1.363&\cellcolor{greenc}     1.245&\cellcolor{greenc}     0.352\tabularnewline
   ~~&\cellcolor{white}   K = 4&\cellcolor{white}   SSVS&\cellcolor{white}   Common&   &\cellcolor{white}   &\cellcolor{redc}   -20.343&\cellcolor{redc}   -24.384&\cellcolor{redc}   -21.758&\cellcolor{white}   &\cellcolor{redc}    -1.955&\cellcolor{greenc}     0.359&\cellcolor{redc}    -0.078\tabularnewline
   ~~&\cellcolor{white}   &\cellcolor{white}   SSVS&\cellcolor{white}   State-specific&   &\cellcolor{white}   &\cellcolor{redc}    -2.911&\cellcolor{redc}    -2.124&\cellcolor{redc}    -3.380&\cellcolor{white}   &\cellcolor{redc}    -2.086&\cellcolor{redc}    -0.007&\cellcolor{redc}    -1.260\tabularnewline
\midrule
{\scshape MS-FT kitchen-sink}&&&&&&&&&&&&\tabularnewline
   ~~&\cellcolor{white}   K = 2&\cellcolor{white}   SSVS&\cellcolor{white}   Common&   &\cellcolor{white}   &\cellcolor{redc}    -0.319&\cellcolor{redc}    -1.792&\cellcolor{redc}    -3.581&\cellcolor{white}   &\cellcolor{greenc}     0.199&\cellcolor{greenc}     2.924&\cellcolor{redc}    -1.127\tabularnewline
   ~~&\cellcolor{white}   &\cellcolor{white}   SSVS&\cellcolor{white}   State-specific&   &\cellcolor{white}   &\cellcolor{redc}    -0.941&\cellcolor{redc}    -3.452&\cellcolor{redc}    -6.323&\cellcolor{white}   &\cellcolor{redc}    -5.674&\cellcolor{redc}    -4.547&\cellcolor{redc}    -9.718\tabularnewline
   ~~&\cellcolor{white}   K = 3&\cellcolor{white}   SSVS&\cellcolor{white}   Common&   &\cellcolor{white}   &\cellcolor{redc}    -9.635&\cellcolor{redc}    -4.012&\cellcolor{redc}    -3.461&\cellcolor{white}   &\cellcolor{redc}    -5.519&\cellcolor{redc}    -1.261&\cellcolor{redc}    -3.492\tabularnewline
   ~~&\cellcolor{white}   &\cellcolor{white}   SSVS&\cellcolor{white}   State-specific&   &\cellcolor{white}   &\cellcolor{redc}    -9.374&\cellcolor{redc}   -10.639&\cellcolor{redc}   -14.715&\cellcolor{white}   &\cellcolor{redc}   -14.169&\cellcolor{redc}   -12.835&\cellcolor{redc}   -14.602\tabularnewline
   ~~&\cellcolor{white}   K = 4&\cellcolor{white}   SSVS&\cellcolor{white}   Common&   &\cellcolor{white}   &\cellcolor{redc}    -2.274&\cellcolor{redc}    -2.466&\cellcolor{redc}    -4.302&\cellcolor{white}   &\cellcolor{redc}    -8.277&\cellcolor{redc}    -5.785&\cellcolor{redc}    -6.437\tabularnewline
   ~~&\cellcolor{white}   &\cellcolor{white}   SSVS&\cellcolor{white}   State-specific&   &\cellcolor{white}   &\cellcolor{redc}   -23.645&\cellcolor{redc}   -25.141&\cellcolor{redc}   -25.895&\cellcolor{white}   &\cellcolor{redc}   -26.066&\cellcolor{redc}   -25.454&\cellcolor{redc}   -26.217\tabularnewline
\midrule
{\scshape Linear}&&&&&&&&&&&&\tabularnewline
   ~~&\cellcolor{white}   Taylor rule&\cellcolor{white}   &\cellcolor{white}   &   &\cellcolor{white}   &\cellcolor{redc}    -4.328&\cellcolor{redc}    -4.140&\cellcolor{redc}    -3.620&\cellcolor{white}   &\cellcolor{redc}    -4.893&\cellcolor{redc}    -3.199&\cellcolor{redc}    -0.890\tabularnewline
   ~~&\cellcolor{white}   Monetary&\cellcolor{white}   &\cellcolor{white}   &   &\cellcolor{white}   &\cellcolor{redc}    -3.632&\cellcolor{redc}    -4.209&\cellcolor{redc}    -7.265&\cellcolor{white}   &\cellcolor{redc}    -2.725&\cellcolor{redc}    -1.626&\cellcolor{redc}    -2.589\tabularnewline
   ~~&\cellcolor{white}   PPP&\cellcolor{white}   &\cellcolor{white}   &   &\cellcolor{white}   &\cellcolor{redc}    -2.508&\cellcolor{redc}    -2.435&\cellcolor{redc}    -4.737&\cellcolor{white}   &\cellcolor{redc}    -0.652&\cellcolor{greenc}     0.335&\cellcolor{redc}    -0.748\tabularnewline
   ~~&\cellcolor{white}   UIP&\cellcolor{white}   &\cellcolor{white}   &   &\cellcolor{white}   &\cellcolor{greenc}     0.275&\cellcolor{redc}    -0.025&\cellcolor{redc}    -0.139&\cellcolor{white}   &\cellcolor{redc}    -0.388&\cellcolor{redc}    -0.373&\cellcolor{greenc}     0.194\tabularnewline
\bottomrule
\end{tabular}
\caption{Cumulative one-, three-, and twelve-step-ahead LBFs (random walk benchmark) at the end of the full hold-out sample summarized for South Korea, Norway, Sweden and the United Kingdom. Values highlighted green are greater than zero, values highlighted red are smaller than zero, indicating a better or a weaker performance compared to the random walk. Best model in bold.}
\label{tab:out2}\end{center}}
\end{table}

\section{Concluding remarks}
In this paper, we propose a Bayesian non-linear time series model for exchange rates. Our framework, a multi process model, allows for dynamically switching between selected theoretical exchange rate models that are used to guide the specific choice of covariates included. As an additional novelty, we assume that the transition probabilities vary over time and depend on a measure of the monetary policy stance at home and abroad. This feature enables us to capture breaks in the policy rule of the central bank that, in turn, could impact the prevailing structural exchange rate model adopted. For instance, our framework entails to dynamically switch between models if short-term interest rates hit the zero lower bound.

We use this framework to predict eight exchange rates vis-\'{a}-vis the US dollar. Considering the transition probabilities, we find considerable evidence of time-variation. The filtered probabilities indicate that especially after interest rates are bound to zero, model evidence shifts in favor of models other than the Taylor rule based models, highlighting the necessity to control for model uncertainty. To assess whether this feature also translates into predictive accuracy gains, we conduct a forecasting exercise. There, we find that results appear to be rather mixed, with point forecasts being only slightly better than the ones obtained from standard models. In terms of density predictions, however, we observe pronounced accuracy increases for selected exchange rates.

\singlespacing
\bibliographystyle{./bibtex/fischer}
\bibliography{./bibtex/References}
\addcontentsline{toc}{section}{References}

\onehalfspacing

\newpage
\begin{appendices}\crefalias{section}{appsec}
\setcounter{equation}{0}
\renewcommand\theequation{A.\arabic{equation}}
\section{MCMC algorithm}\label{append:mcmc}
After specifying appropriate starting values, the Gibbs sampler iterates through the following steps:
\begin{enumerate}
\item Sample parameters of the measurement equation $\bm{\theta} = (\bm \beta, \bm \sigma^2, \bm \delta)$ from $p(\bm{\theta}| \bm{\tilde{S}}_T, \bm{\Delta \tilde{e}}_T)$, with $\bm{\Delta \tilde{e}}_T = (\Delta e_1, \dots \Delta e_T)$.
\begin{enumerate}
\item Conditional on the exchange rate data $\bm{\tilde{e}}_T$ and allocation of the states $\bm{\tilde{S}}_T$, sampling $\bm{\beta}$ and $\bm{\sigma}^2$ can be done in a standard way by drawing the coefficients $\bm{\beta}_{k}$ for $k = 1, \dots, K$ in a block from a multivariate Normal distribution and the variances independently for each state from an inverse Gamma distribution.
\item Conditional on $\bm{\beta}_{k}$ of state $S_t = k$, one is able to sample the elements of $\bm{\delta}_{k}$ from a Bernoulli conditional posterior distribution. 
\end{enumerate}
\item For sampling the unknown states $p(\bm{\tilde{S}}_T|\bm{\Delta \tilde{e}}_T, \bm{\theta}, \bm{\tilde{Z}}_{T}, \bm{\gamma})$ we adapt the filtering algorithm put forth by \citet{kimnelson1999}. 

\item Following \citet{polson2013polya}, we sample multinomial coefficients from $p(\bm{\gamma}|\bm{\tilde{S}}_T, \bm{\tilde{Z}}_{T})$ to construct the time-varying transition probabilities.

\item In case of label switching, we implement an additional permutation step outlined in \citet{fruhwirth2001ident} ensuring an equal probability of each mode appearing in the posterior distribution.
\end{enumerate}

For the second step, we follow \citet{kimnelson1999} and sample $\bm{\tilde{S}}_T$ in a block using a multimove Gibbs sampler. This implies simulating $\bm{\tilde{S}}_T$ in block from the following joint conditional distribution
\begin{equation*}
\begin{aligned}
p(\bm{\tilde{S}}_T|\bm{\Delta \tilde{e}}_T, \bm{\theta}, \bm{\tilde{Z}}_{T}, \bm{\gamma}),  
\end{aligned}    
\end{equation*}
and applying a forward filtering and backward smoothing algorithm \citep{fruhwirth2006}.

Unfortunately for Step 3, there is no closed form of the multinomial logit posterior distribution.
\citet{fruhwirthfruhwirth2010} therefore introduce two auxiliary layers for estimating state-specific utilities. \citet{polson2013polya} take a different approach by representing the likelihood of multinomial logit model as a scale mixture of Gaussians with a P\'{o}lya Gamma  mixing distribution. In a hierarchical form (by introducing a single layer of latent variables) this strategy implies that the multinomial coefficients are drawn from a set of Gaussians, and the auxiliary variables are sampled from a P\'{o}lya Gamma distribution. This approach has the advantage of fast convergence, simple implementation and no need for an additional layer for approximating the error distribution.

Following \citet{polson2013polya}, conditional on the latent states $\bm{\tilde{S}}_T$ and the auxiliary variables $\psi_{jt}$ for $t = 1,\dots, T$ sampled from a $\Pee \G(\psi_{jt}|1,0)$ distribution, the posterior quantities are given by
\begin{equation*}
\begin{aligned}
\bar{\bm{V}}_j =& \left(\bm{\tilde{Z}}_{T}'\bm{\Psi}_j \bm{\tilde{Z}}_{T} + \underline{\bm{V}}^{-1} \right)^{-1} \\
\bar{\bm{\gamma}}_{kj} =& \bar{\bm{V}}_j \left (\bm{\tilde{Z}}_{T}'\bm{\tilde{\kappa}}_j \right) \\
\end{aligned}
\end{equation*}
where $\bm{\gamma}_{kK}$ is zero for $k = 1, \dots, K$ for reasons of identification, $\bm{\Psi}_j =\text{diag}(\psi_{j1}, \dots, \psi_{jT})$ and
\begin{equation*}
\bm{\tilde{\kappa}}_j = \left( \mathcal{I}[S_t = j]-0.5 \right) - \bm{\Psi}_j \bm{C}\bm{e}_j.
\end{equation*}
$\bm{C}$ is a $T \times K-1$ dimensional matrix, with elements being defined as 
\begin{equation*}
C_{tl} = \log\left( \sum_{k \neq l} \exp(\bm{Z}'_{t}\bm{\gamma}_{kj})\right),
\end{equation*}
and $\bm{e}_j$ represents the unit vector with an one at the \textit{j}th position.

\end{appendices}

\end{document}